\documentclass[twocolumn]{test631}

\usepackage{amsmath,amstext}
\usepackage[T1]{fontenc}
\usepackage[figure,figure*]{hypcap}
\usepackage{comment}
\usepackage{graphicx}
\usepackage{natbib}
\usepackage{epsfig}
\usepackage{epstopdf}
\usepackage{enumerate}
\usepackage{tabularx}
\usepackage{multirow}
\usepackage{booktabs}
\usepackage{color}
\usepackage[normalem]{ulem}  
\usepackage{xspace}

\turnoffeditone

\newcommand{\msun}{M$_{\odot}$\xspace}
\newcommand{\kpush}{$k_{\rm push}$\xspace}
\newcommand{\trise}{$t_{\rm rise}$\xspace}
\newcommand{\paperI}{Paper~I}
\newcommand{\paperII}{Paper~II}
\newcommand{\paperIII}{Paper~III}
\newcommand{\paperIV}{Paper~IV}
\newcommand{\Ye}{Y$_e$\xspace}
\newcommand{\BHBlp}{BHB$\lambda\phi$\xspace}

\graphicspath{{./}{figures/}{nucleo/}{0finalfigures/}{1finalfigures/}{2finalfigures/}{3finalfigures/}{4finalfigures/}{5finalfigures/}{6finalfigures/}}

\received{\today}
\revised{}
\accepted{}
\submitjournal{ApJ}

\shorttitle{Pushing 1D CCSNe to explosions: EOS dependency}
\shortauthors{Ghosh et al.}

\begin{document}

\title{PUSHing core-collapse supernovae to explosions in spherical symmetry V:\\ 
Equation of state dependency of explosion properties, nucleosynthesis yields, and compact remnants}

\correspondingauthor{Ghosh and Fr\"ohlich}
\email{sghosh23@ncsu.edu, cfrohli@ncsu.edu}

\author[0000-0001-7868-6944]{Somdutta Ghosh}
\affiliation{Department of Physics, North Carolina State University, Raleigh NC 27695, USA}

\author[0000-0003-2540-3845]{Noah Wolfe}
\affiliation{Department of Physics, North Carolina State University, Raleigh NC 27695, USA}

\author[0000-0003-0191-2477]{Carla Fr\"ohlich}
\affiliation{Department of Physics, North Carolina State University, Raleigh NC 27695, USA}

\begin{abstract}
In this fifth paper of the series, we use the parametrized, spherically symmetric explosion method PUSH to investigate the impact of eight different nuclear equations of state (EOS). We present and discuss the explosion properties and the detailed nucleosynthesis yields, and predict the remnant (neutron star or black hole) for all our simulations. For this, we perform two sets of simulations. One, a complete study of non-rotating stars from 11 to 40 \msun at three different metallicities using the SFHo \edit1{EOS}. And two, a suite of simulations for \edit1{four} progenitors (16 \msun at three metallicities \added{and 25 \msun at solar metallicity}) for eight different nuclear EOS. We compare our predicted explosion energies and yields to observed supernovae and to the metal-poor star HD~84937. We find EOS-dependent differences in the explosion properties and the nucleosynthesis yields. However, when comparing to observations, these differences are not large enough to rule out any EOS considered in this work.
\end{abstract}

\keywords{}

\section{Introduction} 
\label{sec:intro}

When massive stars ($\gtrsim$ 8M$_{\odot}$) reach the end of their hydrostatic lives, they undergo gravitational collapse of their core. This collapse can result in a spectacular explosion, called a core-collapse supernova (CCSN), which disrupts all but the very core of the star and leaves behind a neutron star (NS). In these bright and energetic events many chemical elements are synthesized and subsequently ejected in the explosion, thus enriching the surrounding gas with `metals' (elements heavier than He). For some stars, the gravitational collapse of the core cannot be turned around into a successful explosion. Instead, they fail to explode and ultimately form black holes (BHs). Exactly which stars successfully explode and which stars fail to explode remains an open question. Moreover, the exact explosion mechanism (by which the stalled shock resulting from the core collapse is revived) is still not fully understood, despite decades of efforts. 

The CCSN problem is complex, requiring the inclusion of general relativity, (magneto-) hydrodynamics, neutrino transport, and physics of nuclear matter at high densities in simulations. Additionally, CCSNe need to be simulated in full three-dimensional space. This means that despite considerable progress in model sophistication \citep[c.f.][and references therein for a discussion of the status of explosion engine simulations]{Mueller.LRCA:2020}, CCSN simulations remain computationally expensive endeavors which are not suitable for large-scale investigations of tens to hundreds of models. Here, we use a complimentary approach which allows to more systematically study the effects of one of the crucial --- and poorly known -- ingredient: the nuclear equation of state (EOS).

 \edit1
 {The mass-radius relationship of neutron stars is one method to constrain the nuclear EOS \citep[see][for some recent examples]{Miller2019PSRJ0030,Riley2019NICER,Raaijmakers2020NICERLIGO,Miller2021PSRJ0740,Riley2021PSRJ0740,Pang2021J0740}.} Combining the laboratory measurement of the neutron skin at PREX-II with pulsar timing observations in the radio and X-ray and using a Bayesian statistics approach, led to \edit1{complementary} constraints on the mass-radius relationship \citep{BiswasPREXII}. Additional constraints can be derived from the detection of a binary neutron star by LIGO/VIRGO \citep{Abbott2017GW170817} and from the kilonova and the GRB afterglow \citep{AbbottMultiMessenger2017}.

There exist several studies in the literature with a focus on the nuclear EOS in CCSNe. 
For example, \citet{Schneider2019} constructed a series of finite-temperature EOS for different effective nucleon masses and investigated their impact on the collapse of a $20 M_{\odot}$ star in spherical symmetry and in an octant-3D simulation. They conclude that a larger effective mass \added{--- while keeping all other parameters in the EOS fixed ---} leads to larger neutrino heating and hence an increased likelihood for explosion. A similar results was found by \citet{Yasin2020} also using spherically symmetric simulations with an increased heating factor to achieve explosions. \added{During the revision of this paper, a study conducted in spherical symmetry by \cite{2021Boccioli} found that the EOS which generates higher central entropy soon after bounce produces a faster and stronger explosion.} In axisymmetric simulations with Boltzmann transport, the different nuclear composition due to different EOS models resulted in different energy losses due to photodissociation, leading to an explosion in one case and no explosion in the other case \citep{Harada2020}. In a 3D simulation of a $18.8 M_{\odot}$ star, the shock runaway occurs earlier for a hotter PNS which \edit1{has} higher neutrino luminosities and harder neutrino spectra \citep{bollig2021}. 
The amount of (neutrino-driven) mass ejection in failed supernovae carries the imprint of the stiffness of the nuclear EOS \citep{ivanov2021}. The neutrino signal from failed supernovae can be used to constraint the temperature-dependence of the nuclear EOS \citep{schneider2020}.
The impact of the nuclear EOS can also be seen in the gravitational wave (GW) signal from CCSNe. In particular for fast rotating stars, the early GW signal has some EOS dependency \citep{Richers2017}. 
While all of these studies found some effect due to the properties of the nuclear EOS used, no truly systematic investigations have been undertaken. 

This work builds upon a series of investigations using the PUSH method. The PUSH method is an effective method which is built upon the neutrino-driven mechanism for the central engine of CCSNe. The core idea of the PUSH method is to mimic the enhanced neutrino energy deposition as observed in multi-dimensional simulations in computationally more efficient, spherically symmetric simulations. The PUSH method was first introduced in \cite{push1} (hereafter \paperI) and subsequently refined in \cite{push2} (hereafter \paperII). The PUSH method has been used to study the explosion and remnant properties of solar metallicity progenitors in \paperII~and of low and zero metallicity progenitors in \cite{push4} (hereafter \paperIV). We presented the detailed nucleosynthesis yields from PUSH models in  \cite{push3} (hereafter \paperIII) for solar metallicity models and in \paperIV~for low and zero metallicity models. The light curves and time-dependent spectra have been calculated and analysed for all exploding models across all metallicities in \cite{pushlc}. All of these studies have been performed using the HS(DD2) nuclear equation of state \citep{DD2}.

In this paper, we investigate how the explosion properties, remnant properties, and nucleosynthesis yields depend on the nuclear EOS. For this, we have selected six nuclear EOS models which can accommodate a maximum neutron star mass of $>2M_{\odot}$: 
The SFHo and SFHx from \citet{SFHo},
the HS(DD2), HS(TM1), and HS(NL3) from \citet{hempel2010,hempel2012}, and
the BHB$\lambda\phi$ EOS model from \citet{banik2014}. In addition, we include two EOS models commonly used in the literature: LS220 \citep{ls220} and Shen \citep{shen98_Thphy,shen98_nuphy} for comparison.

This article is organized as follows:
In section \ref{sec:methods} we discuss our setup for this study, including the input models and the nuclear EOS models. 
In section \ref{sec:explosionproperties.sfho} we present the key findings on the explosion properties when using the SFHo EOS model. 
We then investigate the trends for all eight nuclear EOS models for three 16~\msun~ progenitors at different metallicities \added{and a $25M_{\odot}$ progenitor at solar metallicity} in section \ref{sec:exploutcomes.all}. 
Section \ref{sec:nucleo} discusses the nucleosynthesis yields and trends across compactness and progenitor metallicity for the SFHo EOS. 
We compare our results with observations of supernovae and with abundances observed in metal-poor stars in section \ref{sec:observations}. 
Finally, we present the resulting NS and BH distributions for the SFHo EOS. 
We summarize the work in section \ref{sec:summary}.

\section{Models and Input} \label{sec:methods}

\subsection{Hydrodynamics and Neutrino Transport} 
\label{subsec:hydro}

We simulate the collapse, bounce, and post-bounce evolution using the same setup as in our previous work (\paperII, \paperIII, and \paperIV) except for the nuclear equation of state (see section \ref{subsec:EOS}). Hence, we only summarize here the key points relevant to this study.
 We solve the general relativistic hydrodynamic equations in spherical symmetry using AGILE (\citealt{agile}) which features an adaptive grid and implicit time evolution.  We apply the deleptonization scheme from \citet{deleptonizationscheme} during the collapse phase. For electron flavor neutrinos ($\nu_e$) and anti-neutrinos ($\bar{\nu}_e$), we use the isotropic diffusion source approximation (IDSA) scheme of \citet{idsa}. For heavy-lepton flavour neutrinos and anti-neutrinos ($\nu_x$ = $\nu_{\mu}$, $\bar{\nu}_\mu$, $\nu_{\tau}$, $\bar{\nu}_{\tau}$ ) we use an advanced spectral leakage  (ASL) scheme \citep{asl}. More details of the individual components of our code can be found in \citet{push2}.

To achieve explosions in otherwise not exploding models in spherical symmetry, we use the PUSH method \added{introduced in} \citet{push1} \added{and refined in} \citet{push2}. PUSH is a physically motivated, effective method that mimics in spherical symmetry the enhanced heating (due to convection and accretion) observed in multi-dimensional simulations. A fraction of the heavy-flavor neutrino energies is deposited behind the shock via a parametrized heating term $Q^+_{\mathrm{push}}(r,t)$ (cf.\ equation (4) in \paperI). This heating includes a spatial term (so that the extra heating only occurs where electron neutrinos and anti-neutrino are heating), a temporal term (which includes the free parameters \kpush and \trise), and a dependence on the spectral energy flux of a single heavy-lepton neutrino flavor.

For the calibration of the free parameters in the PUSH method, we follow \paperIV~and use the standard calibration presented in \paperII. In this calibration, $t_{\mathrm{rise}}=400$~ms and $k_{\mathrm{push}}(\xi) = a\xi^2 + b \xi + c$ (where $a = -23.99$, $b = 13.22$, and $c = 2.5$) is parametrized as function of the compactness (as introduced in \cite{compactness})
\begin{equation}
    \xi_M = \frac{M/M_{\odot}}{R(M)/1000km}.
    \label{eq:compactness}
\end{equation}

With this setup, we perform hydrodynamic simulations of the collapse, bounce, and post-bounce evolution. We use a total of 180 radial zones, which includes the progenitor star up to the helium shell. With the adaptive grid, a greater number of zones are placed in regions where the thermodynamic quantities show steeper gradients. Hence, in the post-bounce and explosion phases, the surface of the proto-neutron star (PNS) and the shock are the regions that are better resolved. We follow the evolution for up to 5 seconds for exploding models. If the shock reaches the edge of the computational domain before then, the simulation is terminated at that time.

We categorize the outcome of each simulation as `exploding model', `failed SN' or `BH formation'. For the failed SN and BH forming models, the simulation time depends on individual models and is not of much importance in this paper. We differentiate a 'direct collapse' meaning no stalled shock, and 'failed SN' for a stalled shock that never revives by looking at the central density evolution. If the central density rises very rapidly and reaches more than 10$^{15}$ g/cm$^3$ we call it a BH forming model.

The explosion energy and other explosion properties are calculated in the same way as in \cite{push1}. The explosion energy is the sum of thermal, kinetic, and gravitational energy integrated over the mass of the star from the outer layers to the time-dependent mass cut. The mass cut at each time step is set at the mass coordinate which has the highest value of explosion energy. The explosion time is obtained from the time after bounce when the shock reaches 500~km.

\subsection{Nuclear Equation of State} 
\label{subsec:EOS}

\begin{figure}
    \centering
    \includegraphics[width=0.5\textwidth]{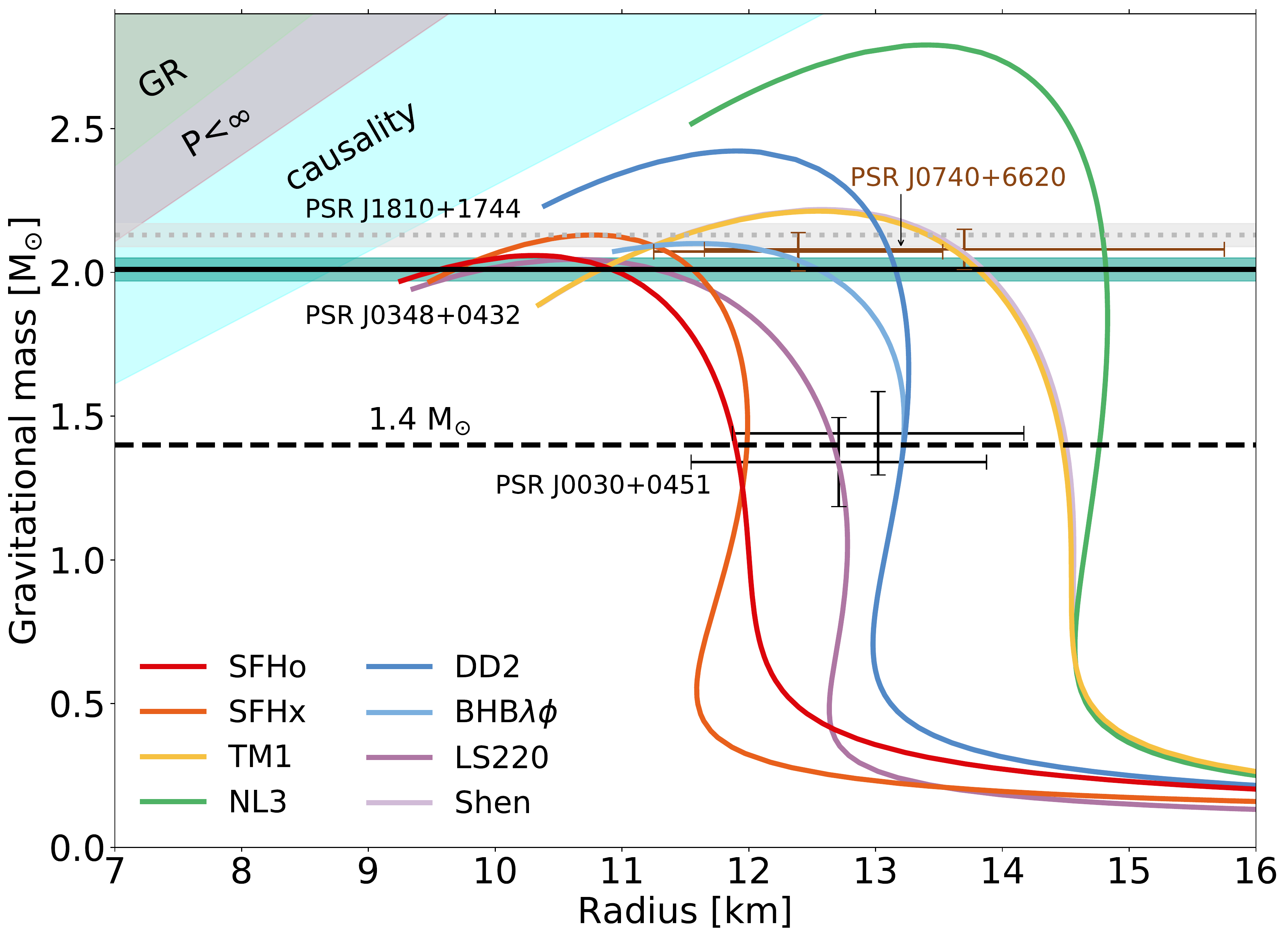}
    \caption{The mass-radius relationship for the EOS models \added{of this work} at $T=0.1$~MeV. 
    The horizontal line with errorbars represents the constraint from pulsar PSR~J0348+0432. The grey shaded line indicates pulsar PSR~J1810+1744. \added{The crosses represents the mass-radius constraints obtained from PSR J0030+0451 (black) and PSR J0740+6620 (brown).}
    The black dashed line is at $1.4 M_{\odot}$. 
    \label{fig:mass-radius}
    }
\end{figure}

In this work, we explore eight different nuclear equations of state (listed in table \ref{tab:EOS}). 
Six of these EOS models -- namely SFHo, SFHx, HS(TM1), HS(NL3), HS(DD2), and BHB$\lambda \phi$ -- are based on the relativistic mean-field (RMF) interaction of nucleons and include a nuclear statistical equilibrium (NSE) distribution of nuclei below saturation densities. These six EOS models differ in the parametrization of the nuclear interaction and the nuclear mass tables used for the experimentally unknown nuclei. All of these EOS models allow for a maximum NS mass above 2~\msun. 
In the following discussions we will reference the EOS models by their names without the HS prefix, e.g.\ HS(DD2) will be referred to as DD2 (see also the first column in Table \ref{tab:EOS}). 
In addition, we include two EOS models commonly used in the literature: the LS220 model \citep{ls220} and the Shen-STOS EOS model \citep{shen98_nuphy,shen98_Thphy}. Both of these EOS models also support a maximum NS mass of $\sim 2 M_{\odot}$. 
The mass-radius relationship \added{at $T=0.1$~MeV} for all 8 EOS models is shown in figure~\ref{fig:mass-radius}. The horizontal lines are the mass measurement for pulsar PSR~J0348+0432 \citep{demorest.nature:2010}. This was the most massive neutron star observed. Only very recently (during the final stages of the work presented here) this has been surpassed by PSR~J1810+1744 for which \citet{Romani2021} derived a mass of $2.13 \pm 0.04 M_{\odot}$ using spectrophotometry. \added{There are no radius constraints for these pulsars. However, recent NICER observations provide a mass and a radius constraint for two pulsars: PSR~J0030+0451 \citep{Riley2019NICER, Miller2019PSRJ0030} PSR~J0740+6620 \citep{Riley2021PSRJ0740, Miller2021PSRJ0740}, shown as crosses in figure~\ref{fig:mass-radius}.}

The SFHo and SFHx parametrizations were developed from the neutron star mass-radius measurements combined with the charge radii and binding energies of $^{208}$Pb and $^{60}$Zn \citep{SFHo}. For the experimentally unknown nuclei, the mass table from the finite range droplet model (FRDM) is used. 
The SFHo EOS is more consistent with the observations of NS mass and radii as well as with theoretical constraints from nuclear experiments on matter near and below saturation density known at the time. The SFHx EOS model is similar to SFHo EOS model, except that it attempts to minimize the radius of low-mass NS while remaining consistent with other constraints. 
The NL3 parametrization uses binding energies and charge radii of ten nuclei and of neutrons and the mass table from \citet{NL3masses}. 
The TM1 parametrization is similar to NL3. It was fit to charge radii and binding energies of heavy nuclei and uses the mass table of \citet{TM1masses}. It differs from NL3 at high densities due to the inclusion of vector self-interactions. 
Finally, the DD2 and the BLB$\lambda \phi$ EOS models are both based on the DD2 parametrization which uses experimental nucleon masses and the same FRDM mass table for experimentally unknown nuclei as SFHo and SFHx. The BHB$\lambda \phi$ EOS additionally includes the lambda hyperon and the repulsive hyperon-hyperon interaction (represented by the $\phi$). 
\added{It is important to note that these six EOS models represent six discrete EOS models. While they can be ordered by increasing values of any of the parameters listed in table~\ref{tab:EOS}, they also differ in their other properties, including the parameterization of the nuclear interaction. Hence, one cannot easily make statements about the effect of any of these parameters alone on the explosion properties and the nucleosynthesis yields.}

The EOS developed by \citet{shen98_nuphy,shen98_Thphy} (``Shen'') uses a RMF approach and the Thomas-Fermi approximation for the nuclei. 
The LS220 EOS developed by \citet{ls220} with the incompressibility parameter $K=220$~MeV is based on a non-relativistic parametrization of the nuclear interactions and the nuclei are calculated in the liquid-drop approach. Both of these EOS models use the single nucleus approximation in which the distribution of nuclei at finite temperatures is represented by a single nucleus (SNA). While this approximation does not affect the EOS much \citep{Burrows.Lattimer:1984}, it only describes the composition in an averaged way, which may affect the supernova dynamics.


We also include an extension of the EOS to non-NSE conditions by introducing a transition region. In this region, we apply some parametrized burning for temperatures between 0.3 and 0.4 MeV and introduce a temperature-dependent burning timescale. This transforms the initial non-NSE composition toward NSE. In the non-NSE region, we describe the nuclear composition using 25 representative nuclei which include neutrons, protons, $\alpha$-nuclei, and a few asymmetric isotopes up to iron-group nuclei. We map the abundances of the progenitor to these nuclei in a way that is consistent with the provided \added{electron fraction} \Ye. To have a consistent description of the two regions, we use the same underlying calculations for NSE and non-NSE regimes (\citealt{DD2}) except for some modifications. We neglect the excited states of nuclei and do not account for excluded volume effects. Also, the nucleons are treated as non-interacting Maxwell-Boltzmann gases. This helps us to get rid of spurious effects in the transition region. However, some differences persist between the two regions because of the limited number of nuclei considered in non-NSE. 

\begin{deluxetable*}{llllllllllll}    
\tablecaption{
    EOS models used in this study.
    \label{tab:EOS}
}
\tablewidth{0pt}
\tablehead{
    \colhead{Label} & 
    \colhead{EOS} & 
    \colhead{Model for uniform} & 
    \colhead{Nuclei} & 
    \colhead{$n_B^0$} &
    \colhead{$K$} &
    \colhead{$J$} &
    \colhead{$L$} &
    \colhead{$m^*_n/m_n$} & 
    \colhead{$m^*_p/m_p$} & 
    \colhead{$M_{\mathrm{max}}$} & 
    \colhead{Ref.} \\
    \colhead{} &
    \colhead{} &
    \colhead{nuclear matter} &
    \colhead{} &
    \colhead{(fm$^{-3}$)} &
    \colhead{(MeV)} &
    \colhead{(MeV)} &
    \colhead{(MeV~fm$^{-3}$)} &
    \colhead{} &
    \colhead{} &
    \colhead{(\msun)} &
    \colhead{} 
}
\startdata
SFHo & SFHo & RMF, SFHo & NSE & 
    0.1583 &
    $245.4$ & 
    31.57 &
    47.10 &
    $0.7609$ & 
    0.7606 &
    $2.06$ & 1 \\
SFHx & SFHx & RMF, SFHo & NSE & 
    0.1602 &
    $238.8$ & 
    28.67 &
    23.18 &
    $0.7179$ & 
    0.7174 &
    $2.13$ & 1 \\
TM1 & HS(TM1) & RMF, TM1 & NSE & 
    0.1455 &
    $281.6$ & 
    36.95 &
    110.99 &
    $0.6343$ & 
    0.6338 &
    $2.21$ & 2, 3 \\
NL3 & HS(NL3) & RMF, NL3 & NSE & 
    0.1482 &
    $271.5$ & 
    37.39 &
    118.49 &
    $0.5954$ & 
    0.5949 &
    $2.79$ & 2, 3 \\
DD2 & HS(DD2) & RMF, DD2 & NSE & 
    0.1491 &
    $242.7$ & 
    31.67 &
    55.03 &
    $0.5628$ & 
    0.5622 &
    $2.42$ & 2, 3 \\
BHB$\lambda \phi$ & BHB$\lambda \phi$ & RMF, DD2, hyperons & NSE & 
    0.1491 &
    $242.7$ & 
    31.67 &
    55.03 &
    $0.5628$ & 
    0.5622 &
    $2.10$ & 4 \\
LS220 & LS-EOS & Skyrme & CLD, SNA & 
    $0.155$ &
    $220$ &
    $29.6$ &
    $73.7$ &
    $1.0$ & 
    $1.0$ &
    $2.06$ & 5 \\
Shen & Shen-EOS & RMF, TM1 & RMF, TFA, SNA & 
    $0.145$ &
    $281$ &
    $36.9$ &
    $110.8$ &
    $0.634$ & 
     &
    $2.18$ & 6,7 \\
\enddata
\tablecomments{
    Saturation density $n_B^0$, incompressibility $K$, symmetry energy $J$, symmetry energy slope coefficient $L$, effective neutron mass $m_n^*$, effective proton mass $m_p^*$, and maximum mass $M_{\mathrm{max}}$ of a cold neutron star.
    RMF: relativistic mean field. 
    NSE: nuclear statistical equilibrium. 
    CLD: compressible liquid drop. 
    SNA: single nucleus approximation. 
    TFA: Thomas-Fermi approximation.
}
\tablerefs{
    (1)~\citet{SFHo}
    (2)~\citet{DD2}
    (3)~\citet{hempel2012}
    (4)~\citet{banik2014}
    (5)~\citet{ls220}
    (6)~\citet{shen98_Thphy}
    (7)~\citet{shen98_nuphy}
}
\end{deluxetable*}


\subsection{Initial models} 
\label{subsec:models}

We use four series of spherically symmetric, non-rotating progenitor models having three different values of metallicity ($Z=Z_{\odot}$, $Z=10^{-4}Z_{\odot}$, and $Z=0$) and spanning zero-age main sequences (ZAMS) masses from $\sim11$ -- 40~$M_{\odot}$. These pre-explosion models are taken from \cite{Woosley02,Woosley07} and were generated using the KEPLER evolution code. The progenitor models are labelled by their ZAMS masses with a letter prefix representing the series, and hence the metallicity of the models. For solar metallicity, we use the \cite{Woosley02} `s-series' model and \cite{Woosley07} `w-series' model. For low metallicity, we use the models from \cite{Woosley02} which have a metallicity of $10^{-4}$Z$_{\odot}$ (`u-series'). For zero metallicity we use the models from \cite{Woosley02} with zero metallicity (`z-series'). All of these models have also been used in \paperII -- \paperIV. 
A complete list of all the pre-explosion models used in this study is given in table \ref{tab:progenitors}.

\begin{table}    
\begin{center}
	\caption{Pre-explosion models used in this study.
    	\label{tab:progenitors}
	}
	\begin{tabular}{lllllc}
	\tableline \tableline 
Series & Label & Min Mass & Max Mass & $\Delta m$ & Ref. \\
 &  & ($M_{\odot}$) & ($M_{\odot}$) & ($M_{\odot}$) &  \\
	\tableline
s-series & s & $10.8$ & $28.2$ & $0.2$ & 1 \\
         &   & $29.0$ & $40.0$ & $1.0$ & 1 \\
         
w-series & w & $12.0$ & $35.0$ & $1.0$ & 2 \\
         &   & $35.0$ & $40.0$ & $5.0$ & 2 \\

u-series & u & $11.0$ & $40.0$ & $1.0$ & 1 \\
z-series & z & $11.0$ & $40.0$ & $1.0$ & 1 \\
	\tableline
	\end{tabular}
\end{center}
\tablecomments{
    The s-series and w-series has solar metallicity ($Z=Z_{\odot}$). The u-series has sub-solar metallicity ($Z=10^{-4}Z_{\odot}$). The z-series has zero metallicity ($Z=0$). 
}

\tablerefs{
    (1)~\citet{Woosley02}
    (2)~\citet{Woosley07}
}
\end{table}

\subsection{Nucleosynthesis post-processing} 
\label{subsec:postpr}

For the successfully exploding models, we calculate the detailed nucleosynthesis in a post-processing approach using the nuclear reaction network CFNET \citep{CF}, as in \paperIII~and \paperIV. 
The network includes 2902 isotopes from free nucleons to neutron-rich and neutron-deficient isotopes up to $^{211}$Eu. We use the reaction rate library REACLIB (\citealt{REACLIB}), which uses experimentally measured reaction rates. For those reactions where experimental rates are not known, REACLIB uses n, p and $\alpha$ capture reaction rates from theoretical predictions of \cite{Rauscher}. Our network also includes weak interactions where the electron and positron capture rates are taken from \cite{Laganke}. 
The $\beta^-$/$\beta^+$ decay rates are obtained from the nuclear database NuDat2 (if available) and \cite{Moller}. The capture reactions of $\nu$ and $\bar{\nu}$ on free nucleons are also included in our network.

For the post-processing, we take the ejecta from our hydrodynamic simulations and divide them into mass elements of equal mass ($10^{-3} M_{\odot}$). We call each of these mass elements `tracer'. Each tracer has a mass of $10^{-3} \textrm{M}_\odot$. The thermodynamic evolution of each tracer particle is known for the duration of the hydrodynamic simulation. 
As in \paperII~and \paperIV), we post-process only those tracers which reach a peak temperature $\geq 1.75$~GK. 

For the innermost tracers, the peak temperature is $\geq 10$~GK. We assume NSE condition for these tracers and start the post-processing when the temperature drops below 10~GK during the expansion. For the tracers that do not reach such high temperatures, we post-process them from the beginning of the hydrodynamical simulation. In both cases, the initial electron fraction \Ye is taken to be the same as the value in the hydrodynamical simulation and then it is evolved in our network consistent with the nuclear reactions.

For tracers where at the end of the hydrodynamic simulation the temperature and density are high enough for nucleosynthesis to occur, we extrapolate these using a free expansion for the density $\rho$ and an adiabatic expansion for the temperature, as in \paperIII~and \paperIV:
\begin{eqnarray}
r(t) &=& r_{\rm final}  + t v_{\rm final} \label{eq:extrapol_rad}, \\
\rho(t) &=& \rho_{\rm final} \left( \frac{t}{t_{\rm final}} \right)^{-3}, \\
T(t) &=& T[s_{\rm final},\rho(t),Y_e(t)], 
\label{eq:extrapol_t9}
\end{eqnarray}
where $r$ is the radial position, $v$ the radial velocity, $\rho$ the density, $T$ the temperature, $s$ the entropy per baryon, and $Y_e$ the electron fraction of the tracer.  The subscript `final' corresponds to the end time of the hydrodynamical simulation.
We calculate the temperature at each time step using the EOS from \cite{Timmes}. We end the nucleosynthesis calculation when the temperature of the tracer drops below 0.05~GK.

\section{Systematic Explosion properties from the SFHo EOS} 
\label{sec:explosionproperties.sfho}

In this section, we present and discuss the explosion properties of our simulations using the SFHo EOS model. As mentioned earlier in section \ref{subsec:models}, we use progenitors from four different series with different metallicities, namely the s, w, u, and z series.

\subsection{Explosion outcomes} 
\label{sec:exploutcomes.sfho}

\begin{figure}
    \centering
    \includegraphics[width=0.5\textwidth]{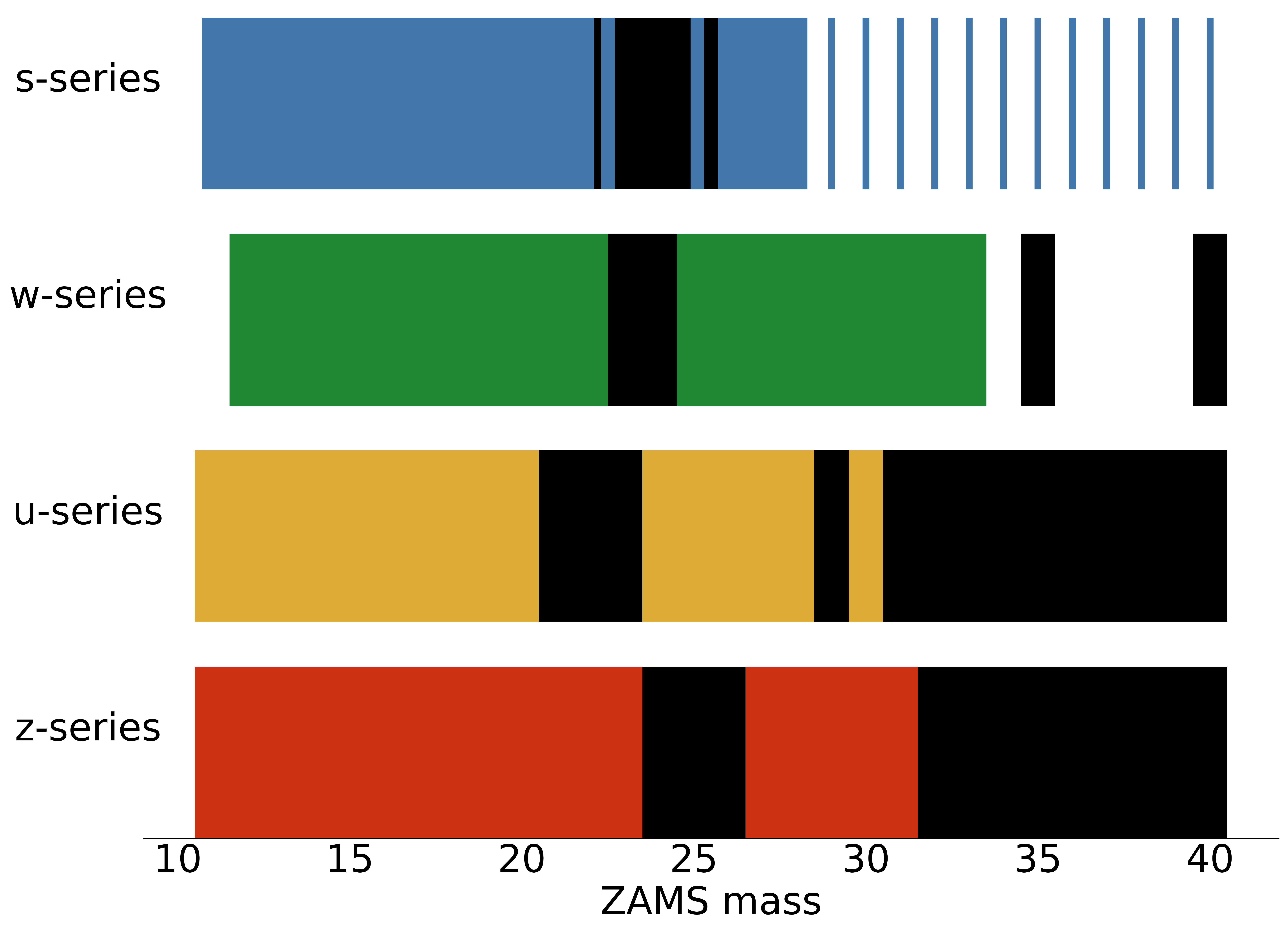}
    \caption{Explosion outcomes for simulations using the SFHo EOS for all four series of progenitor models: s-series (Z=Z$_{\odot}$, blue), w-series (Z=Z$_{\odot}$, green), u-series (Z=$10^{-4}$Z$_{\odot}$, yellow), and z-series (Z=$0$, red). Each colored bar represents an exploding model (leaving behind a neutron star) and each black bar corresponds to a non-exploding model (leaving behind a black hole).}
    \label{fig:rectangle_plots}
\end{figure}

Figure \ref{fig:rectangle_plots} shows the outcome of our simulations using the SFHo EOS for all four progenitor series. This gives a summary of which models explode and which do not explode. It can be seen from the figure that there is no ZAMS mass which divides the exploding models from the non-exploding models. Instead, there are regions of non-exploding models lying in between exploding ones. This is in agreement with other similar studies using a different nuclear EOS with our PUSH setup \citep{push2,push4} or using other effective simulation setups \citep{ugliano12,pejcha2015,ertl16,sukhbold16,mueller2016,couch19,murphy19}.

\begin{figure*}
\begin{center}
    \begin{tabular}{cc}
    \includegraphics[width=0.5\textwidth]{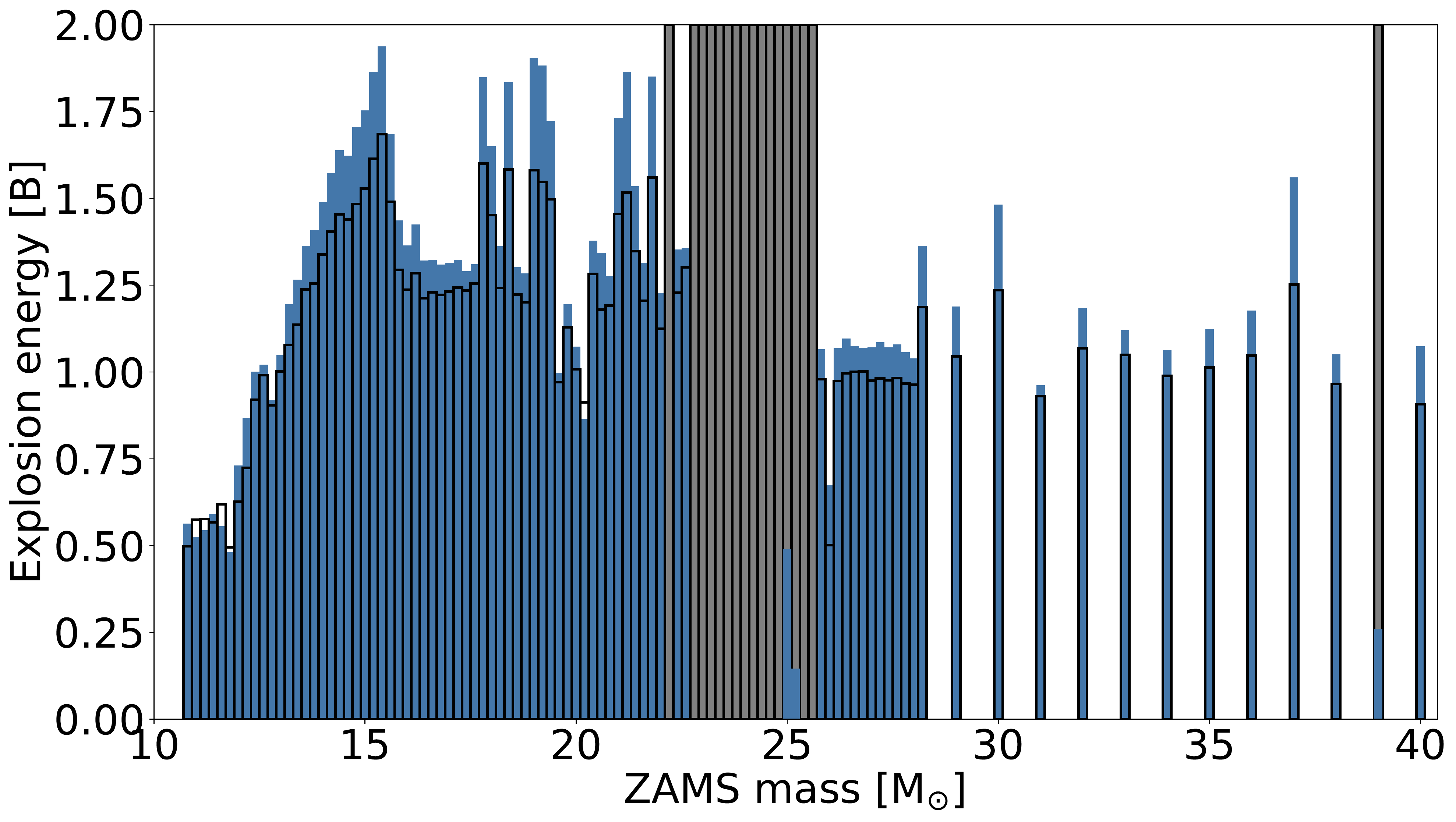} &
    \includegraphics[width=0.5\textwidth]{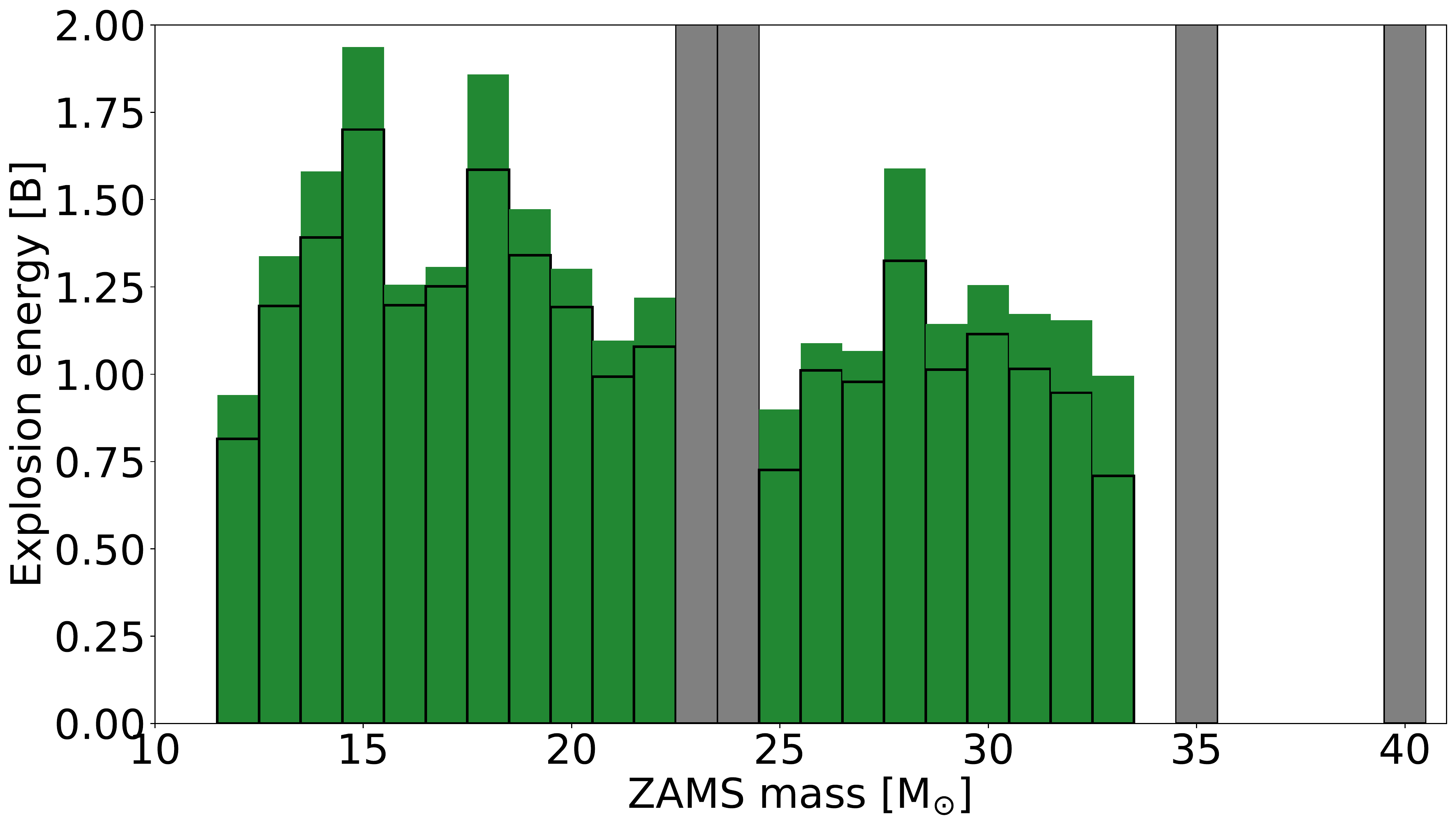} \\
    \includegraphics[width=0.5\textwidth]{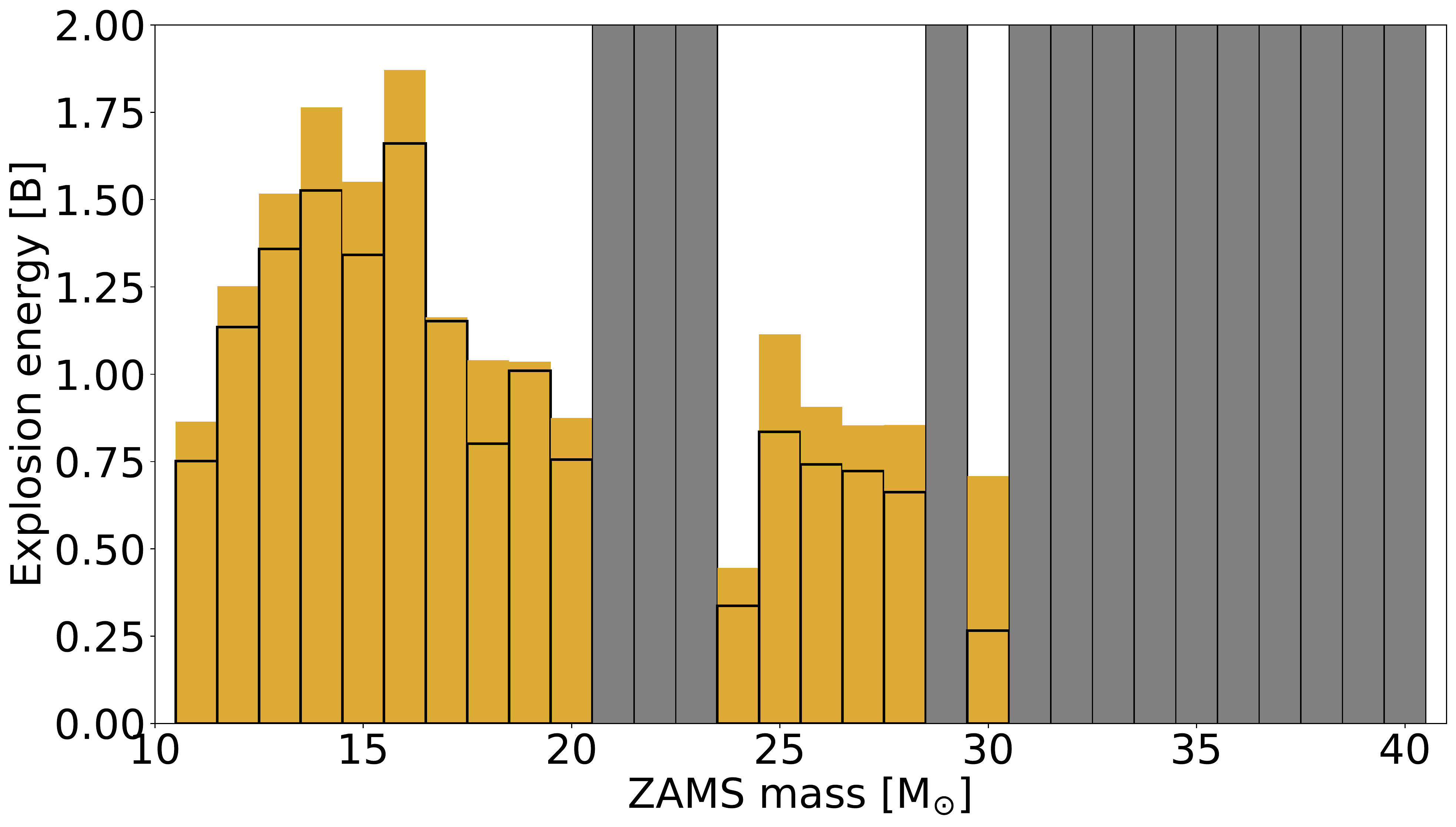} &
    \includegraphics[width=0.5\textwidth]{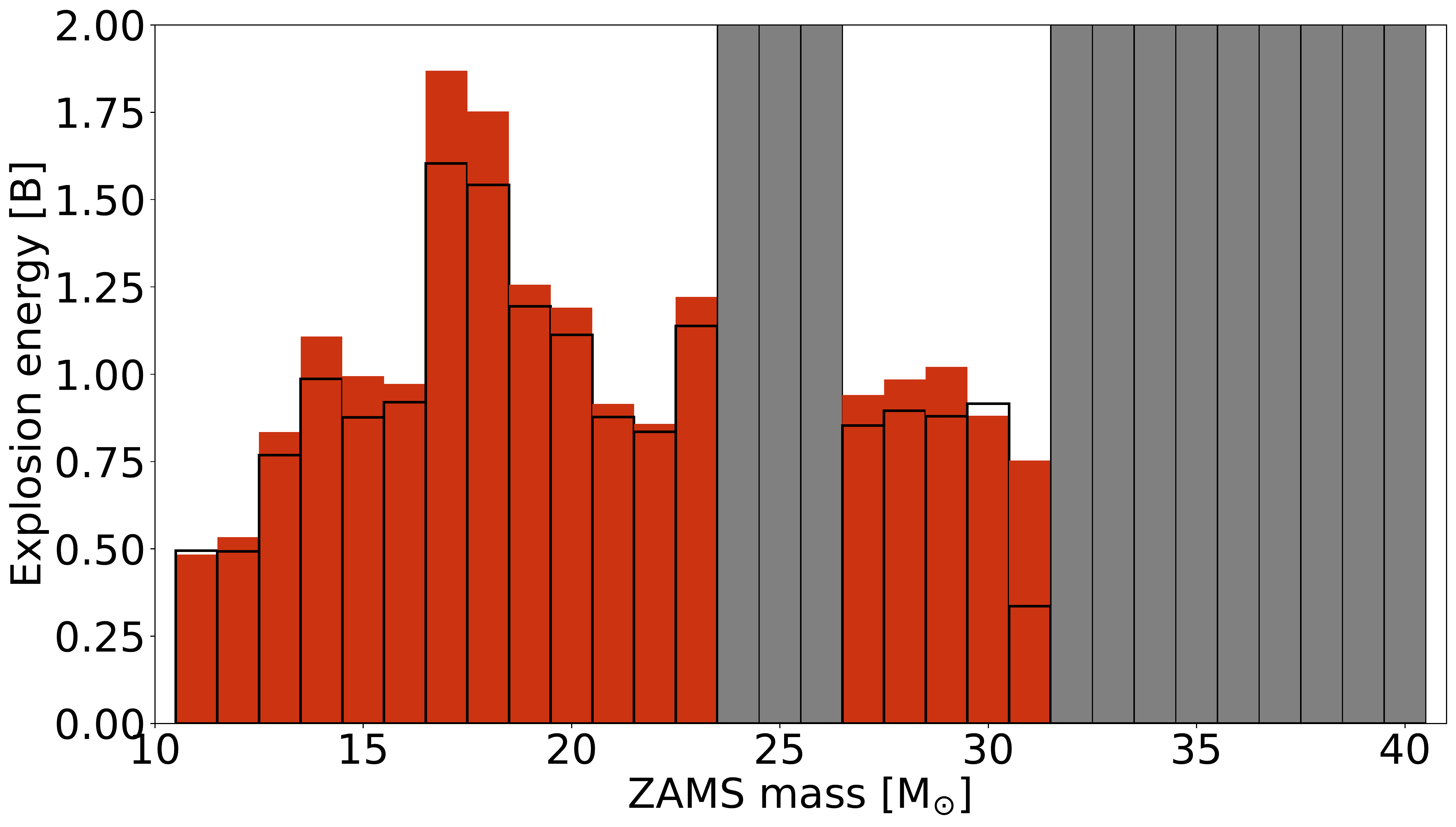} \\
    \includegraphics[width=0.5\textwidth]{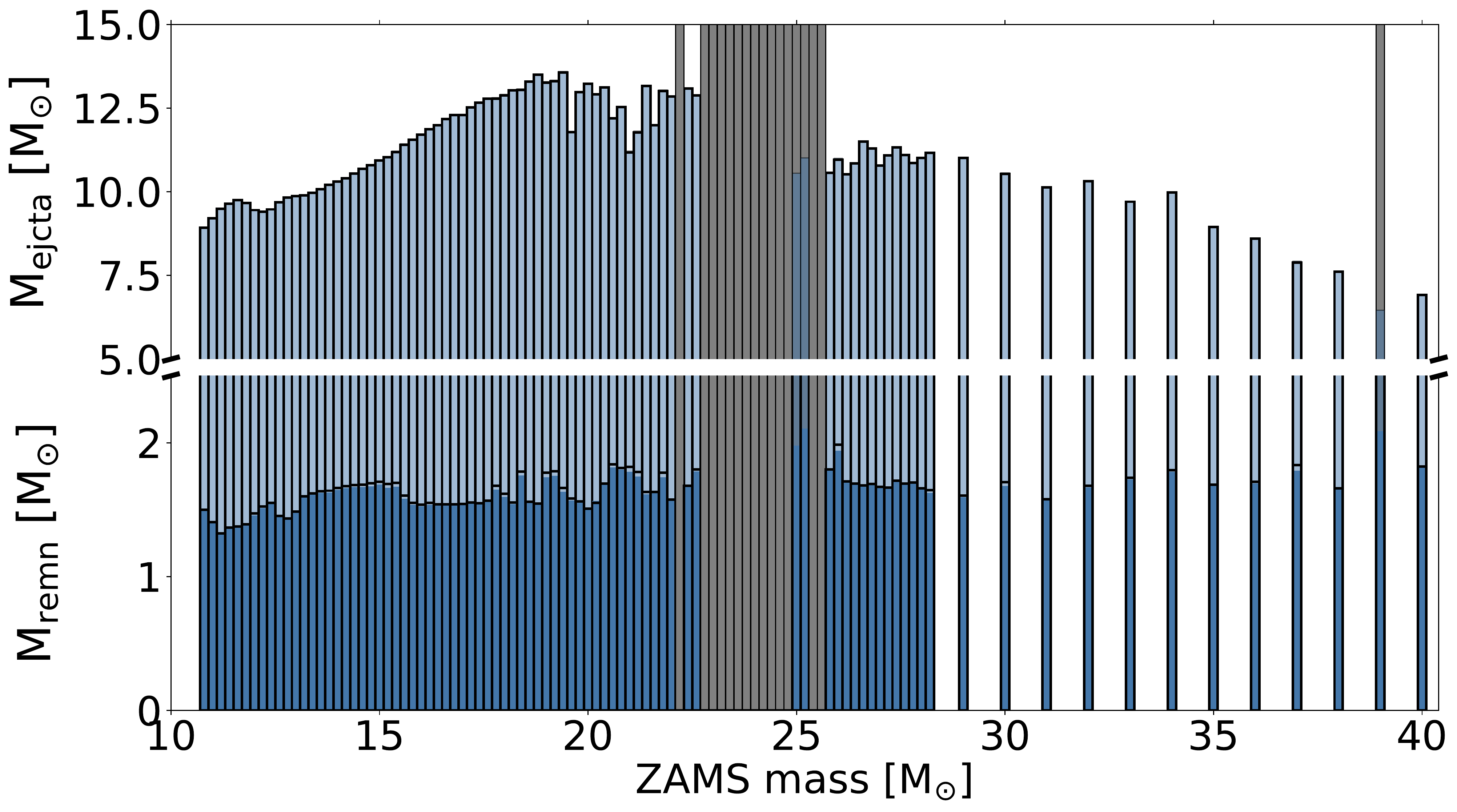} &
    \includegraphics[width=0.5\textwidth]{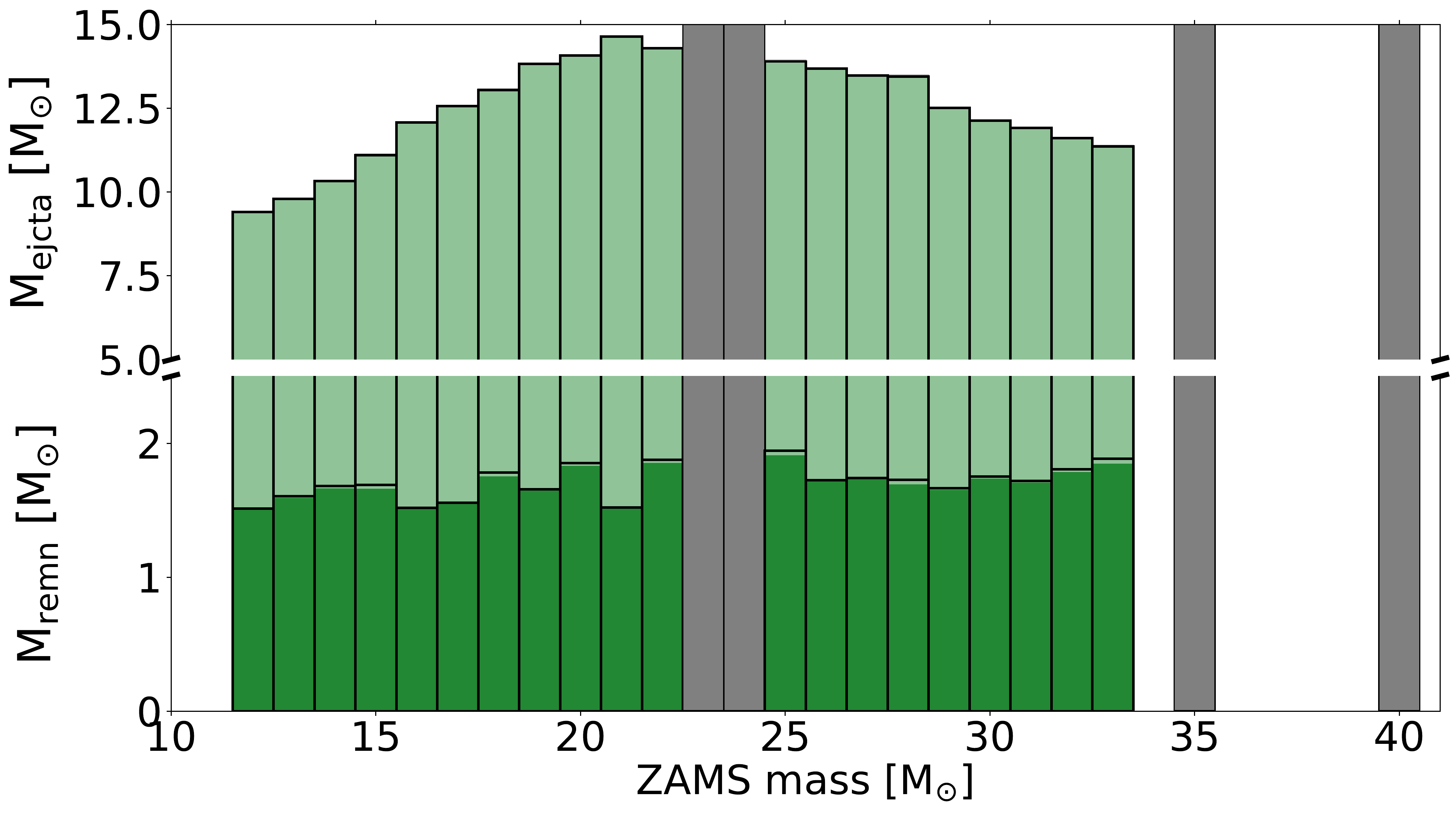} \\
    \includegraphics[width=0.5\textwidth]{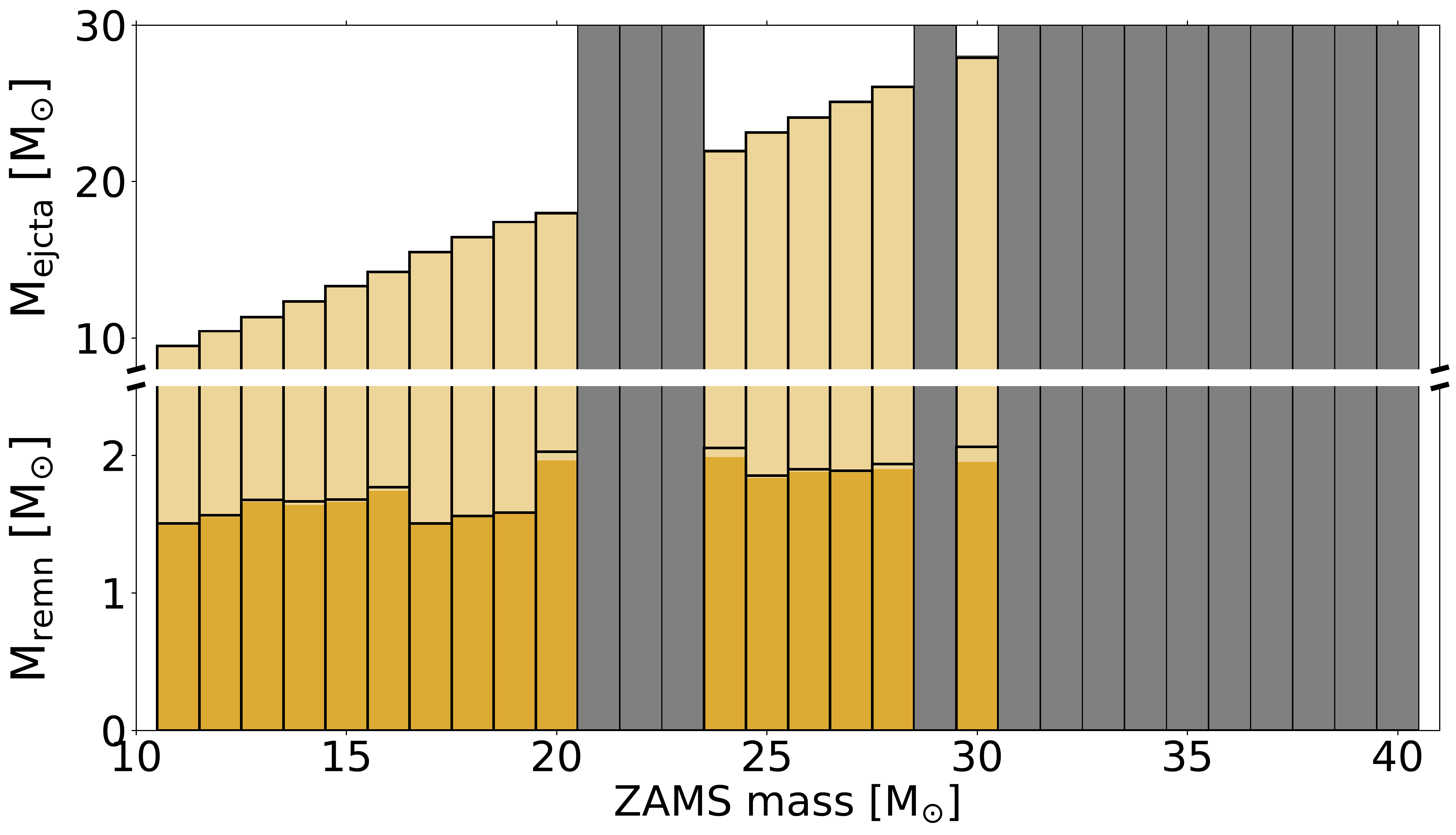} &
    \includegraphics[width=0.5\textwidth]{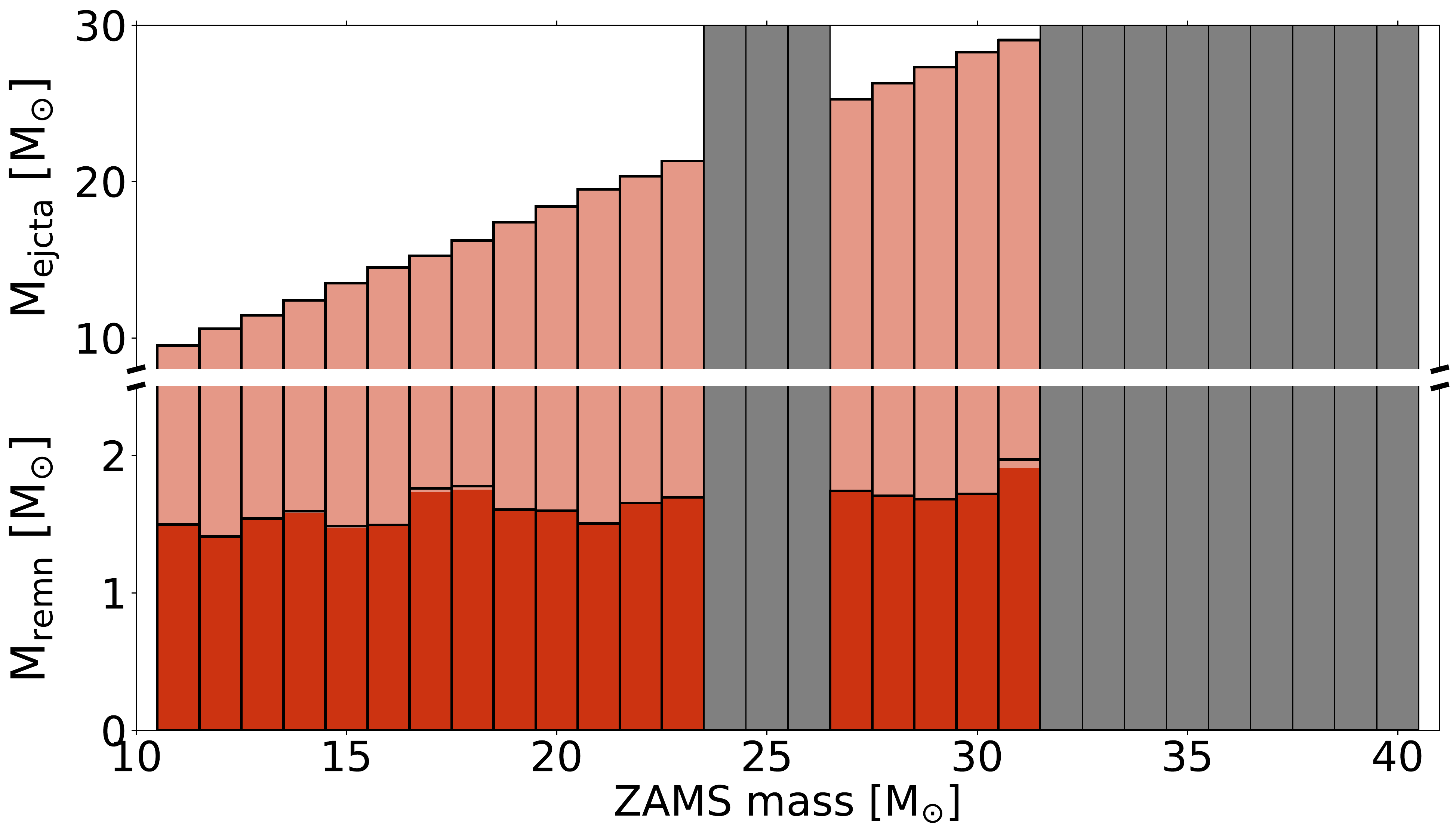}
    \end{tabular}
  
    \caption{\edit1{Top four panels: Explosion energy as a function of the ZAMS mass. Bottom four panels: Remnant mass (solid color) and ejecta mass (lighter shade) as a function of the ZAMS mass for all four series of pre-explosion models.} The s-series is represented in blue, the w-series is in green, the u-series is in yellow and the z-series is in red. \added{Note there is a scale break in the y-axis in the bottom four panels to accommodate the different scales of remnant mass and ejecta mass.} The colored bars are for SFHo EOS and the histograms given by the black line are for the DD2 EOS from \paperII~ and \paperIV~. The gray bars correspond to non-exploding models.
    \label{fig:barplot_SFHo}
    }
\end{center}
\end{figure*}

In figure \ref{fig:barplot_SFHo} we show the explosion energy, remnant mass, and ejecta mass for all the progenitor models used in this work. The colored bars represent the simulations of this work using the SFHo EOS. The histogram indicated by the black line represents the simulations with the DD2 EOS, taken from \paperII~ (s-series and w-series) and from \paperIV~ (u-series and z-series), which we include for comparison. 

The top four panels show the explosion energy as a function of the ZAMS mass. We obtain explosion energies of $\sim 0.15$ to 1.9~Bethe. Progenitors with very low and very high ZAMS masses and also those which lie next to non-exploding models have the lowest explosion energies. This is in agreement with the results obtained using the DD2 EOS (black lines; see also figure 12 in Paper~II and figure 3 in Paper~IV). Pre-explosion models with ZAMS masses around 15~M$_{\odot}$ result in the highest explosion energy for the w-, s-, and u-series. For the z-series, the highest explosion energy is obtained for ZAMS masses around 17~M$_{\odot}$. The highest explosion energies obtained for the SFHo EOS (1.9~Bethe) is slightly higher than the highest explosion energy for the DD2 EOS. Overall, the simulations using SFHo result in higher explosion energies (up to 15\%). This is due to the neutrino luminosities being systematically higher for SFHo compared to DD2.

The bottom four panels correspond to the ejecta masses (lighter color) and remnant masses (darker color). For all the four series, we obtain remnant mass around 1.3 to 2.0~M$_{\odot}$. Overall we make less massive remnants with the SFHo EOS than with DD2. The ejecta masses are computed from the stellar mass at collapse minus the remnant mass, hence they are very similar between simulations with SFHo and simulations with DD2. For all four series, the ejecta mass increases up to 20 \msun, beyond which the ejecta mass decreases for the s- and w-series due to line-driven mass loss at solar metallicity. For the u- and z- series the ejecta mass continues to increase with increasing ZAMS mass.

For almost all progenitors, the outcome of stellar collapse is the same for the SFHo EOS and for the DD2 EOS. The exception to this are three progenitors (s25.0, s25.2, s39.0) which have a drastically different outcome depending on the nuclear EOS used. We discuss these three models in detail in section \ref{subsec:mixedmodels}.

\subsection{Trends with compactness} 
\label{sec:compactnesstrends.sfho}

As we have seen in the previous section, the ZAMS mass is not a good indicator for whether a model explodes or not. Here, we look at the outcome of our simulations as a function of compactness $\xi_{2.0}$ at bounce to identify any emerging trends. 

\begin{figure}
    \centering
    \includegraphics[width=0.5\textwidth]{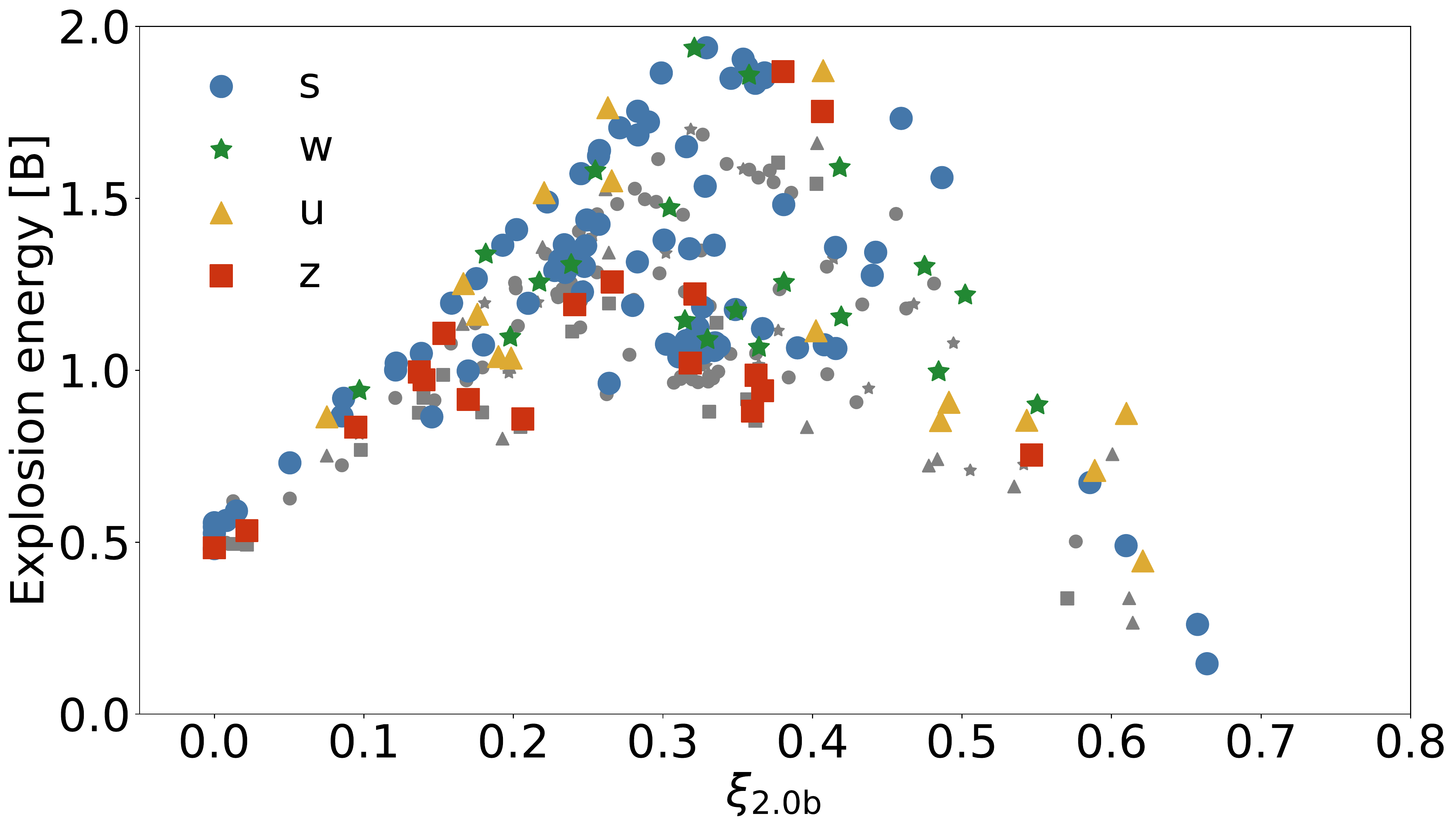}
    \includegraphics[width=0.5\textwidth]{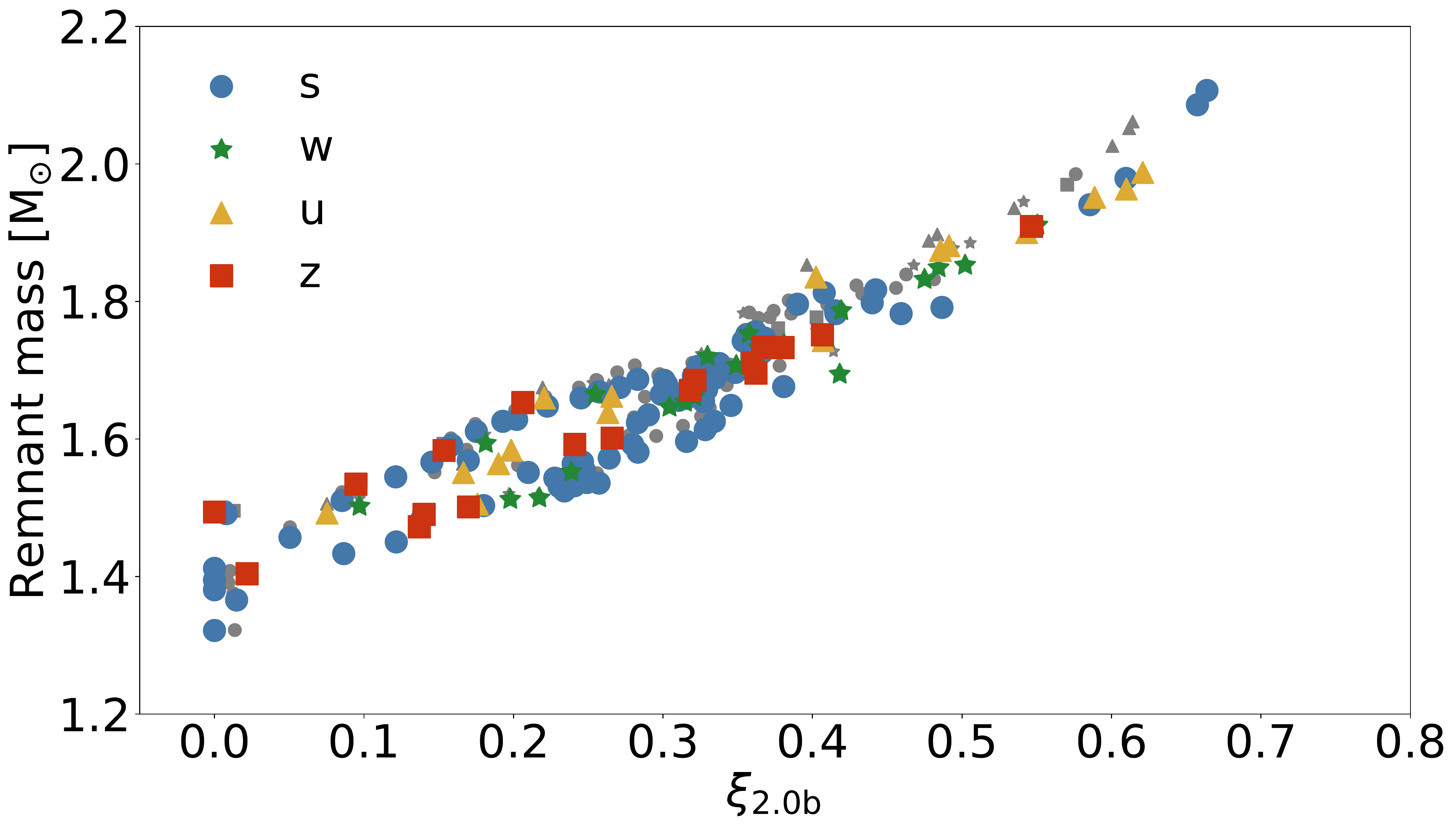}
    \includegraphics[width=0.5\textwidth]{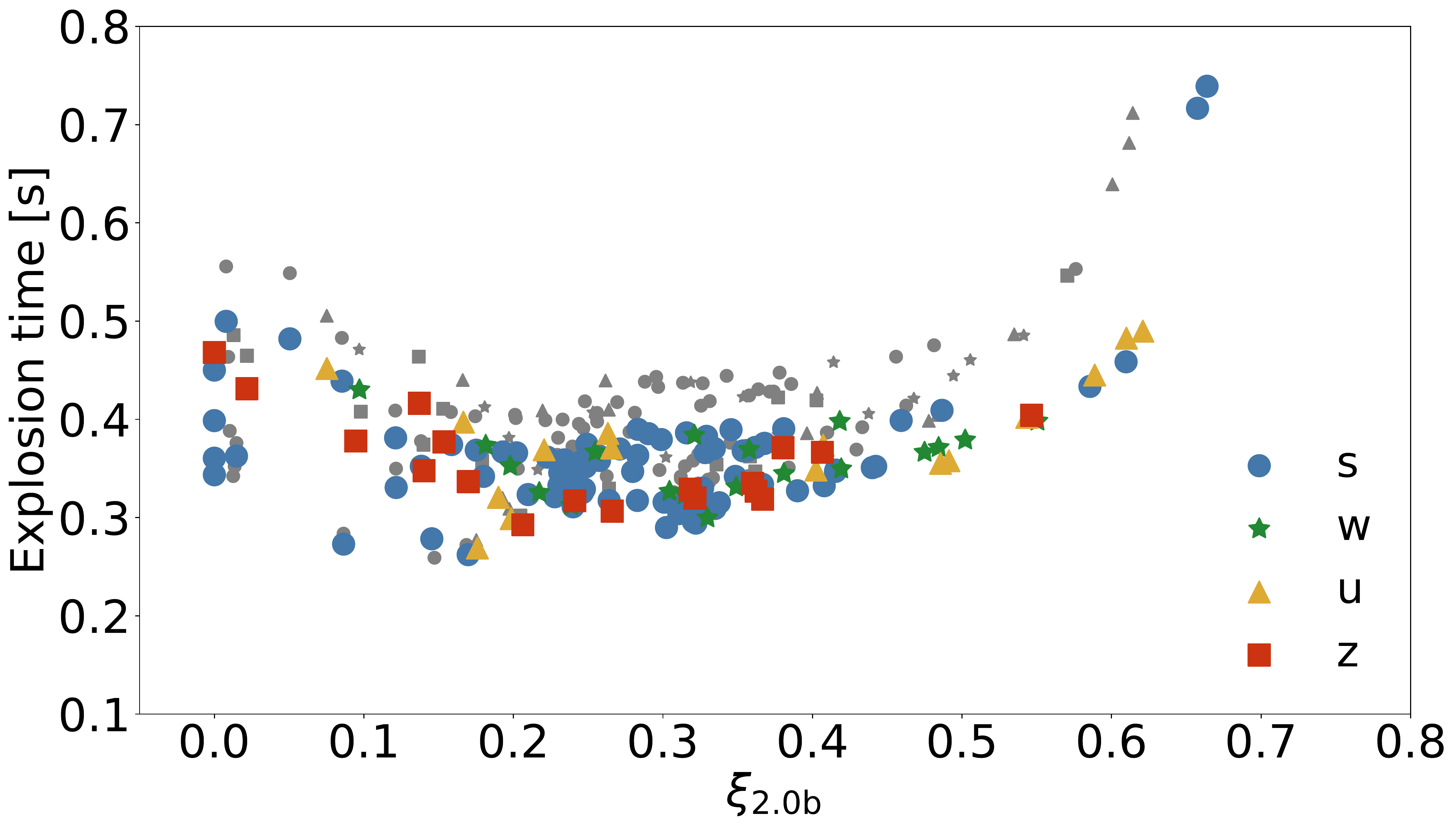}
    \caption{From top to bottom: explosion energy, neutron star (baryonic mass) and explosion time for the s-series (blue circles), w-series (green stars), u-series (yellow triangles), and z-series (red squares) as a function of the compactness $\xi_{2.0b}$ at bounce. \added{The colored symbols represent results from the SFHo EOS.} The results for simulations using the DD2 EOS with the same progenitor sets are shown in gray.
    \label{fig:correlation_w_compactness}
    }
\end{figure}

In figure \ref{fig:correlation_w_compactness}, we show the correlation between explosion properties (explosion energy, remnant mass, and explosion time) and the compactness of the progenitor. The colored points present the simulations of this work using the SFHo EOS. The grey points are the equivalent simulations using the DD2 EOS and are taken from \paperII~and \paperIV.
 The explosion energy is highest for models with compactness around 0.2 to 0.4 (for SFHo models), and it is lowest for the models with the lowest and the highest compactness values. This trend is the same for all four series of models, and hence independent of the metallicity. Overall, the simulations using the SFHo EOS have slightly higher explosion energies than the simulations using the DD2 EOS (where the peak explosion energies are obtained for models with compactness also between 0.2 and 0.4).
 
For exploding models, the remnant mass is directly correlated to the compactness of the progenitor. As the compactness increases, so does the mass of the neutron star formed after core collapse. This trend is independent of which nuclear EOS is used in the simulation. However, the remnant masses from the SFHo EOS are slightly lower than those from the DD2 EOS.
Most models that do explode, explode within 0.3 to 0.5~s after bounce. The explosion time has a mild parabolic dependence on the compactness which is inverse of that of the explosion energy. Models with the highest compactness, and also models with the lowest compactness, take the longest to explode. This trend in explosion time is related to our calibration of the PUSH method, and hence is not a true prediction from the simulations. Moreover, the models using SFHo explode \edit1{earlier} than the models using DD2. This, combined with the fact that the models using SFHo explode more energetically than the models using DD2 indicates that the stiffer EOS (DD2) makes it more difficult for models to explode, and hence if the model does explode, it takes more time. For all three quantities, there is no obvious trend with metallicity.

\section{Explosions properties from eight different nuclear EOS}
\label{sec:exploutcomes.all} 

In the previous section, we presented the overall trend of explosion properties of four progenitor series for simulations using the SFHo nuclear equation of state and compared them to simulations from previous work using the DD2 nuclear EOS.  
In this section, we present the results from simulations using a total of eight different nuclear EOS models. Six EOS models (SFHo, DD2, SFHx, TM1, NL3, BHB$\lambda\phi$) are described in section~\ref{subsec:EOS} and two models (LS220 and Shen) are included for comparison as they are widely used in the literature. Some aspects of the simulations using SFHo are discussed in the previous section. The simulations using DD2 are taken from \paperII, \paperIII, and \paperIV. \edit1{In section~\ref{subsec:explodingmodel}}, we focus on a 16~M$_{\odot}$ progenitor at three different metallicities (s16.0, u16.0, and z16.0) \added{In \ref{subsec:mixedmodels} we discuss the interesting case of the 25~\msun progenitor at solar metallicity (s25.0).}

\subsection{Trends with nuclear EOS}
\label{subsec:explodingmodel}

\begin{figure}
    \centering
    \includegraphics[width=0.5\textwidth]{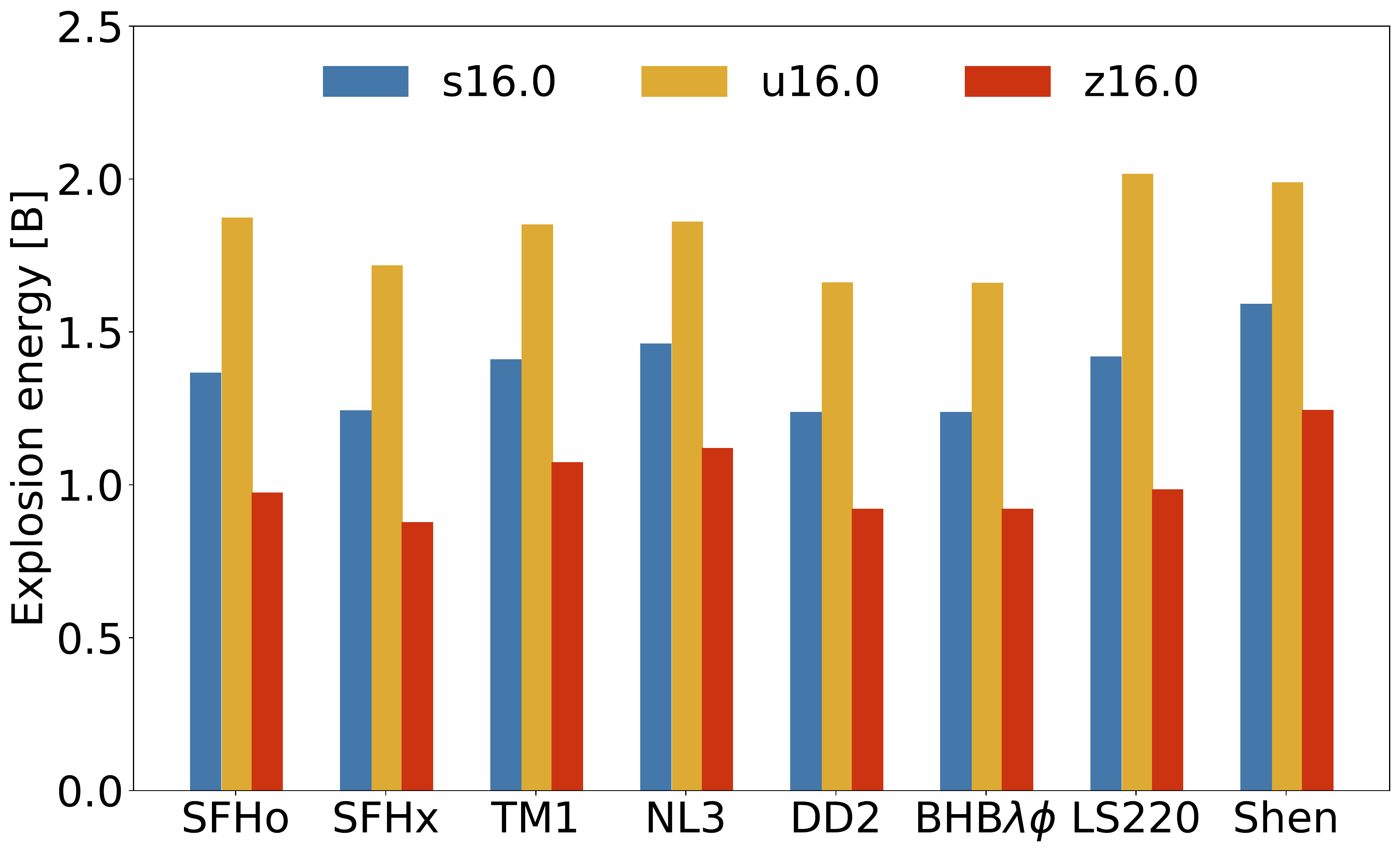}
    \includegraphics[width=0.5\textwidth]{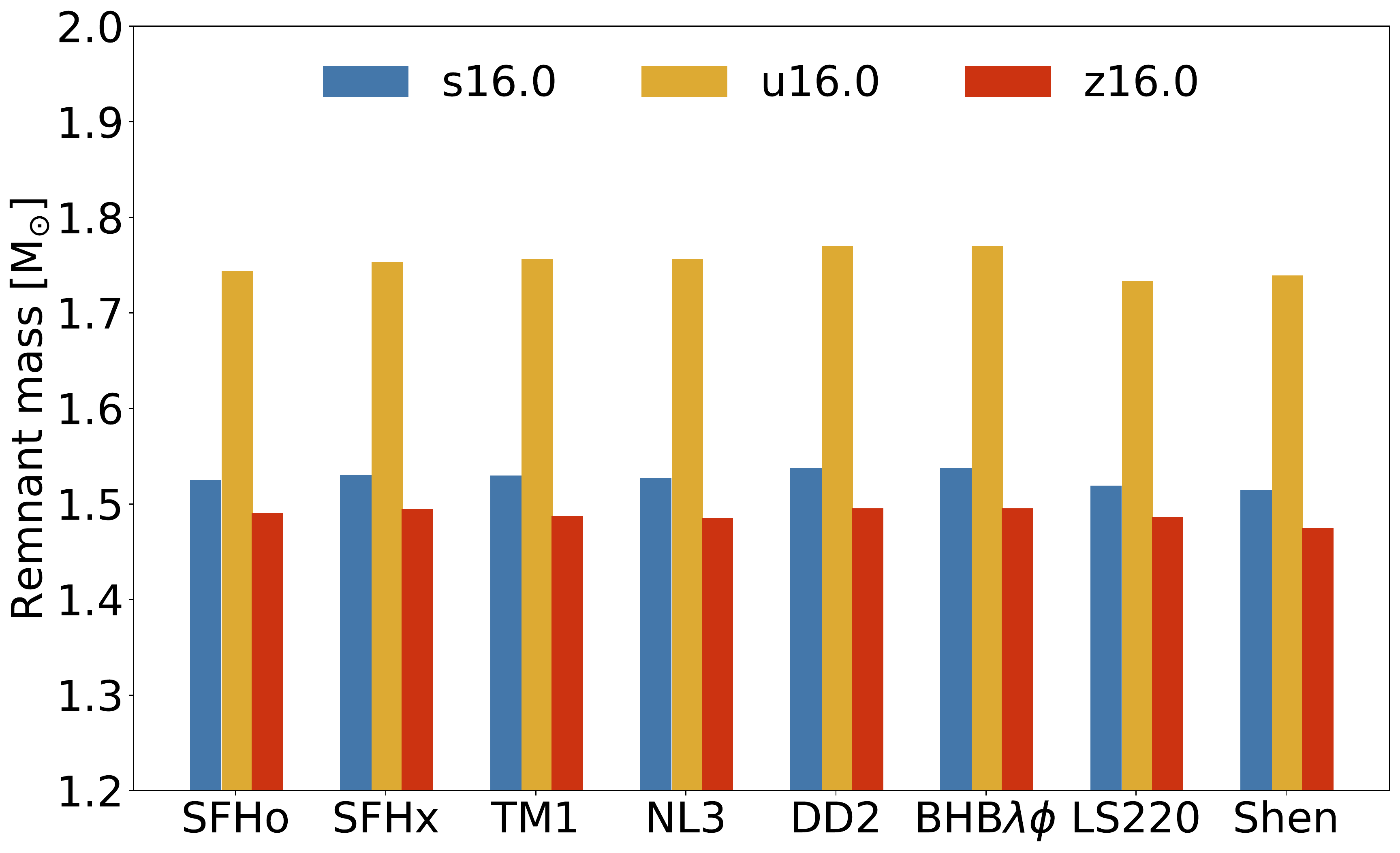}
    \caption{Explosion energy (top) and remnant mass (bottom) for a 16~\msun model at three different metallicities (s16.0 in blue, u16.0 in yellow, and z16.0 in red). 
    \label{fig:bar_plot_all_eos}
    }
\end{figure}

In figure \ref{fig:bar_plot_all_eos} we compare the explosion energy (top) and remnant mass (bottom) for all eight EOS models and 3 progenitor models. 
\edit1
{The explosion energy is not correlated with any single parameter of the EOS as given in Table \ref{tab:EOS}. \citet{Yasin2020} found in a study using one parametrization for the nuclear interaction and varying one parameter at the time that --- of the parameters they studied --- the effective mass has the largest impact on the explosion properties. For our setup, where the EOS differ in more than a single parameter, we do not find a correlation of the explosion energy with the effective mass, which is not surprising given the different setup. To do a quantitative analysis of the uncertainty from each parameter in the EOS on the explosion properties one would have to use a quite different setup with a continuous parametrization of the EOS \citep[(for example as suggested in][]{du.steiner.2021}.}
For a given progenitor, the explosion energy varies by 15\% for the s16.0 model, by 17\% for the u16.0 model, and by 18\% for the z16.0 model across all nuclear EOS used here. 
The remnant mass, however, varies by only $\sim 1$\% in all three cases.

\subsection{Models with EOS-dependent outcomes} 
\label{subsec:mixedmodels} 

In this section, we discuss the models for which the outcome (explosion or no explosion) depends on the nuclear EOS used in the simulation.
As mentioned in section \ref{sec:exploutcomes.sfho} these are three solar metallicity progenitors (s25.0, s25.2, s39.0) with ZAMS mass close to non-exploding models (s24.8, s25.4).  All three models have a relatively high compactness ($\xi_{2.0}> 0.6$).
These progenitors explode with four of the nuclear EOS models (SFHo, SFHx, NL3 and TM1) and with the two commonly used EOS models (LS220 and Shen), while they do not explode with DD2. The evolution is very similar for all three progenitors with mixed outcome. The central density evolves rapidly and then flattens out for all EOS models which result in a successful explosion. In the non-exploding case (DD2 EOS), the central density keeps rising as a function of time without any hint of turnover. 
For the BHB${\lambda \phi}$ model, the core undergoes a small second collapse around 1.2~s. \added{Since BHB$\lambda\phi$ is the only EOS considered here which includes hyperons, it is not surprising that the evolution is different compared to the other EOS.}

\begin{figure}
    \centering
    \includegraphics[width=0.5\textwidth]{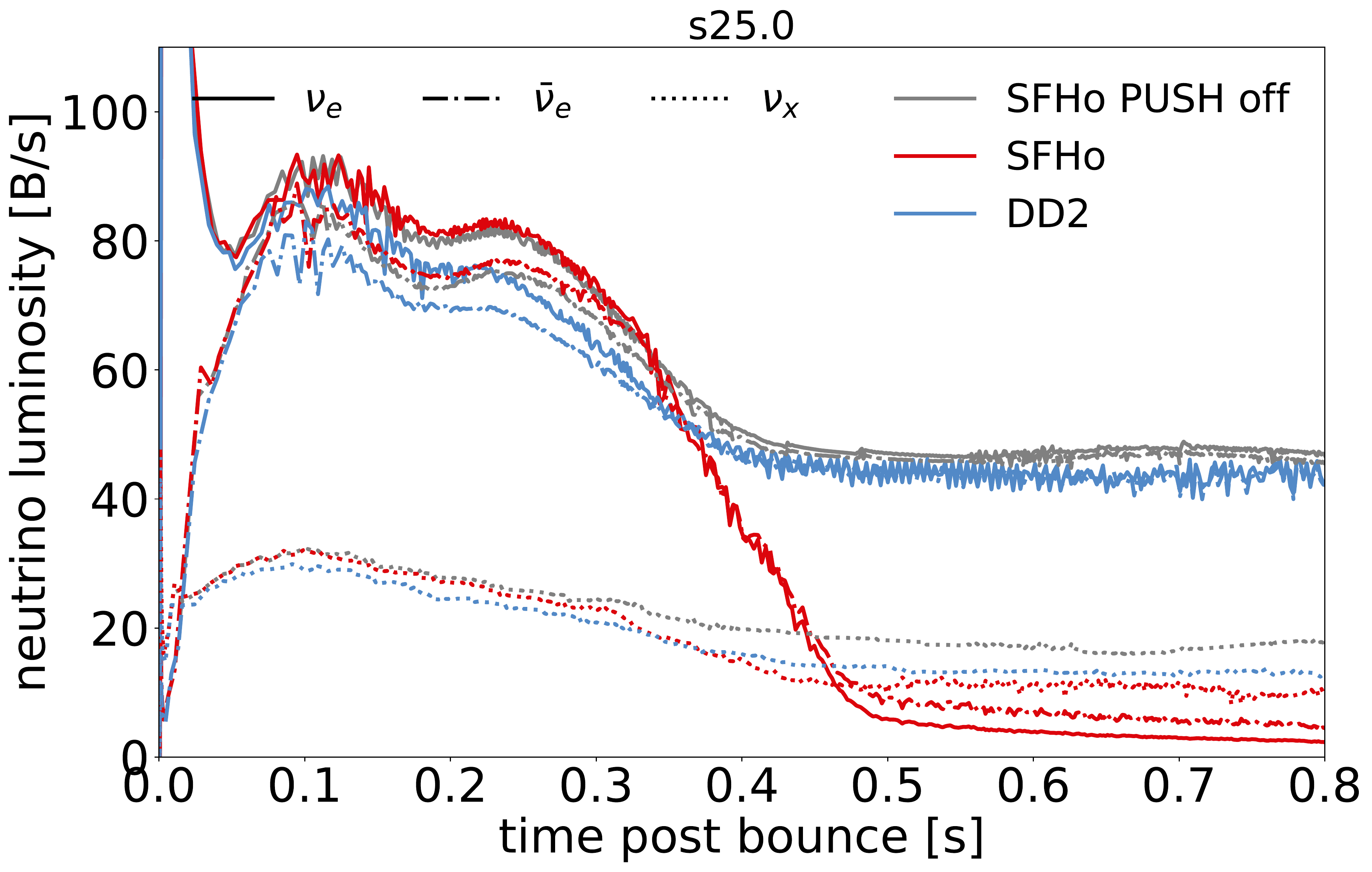}
    \includegraphics[width=0.5\textwidth]{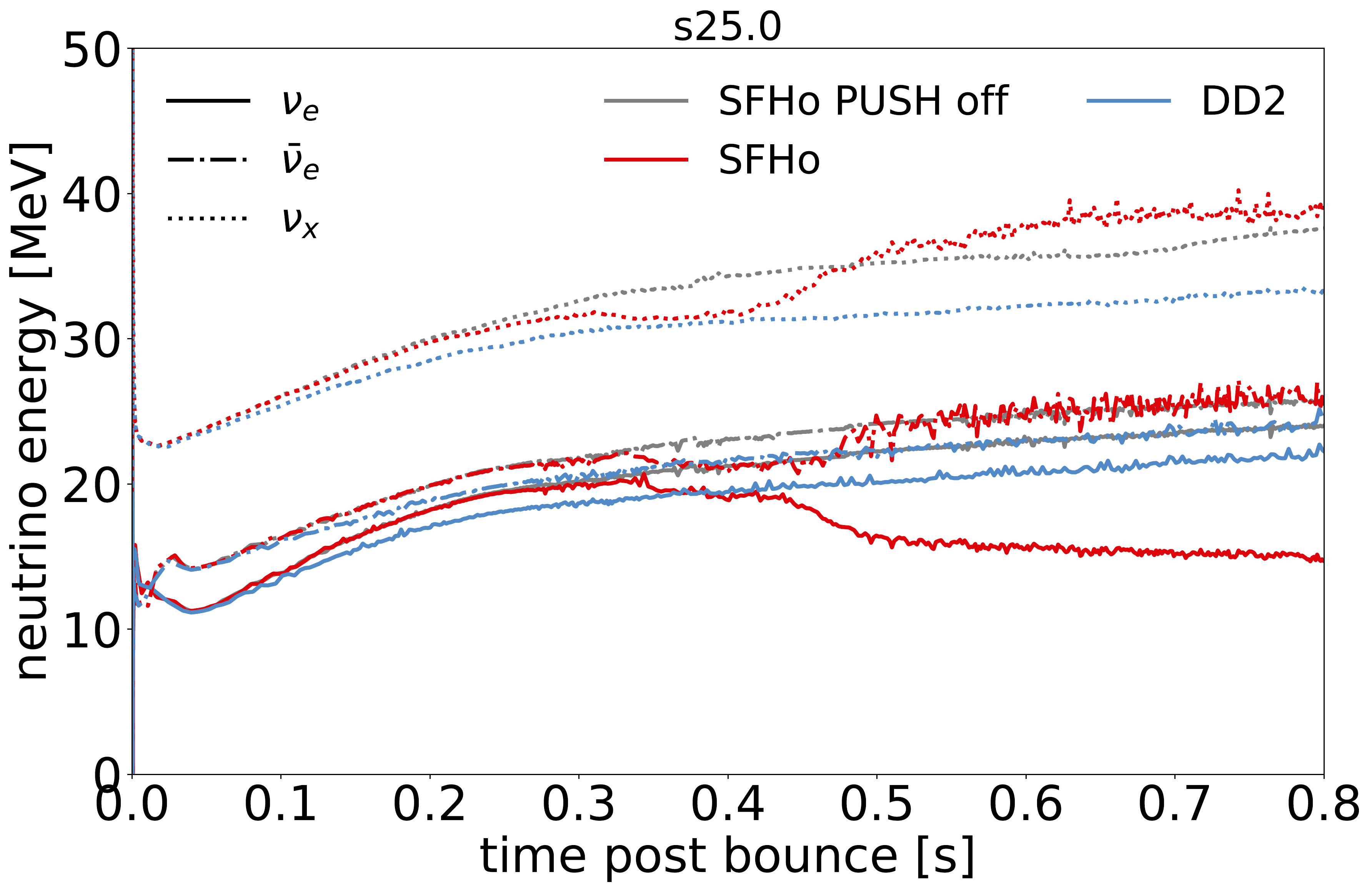}
    \caption{Temporal evolution of the neutrino luminosities (top) and mean energies (bottom) for s25.0 with two different nuclear EOS: SFHo (red) and DD2 (blue). \added{For comparison, the gray lines represent the model using the SFHo EOS without PUSH.}
    \label{fig:neutrino}
    }
\end{figure}

To answer the question what causes a model to explode or not to explode, we look into the neutrino luminosities  and energies for an exploding \added{case} (SFHo \added{with PUSH}) and \edit1{two} non-exploding \edit1{cases (DD2 with PUSH; SFHo without PUSH)}. We choose the 25 \msun progenitor as a representative model. The top panel of fig \ref{fig:neutrino} shows the temporal evolution of the electron neutrino, electron anti-neutrino and the heavy flavor neutrino luminosities. In \edit1{all three} cases we see an initial peak around 0.1~s post bounce followed by a plateau.

Up to about 0.4~s post bounce, the luminosities are qualitatively similar \edit1{in all three cases}. However, the case using the SFHo EOS has consistently higher luminosities in all three neutrino flavors when compared to the case using the DD2 EOS. Beyond 0.4~s, the luminosities are fundamentally different. In the \added{exploding} SFHo case the neutrino and anti-neutrino luminosities drop as the explosion sets in. In the DD2 case (which fails to explode) \added{and the non-exploding SFHo case} the neutrino and anti-neutrino luminosities remain relatively high due to the continued accretion onto the central object. The most important factor in driving the explosion in the SFHo case is the second bump in the neutrino and anti-neutrino luminosities around 0.25~s post bounce. This provides extra heating behind the shock and aids in achieving a successful explosion with the SFHo EOS.
This second bump in the neutrino luminosity is an inherent property of the SFHo nuclear EOS and is not a relic of the PUSH method. When we use the SFHo EOS \edit1{without} PUSH, the model does not explode. In this case, the electron neutrino and anti-neutrino luminosities (gray lines) follow the same behavior up to $\sim 0.4$~s p.b.\, including the second bump. After $\sim 0.4$~s, the electron neutrino and anti-neutrino luminosities remain high, following the non-exploding case using the DD2 EOS (as is expected from continued accretion in non-exploding models.

In addition, up to $\sim 0.4$~s p.b.\ the mean energies of all three neutrino flavors is a few MeV higher for the SFHo EOS when compared to the case using the DD2 EOS (as shown in the bottom panel of fig \ref{fig:neutrino}). The higher neutrino energies and luminosities results in slightly earlier explosions. This is accompanied by faster shock evolution and faster PNS contraction for SFHo as compared to DD2.

\section{Nucleosynthesis yields} 
\label{sec:nucleo} 

Our setup has several important features for the nucleosynthesis predictions, in particular for the innermost layers of the ejecta. In these layers, iron-group elements are synthesized through explosive burning (complete and incomplete silicon burning) and some isotopes beyond iron can be formed via a $\nu$p-process \citep{cf06b}. In the PUSH setup, the mass cut and the explosion energy evolve simultaneously and are \emph{not} independent of each other. Moreover, the neutrino-matter interactions are treated self-consistently, thereby setting the electron fraction \Ye (especially of the innermost ejecta) consistent with the evolution of the explosion and the neutrino emission from the PNS. In our models, the \Ye of the very innermost ejected tracer is very low, sometimes allowing for the production of elements up to mass number $A \sim 130$. However, the mass resolution in our models is quite coarse for the neutrino-driven wind and hence we do not further investigate this in the present work.

For different EOS we expect to see changes in the abundances of elements synthesized in the innermost ejecta, i.e.\ of iron group and trans-iron group elements. While the peak temperatures and densities (setting the conditions under which freeze-out happens) are mostly set by the explosion strength, the nuclear EOS determines the PNS evolution and with it the properties of the neutrino and anti-neutrino fluxes (which are setting the electron fraction \Ye).

In this section we present and discuss the nucleosynthesis yields of all exploding models, which we identify by the progenitor model and the nuclear EOS used in the explosion simulation. The setup for the nucleosynthesis is identical in all cases. We do not include those models which failed to explode as they are expected to have very little ejecta \citep{Lovegrove13,Lovegrove17}. We identify trends among the models using the SFHo EOS (section \ref{subsec:nucleo.sfho}) and discuss in more details the simulations of three representative progenitors (s16.0, u16.0, z16.0) using all eight nuclear EOS of this work (section \ref{subsec:nucleo.alleos}).

\subsection{General trends for models using the SFHo EOS}
\label{subsec:nucleo.sfho}

\begin{figure}
\centering
    \includegraphics[width=0.45\textwidth]{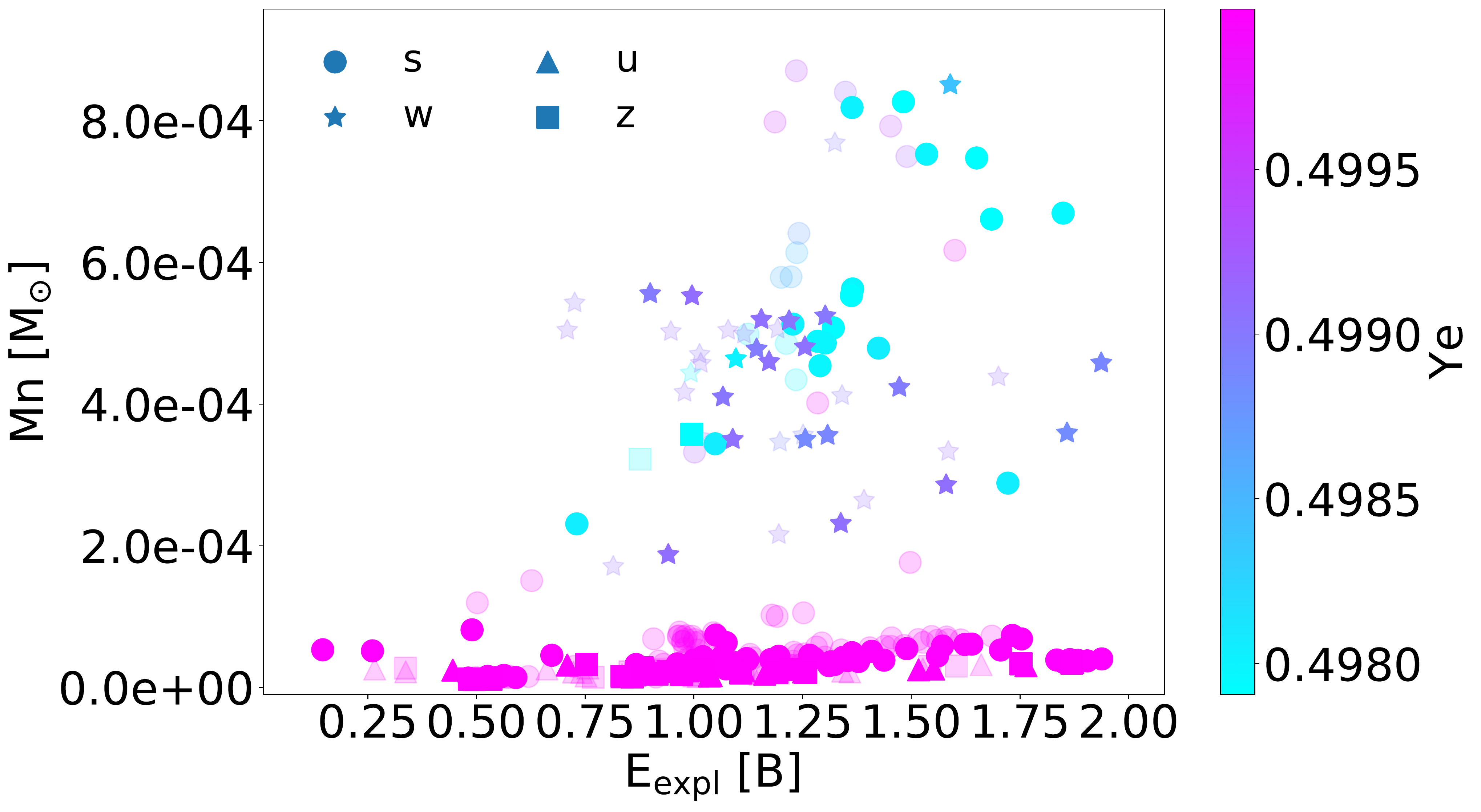}
    \includegraphics[width=0.45\textwidth]{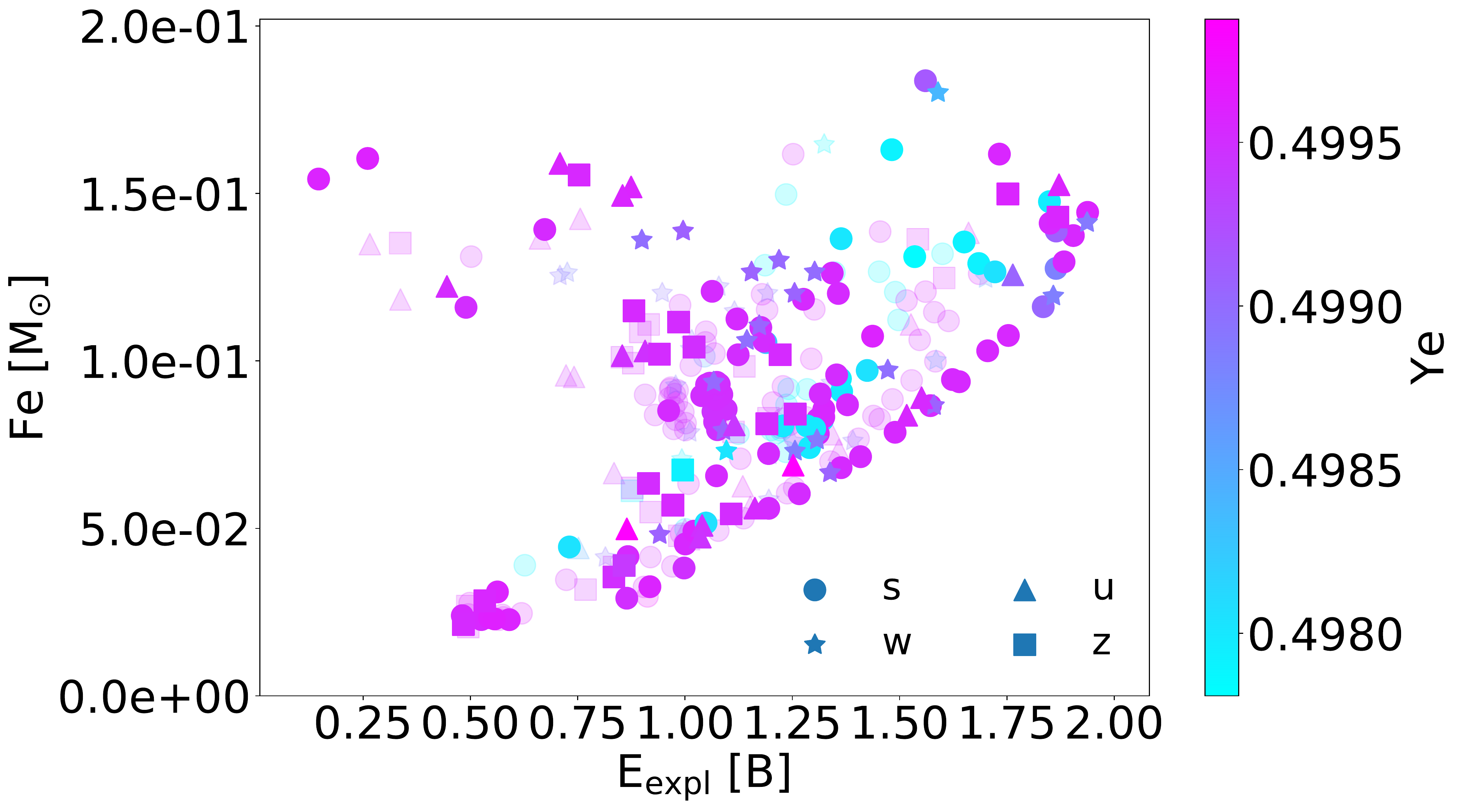}
    \includegraphics[width=0.45\textwidth]{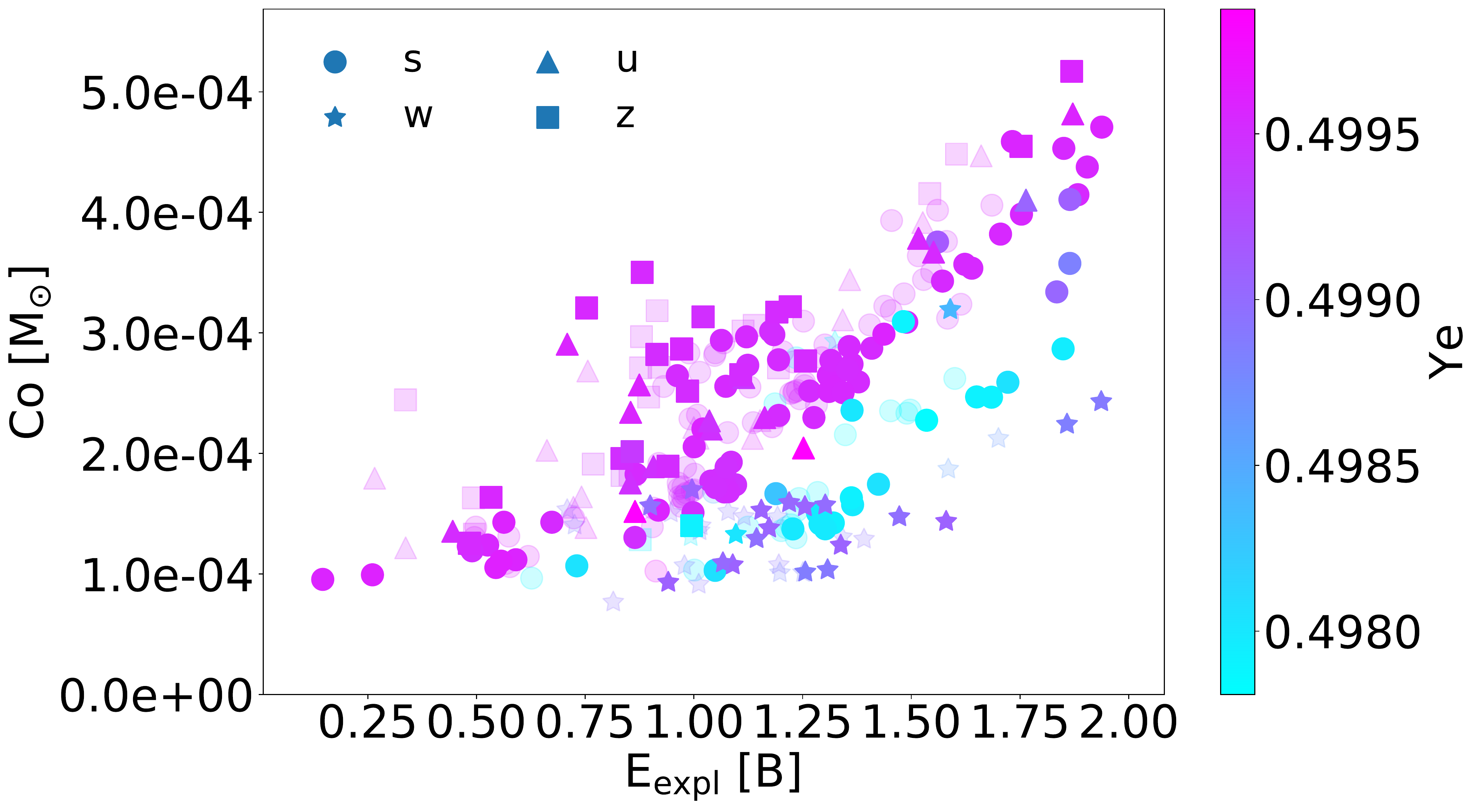}
  \includegraphics[width=0.45\textwidth]{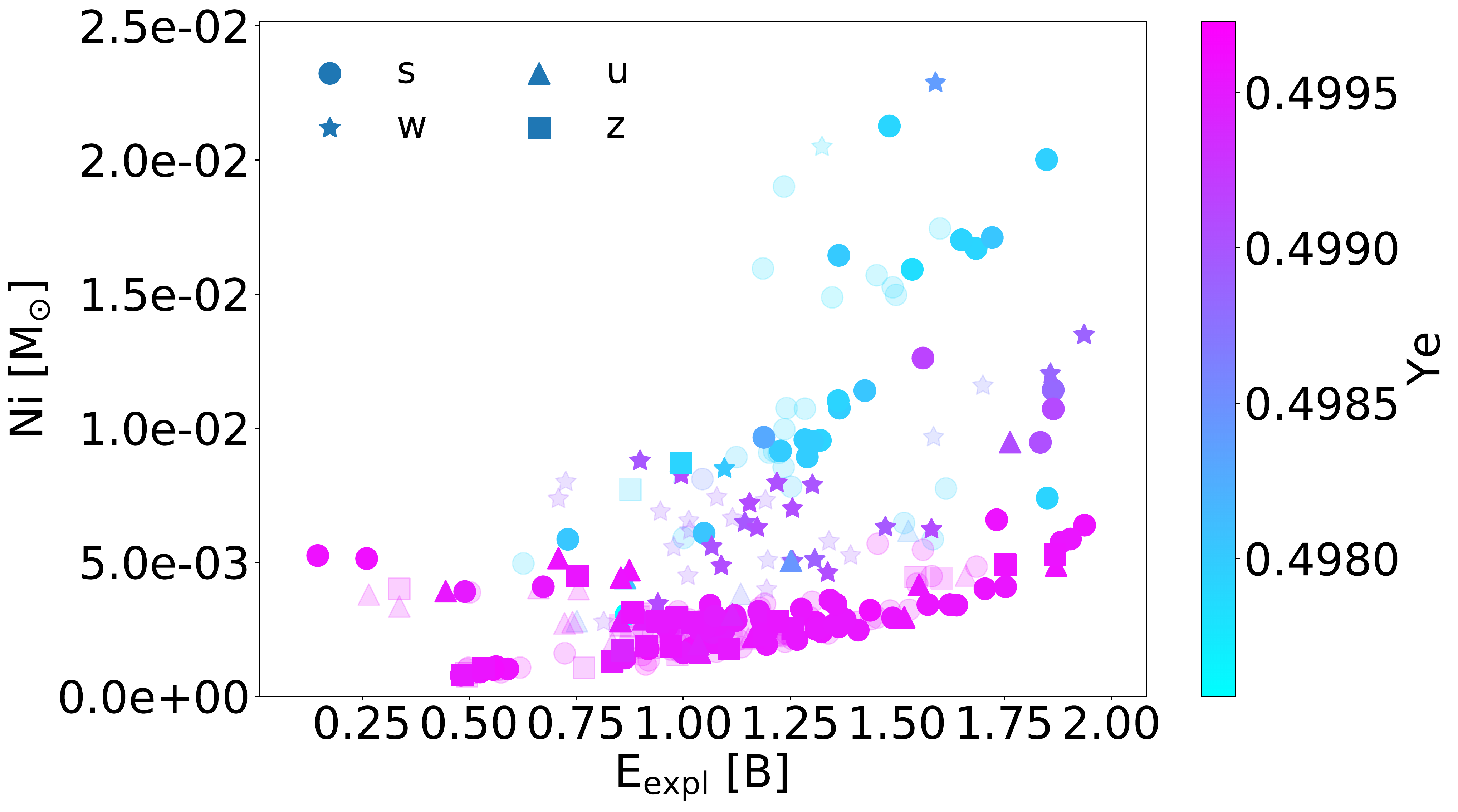}
    \caption{From top to bottom: Elemental yields of Mn, Fe, Co, and Ni after the explosion as a function of the compactness for the s-series (circles), w-series (stars), u-series (triangles), and the z-series (squares). The colored symbols are for simulations with the SFHo EOS, where the color represents the \Ye, and the transparent symbols are simulations with the DD2 EOS (taken from \paperIII~and \paperIV).
    \label{fig:elemental_explvsyields}
    }
\end{figure}

In figure \ref{fig:elemental_explvsyields} we present the elemental yields of four iron group elements -- manganese, iron, cobalt and nickel -- as a function of the explosion energy for simulations with the SFHo EOS (solid points). In addition, the corresponding results for simulations with the DD2 EOS (taken from \paperIII~and \paperIV) are included as transparent points. The color of each point corresponds to the mean \Ye of the region where most of the element is made. First, we focus on how the nuclear EOS affects the yields and the general trends with explosion energy.

Production of elemental manganese is highly \Ye dependent. Models with a \Ye close to 0.5 have low Mn yields, and vice versa (see also \paperIII). Most of the s-series progenitors have a relatively high \Ye in the relevant layers, and therefore low manganese yields. This is also the case for the u-series and z-series progenitors. The w-series is different from the s-, u- and z-series. The network used during the the hydrostatic evolution is much larger for the w-series, which results in the realization of a large range of \Ye values in the relevant layers of the pre-collapse star and therefore intermediate Mn yields.

The only stable isotope of Mn ($^{55}$Mn) is produced as $^{55}$Co in proton-rich layers ($Y_e > 0.5$) and incomplete Si burning zones. The amount of proton-rich ejecta is the same for SFHo models and DD2 models ($\sim 0.007$ \msun or about 10\% of the ejecta where iron-group elements are made). 
Hence, we expect little difference between the Mn yields from SFHo models and from DD2 models. We find about 5\% difference in Mn yields between SFHo and DD2 for a given progenitor model. All u- and z-models, as well as almost all w- models have higher Mn yields with SFHo than their DD2 counterparts. Generally, the s-models have lower Mn yields with SFHo than with DD2, which has higher $Y_e$ values in the relevant layers.

Next, we discuss elemental iron which has a strong correlation with the explosion energy (higher explosion energy means higher Fe yields -- the exceptions to this (s25.0, s25.2, and s39.0) are discussed in section \ref{subsec:mixedmodels} and also the `almost failing models' (see \paperIV). The main contribution to elemental iron comes from $^{56}$Fe which is made as $^{56}$Ni primarily in Si burning zones. Hence, we expect only small differences in the Fe yields between SFHo and DD2 models, primarily due to the differences in the explosion energies. We find about half the models have differences in the Fe yields of less than 5\%, the other half of the models has differences of less than 12\%. Only a small number of models (s11.0, s11.2, s11.6, s11.8, s19.6, s20.2, u17.0) have lower explosion energies with SFHo compared to DD2. These models accordingly have lower Fe yields than their DD2 counterparts.

The behavior of elemental cobalt is a combination of what we have seen for Mn and for Fe. The Co yields show a correlation with explosion energy as well as a \Ye dependence. Overall, a higher explosion energy means a higher Co yield. In addition, for a fixed  explosion energy, a higher \Ye implies a higher Co yield. The w-series has intermediate \Ye values, but the lowest Co yields. The lowest \Ye values (found mostly in s-series models) also have low Co yields. The models of all series with the highest \Ye values in the relevant zones have the highest Co yields, across all explosion energies. Additionally, as for Mn and Fe, SFHo models have up to 15\% higher Co yields than DD2 models.

Finally, elemental nickel also has strong correlation with explosion energy and has significant \Ye dependence. The Ni yields have two distinct branches, one with low \Ye and high Ni yields, and the other with high \Ye and lower Ni yields. The u- and z-series mostly populates the latter branch (high \Ye, low Ni yields). The w-series has intermediate \Ye values and corresponding intermediate Ni yields. The highest Ni yields are in models of the s-series with the lowest \Ye in the relevant layers. 
The two notable exceptions to this trend are the two models at the lowest explosion energies with $\sim 5\times 10^{-3} M_{\odot}$ of Ni. These are once again the models with EOS-dependent outcomes. 
The \Ye-dependence of the elemental Ni yields originate from the contributions of non-symmetric Ni isotopes, i.e.\ from $^{58}$Ni and $^{60}$Ni. $^{58}$Ni is formed in Si burning zone as itself while $^{60}$Ni is produced as $^{60}$Cu and $^{60}$Zn in the complete Si burning zone. 
In general, the SFHo models make up to 20\% more Ni than the corresponding DD2 models.

Next, we discuss the yields of the four isotopes $^{56}$Ni, $^{57}$Ni, $^{58}$Ni and $^{44}$Ti for which we have observational constraints from SN~1987A and a few other supernovae.
%
$^{56}$Ni is produced from Si burning in the inner layers of the star. The temperature and density conditions in these layers are very similar for SFHo and DD2 EOS. Therefore, the $^{56}$Ni yields are also similar for both EOS, with the SFHo models generally having $\sim 20$\% higher yields than the DD2 models (due to the higher explosion energies). 
The small number of models which have a lower explosion energy with SFHo compared to with DD2, also have lower $^{56}$Ni yields compared to their DD2 counter part. There is one extreme exception (w22.0) which has a higher explosion energy with SFHo, but a much lower $^{56}$Ni yield compared to DD2.

Generally, the correlation of $^{56}$Ni yields and explosion energy are quite strong and independent of the progenitor metallicity. The points in the top left (high $^{56}$Ni and low explosion energy) are the models with EOS-dependent outcomes (see section \ref{subsec:mixedmodels} and the `almost failing models' in \paperIV).

The $^{57}$Ni yields strongly depend on the \Ye of the model. Models with high \Ye make less $^{57}$Ni and vice versa. The s-series progenitors group into two distinct \Ye branches. Most of the s-series models and the u- and z-series models have high \Ye and therefore lower $^{57}$Ni yields. Some of the s-series models have low \Ye and hence comparatively higher $^{57}$Ni yields. Lastly, the w-series models have intermediate \Ye values (see above) and produce intermediate amounts of $^{57}$Ni. The SFHo yields are generally up to $\sim 20$\% higher compared to their DD2 counterpart (similar to $^{56}$Ni). The exceptions to this are the w22.0 model (which makes twice as much $^{57}$Ni as its DD2 counterpart) and a handful of models with very low explosion energy.

Similar to $^{57}$Ni, the $^{58}$Ni yields also have a strong \Ye dependence. 
For the bulk of our models the $^{58}$Ni production occurs at \Ye values of around 0.498. For only five models (s16.8, s25.2, s27.2, s27.8) the dominant production site of $^{58}$Ni is in a proton-rich environment. Without those 5 models, we see a similar behavior as for $^{57}$Ni, where higher \Ye values result in lower $^{58}$Ni yields. Again, the w-series has intermediate \Ye values and hence intermediate $^{58}$Ni yields. The u-, z-, and some s-series models have high \Ye values and hence low $^{58}$Ni yields. The SFHo models have up to 50\% higher $^{58}$Ni yields than the DD2 models.

Lastly, we discuss $^{44}$Ti for which the yields are somewhat correlated with the explosion energy. Here, the comparison between SFHo and DD2 models is more interesting than for the Ni isotopes. For SFHo models where the explosion energy is less than 10\% larger than in the DD2 case, about half the SFHo models have lower $^{44}$Ti yields and the other half have higher $^{44}$Ti yields than in the DD2 case. For SFHo models where the explosion energy is more than 10\% higher than for the corresponding DD2 model, the $^{44}$Ti yields are up to 10\% higher than for DD2.

\begin{figure}
    \centering
    \includegraphics[width=0.45\textwidth]{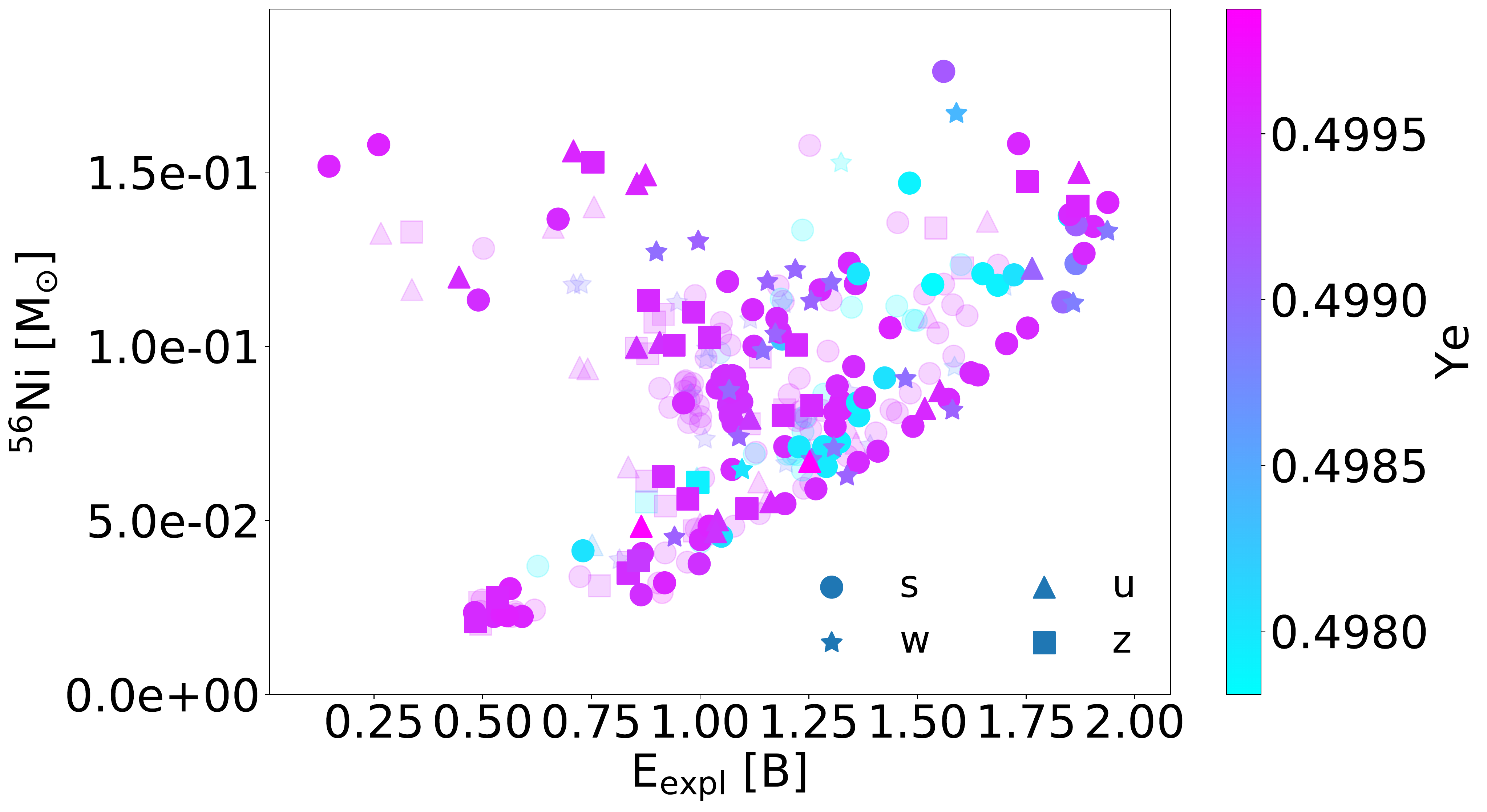}
    \includegraphics[width=0.45\textwidth]{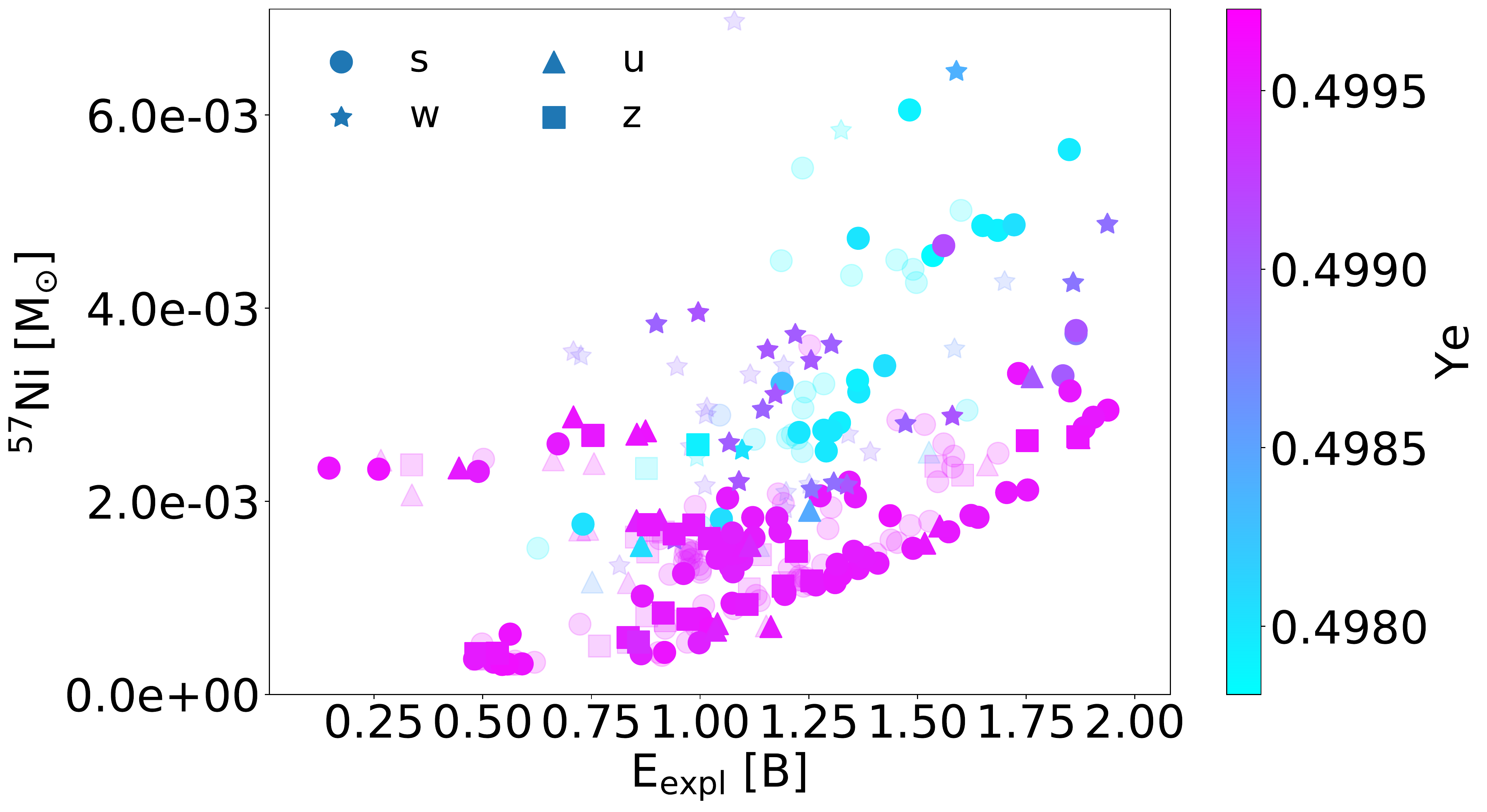}
    \includegraphics[width=0.45\textwidth]{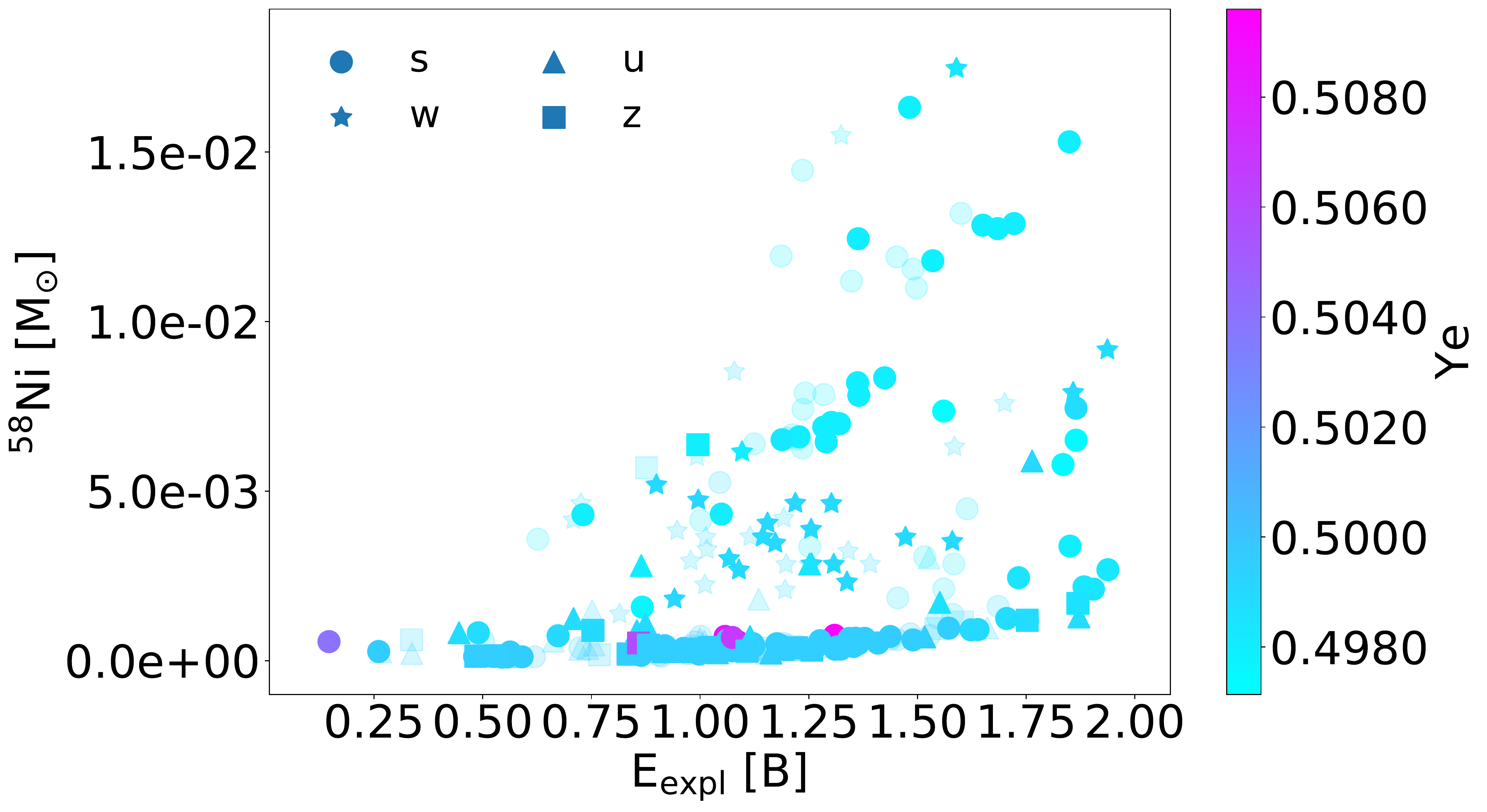}
    \includegraphics[width=0.45\textwidth]{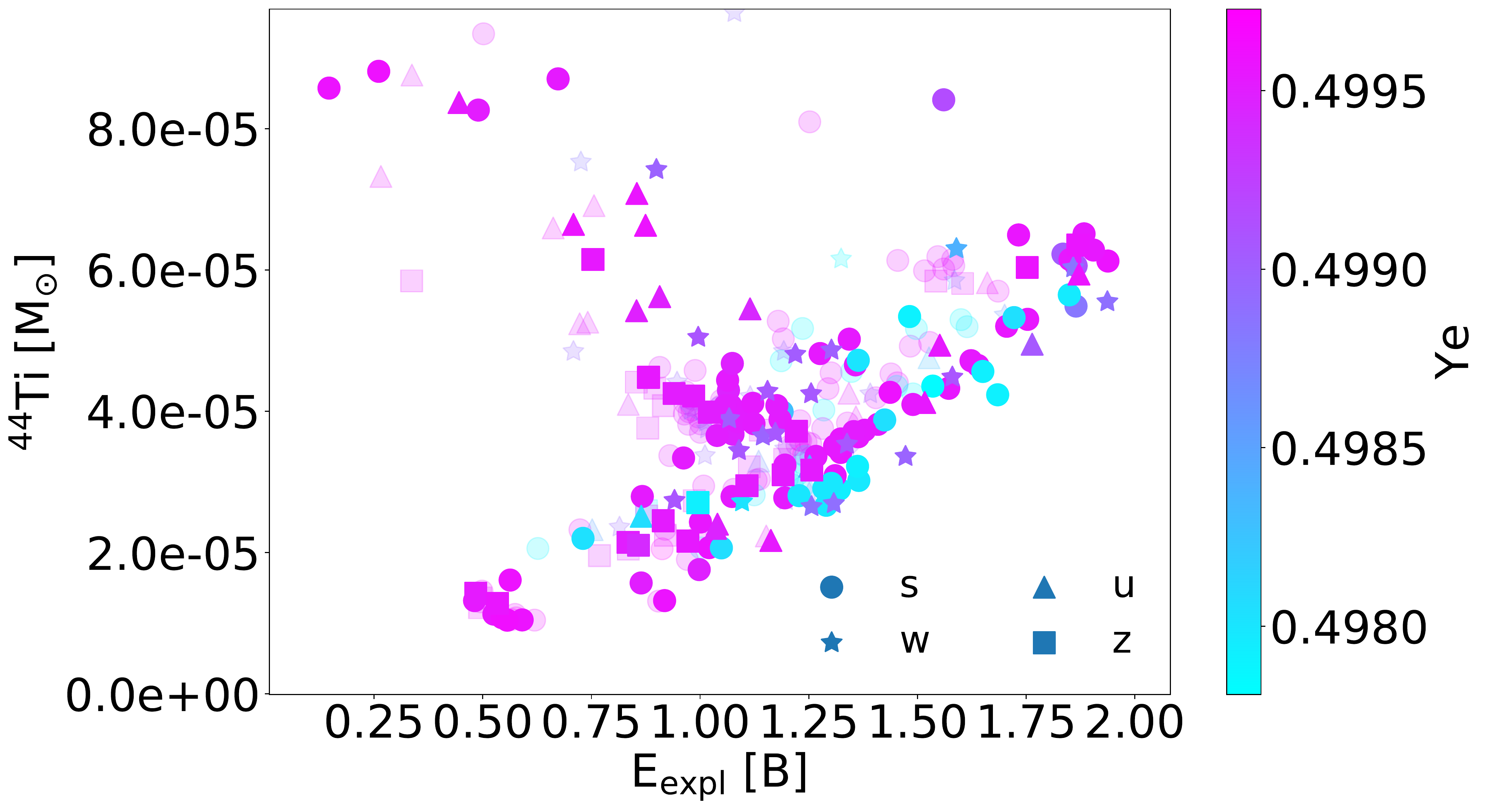}
    \caption{
    \edit1{Same as figure~\ref{fig:elemental_explvsyields}, but for isotopic yields of $^{56}$Ni, $^{57}$Ni, $^{58}$Ni and $^{44}$Ti (from top to bottom).}
    \label{fig:isotopic_explvsyields}
    }
\end{figure}

\subsection{Nucleosynthesis results from all eight nuclear EOS}
\label{subsec:nucleo.alleos}

In this section, we expand our discussion to include all eight nuclear EOS models considered in this work. We will use three models (s16.0, u16.0, and z16.0 -- one ZAMS mass at each metallicity) for this.

\begin{figure}
\centering
    \includegraphics[width=0.5\textwidth]{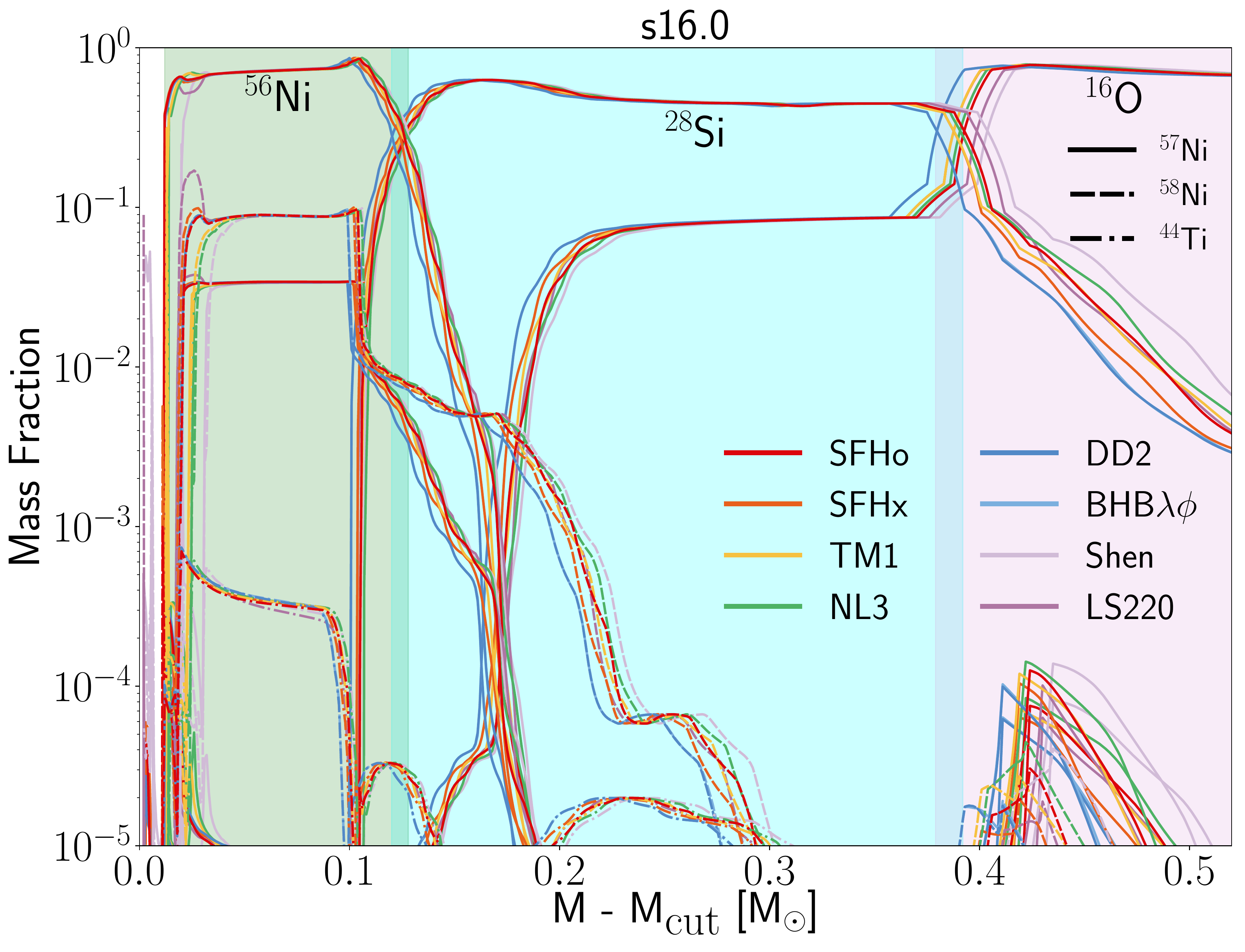} 
    \includegraphics[width=0.5\textwidth]{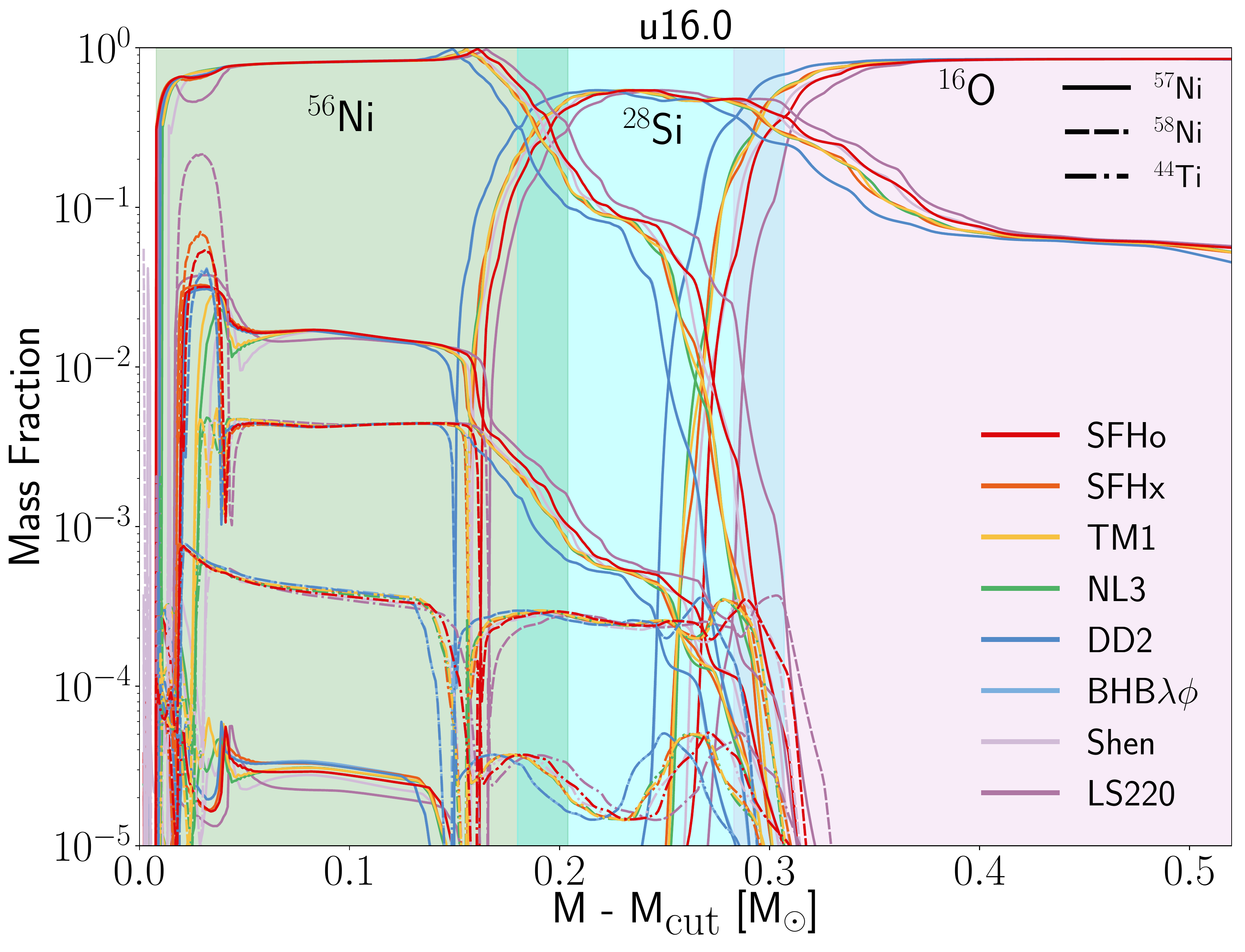}
    \includegraphics[width=0.5\textwidth]{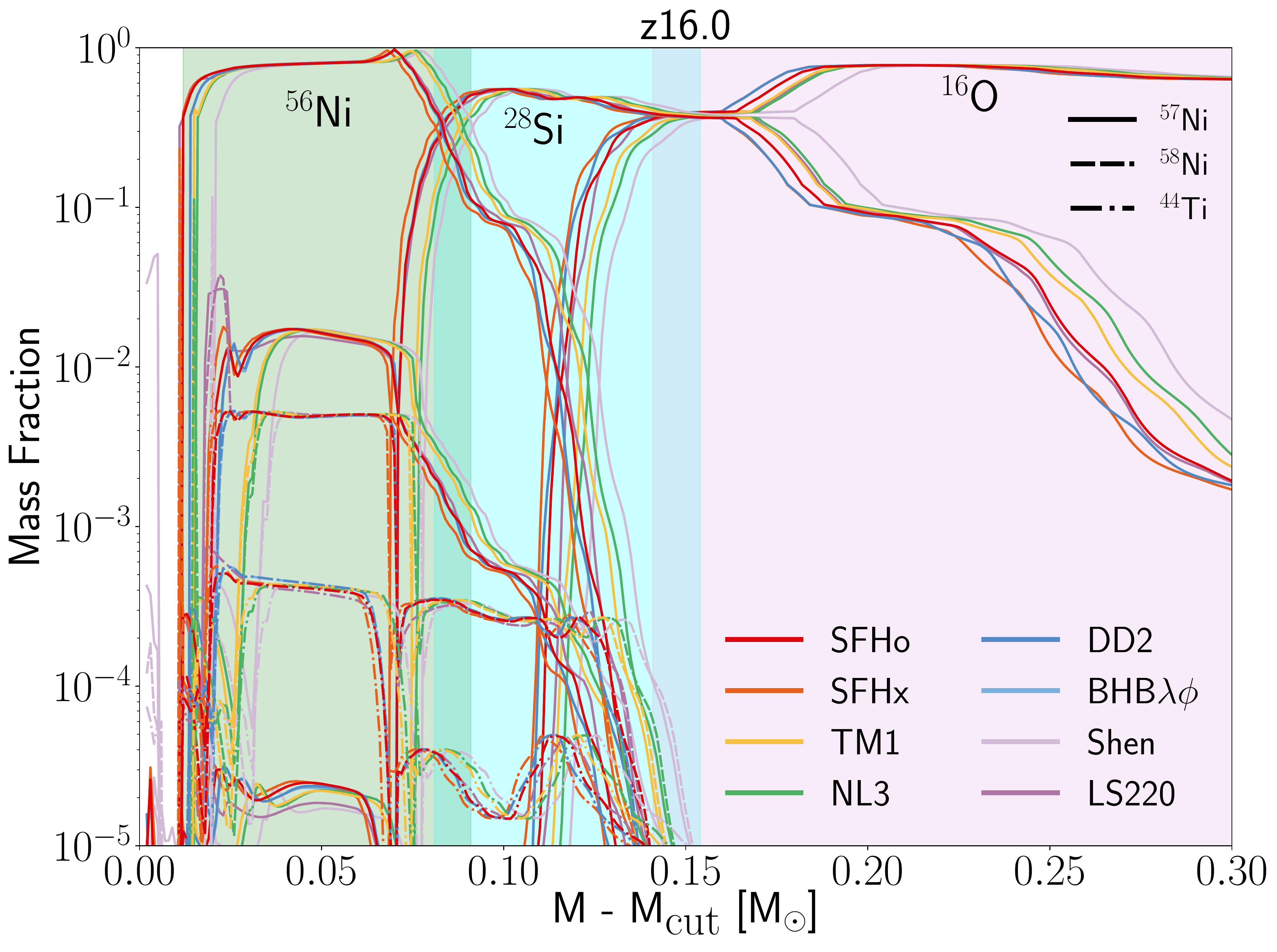}
    \caption{Post-explosion composition profile as a function of the mass coordinate outside the mass cut for the 16 \msun progenitor from the s-series (top), the u-series (middle), and the z-series (bottom). The background shading denotes the $^{56}$Ni, $^{28}$Si, and $^{16}$O layers. The overlap in shading indicates the differences due to different EOS models used.
    \label{fig:post_expl}    
    }
\end{figure}

We start by continuing the discussion of the isotopic yields of $^{56}$Ni, $^{57}$Ni, $^{58}$Ni and $^{44}$Ti.
In figure \ref{fig:post_expl} we plot the post-explosion mass fraction of $^{56}$Ni, $^{57}$Ni, $^{58}$Ni, and $^{44}$Ti (together with $^{16}$O and $^{28}$Si for reference) as a function of the mass coordinate outside of the mass cut. The shaded background in each plot marks the combined width of the $^{56}$Ni, $^{16}$O and $^{28}$Si layers for all EOS models. The overlap between the different layers indicates the maximum and minimum width among all eight cases. 

Overall, the $^{56}$Ni layer follows a quite similar behavior for all EOS models on a progenitor by progenitor basis. Among the three progenitors, the low-metallicity u16.0 model has the broadest $^{56}$Ni layer and also the largest difference in the location of the transition to the $^{28}$ Si layer between different EOS models (0.178 - 0.204 \msun). The same trend is also present in the transition between the $^{28}$Si and $^{16}$O layer. The solar metallicity model s16.0 has a thinner $^{56}$Ni layer, with the Ni layer extending to $\sim 0.125 - 0.127 M_{\odot}$ (the exceptions are DD2 and BHB$\lambda\phi$ at 0.119 - 0.120 \msun). The zero-metallicity model z16.0 has the thinnest $^{56}$Ni layer of the three progenitors discussed here. The different EOS models fall into two groups with respect to the transition from $^{56}$Ni to $^{28}$Si. For most EOS models, this transition occurs 0.08 - \edit1{0.082} \msun outside of the mass cut, only for TM1, NL3, and the Shen EOS it occurs at a slightly larger mass coordinate (0.085 - 0.087 \msun outside of the mass cut).

$^{44}$Ti is co-produced with $^{56}$Ni, but at about three orders of magnitude lower levels. As for $^{56}$Ni, the $^{44}$Ti mass fraction is similar between different EOS models. The effect of using different EOS models is most visible at the edges of the $^{44}$Ti-rich zone. For the s16.0 models, these differences are small. They are largest for u16.0, where for SFHo and LS220 the (relative) $^{44}$Ti-rich region extends to furthest towards the $^{28}$Si layer. 

$^{57}$Ni (and $^{58}$Ni) are co-produced with $^{56}$Ni. The peak mass fractions are very similar among different EOS models within the same progenitor. However, at the inner edge of the $^{57}$Ni-rich zones, we see variations (up to a factor of 2) between different EOS models. This is due to EOS-induced differences in the local electron fraction in these innermost ejecta layers. The $^{58}$Ni mass fractions show the same behavior, but even more pronounced than for $^{57}$Ni (up to a factor of 10). For both $^{57}$Ni and $^{58}$Ni, the mass fractions at the inner edge are enhanced as compared to the main value throughout the $^{57,58}$Ni-rich zones. Whether $^{57}$Ni or $^{58}$Ni are produced at higher levels also depends on the local electron fraction. In the case of s16.0 and z16.0, these layers have a slightly lower \Ye (0.4978 - 0.4980) than the corresponding layer in u16.0 (which have a \Ye of 0.4984 - 0.4988). Hence, the more neutron-rich isotope $^{58}$Ni dominates over $^{57}$Ni in s16.0 and z16.0.

\begin{figure}
    \centering
    \includegraphics[width=0.5\textwidth]{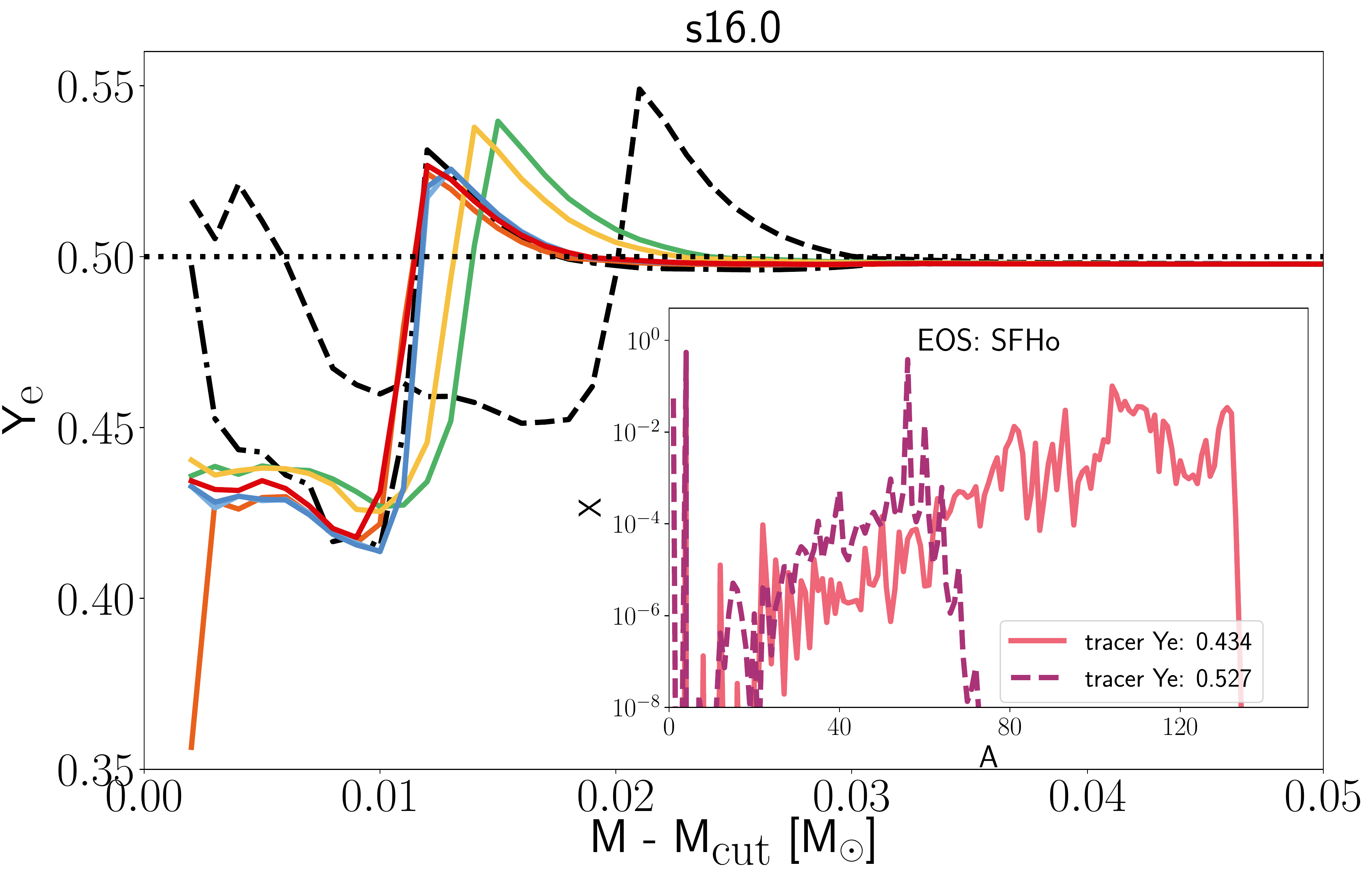}
    \includegraphics[width=0.5\textwidth]{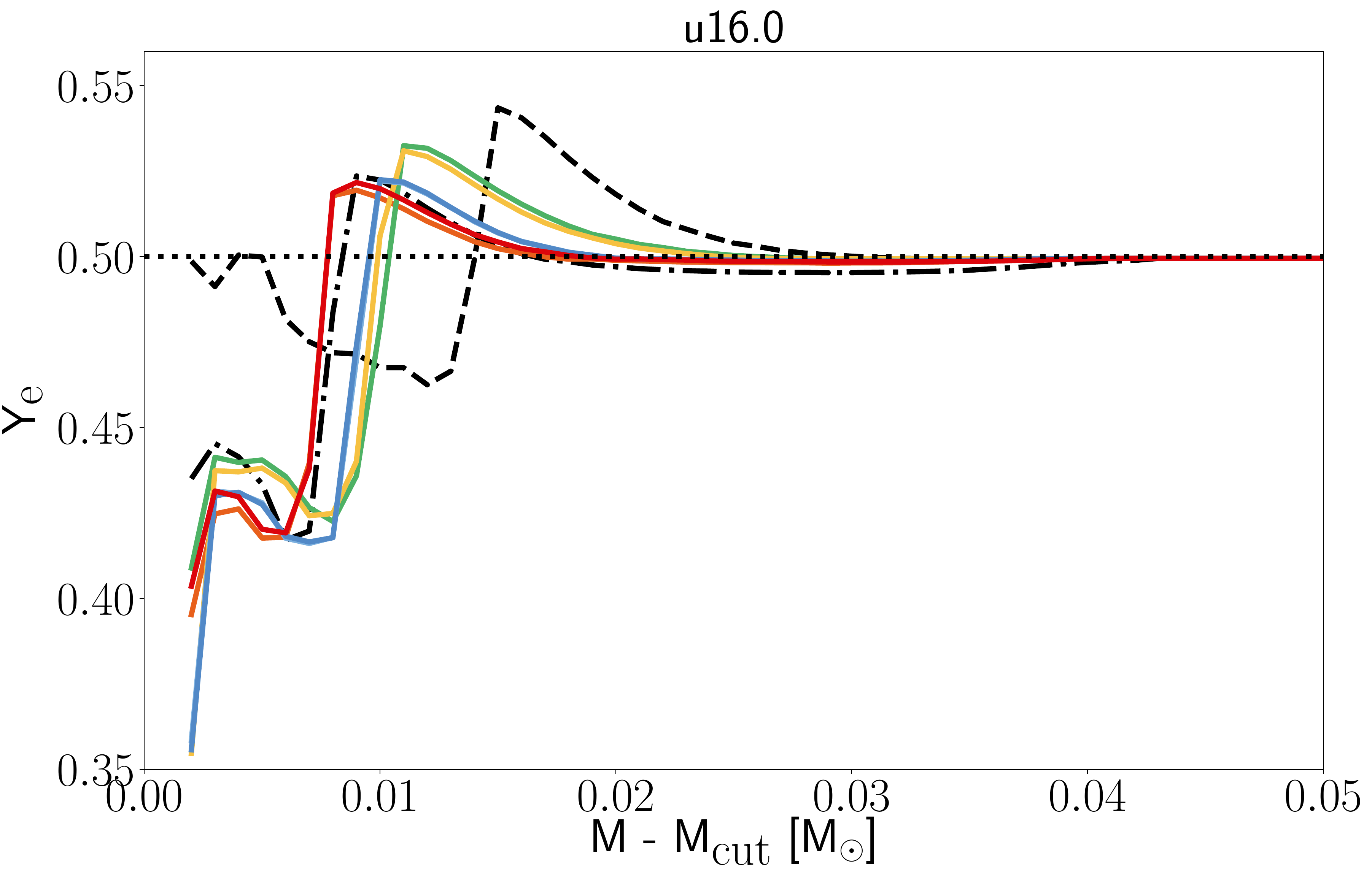}
    \includegraphics[width=0.5\textwidth]{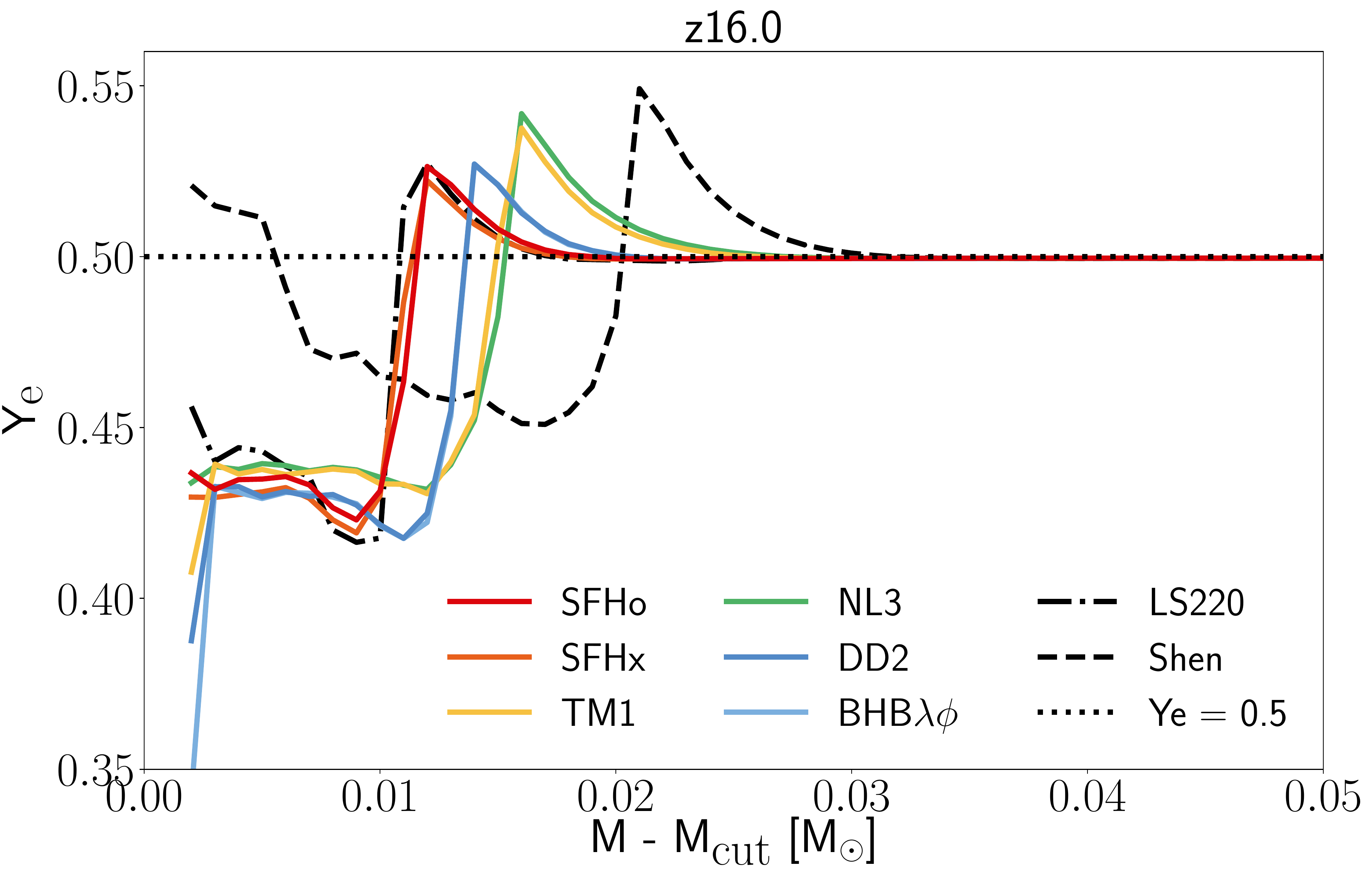}
    \caption{Post explosion \Ye ~evolution for the 16 M$_{\odot}$ progenitor of the s (top), u (middle) and z-series (bottom) as a function of the mass coordinate outside of the mass cut. 
    The insert shows the final mass fractions from two representative tracers, one at $Y_e=0.434$ and one at $Y_e=0.527$.
    \label{fig:ye_evol}
    }
\end{figure}

For a given progenitor, the conditions in the explosive Si-burning layers are very similar between simulations with different EOS models. Hence, any difference in the abundances synthesized in those layers is due to differences in the local \Ye value. In figure \ref{fig:ye_evol}, we plot the \Ye for the different EOS models as a function of the mass coordinate outside the mass cut. We omit our innermost ejected tracer which represents the neutrino-driven wind. Our mass resolution of $10^{-3} M_{\odot}$ is not sufficiently fine to resolve the wind properly.

\begin{figure}
    \centering
    \includegraphics[width=0.5\textwidth]{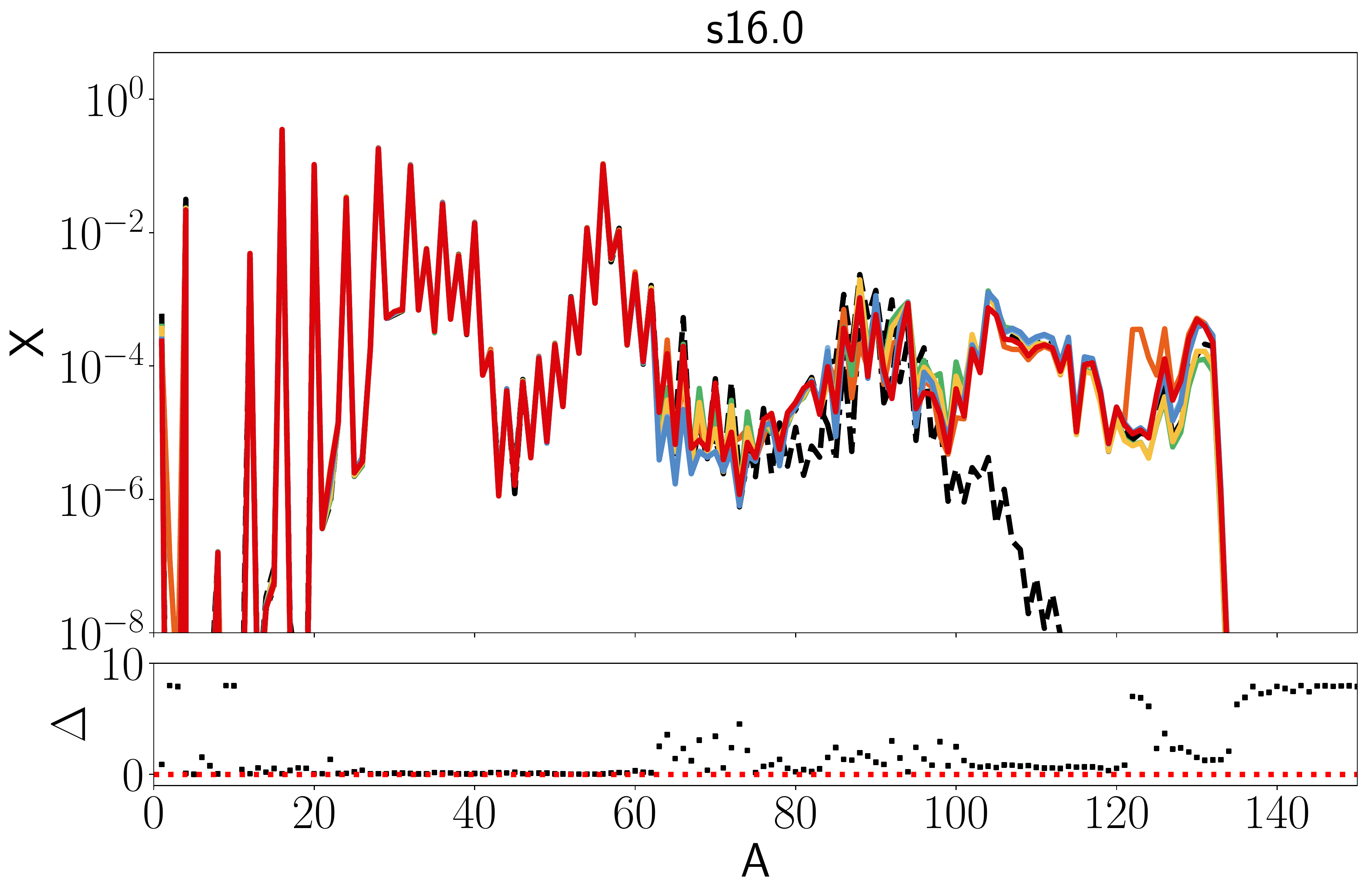}
    \includegraphics[width=0.5\textwidth]{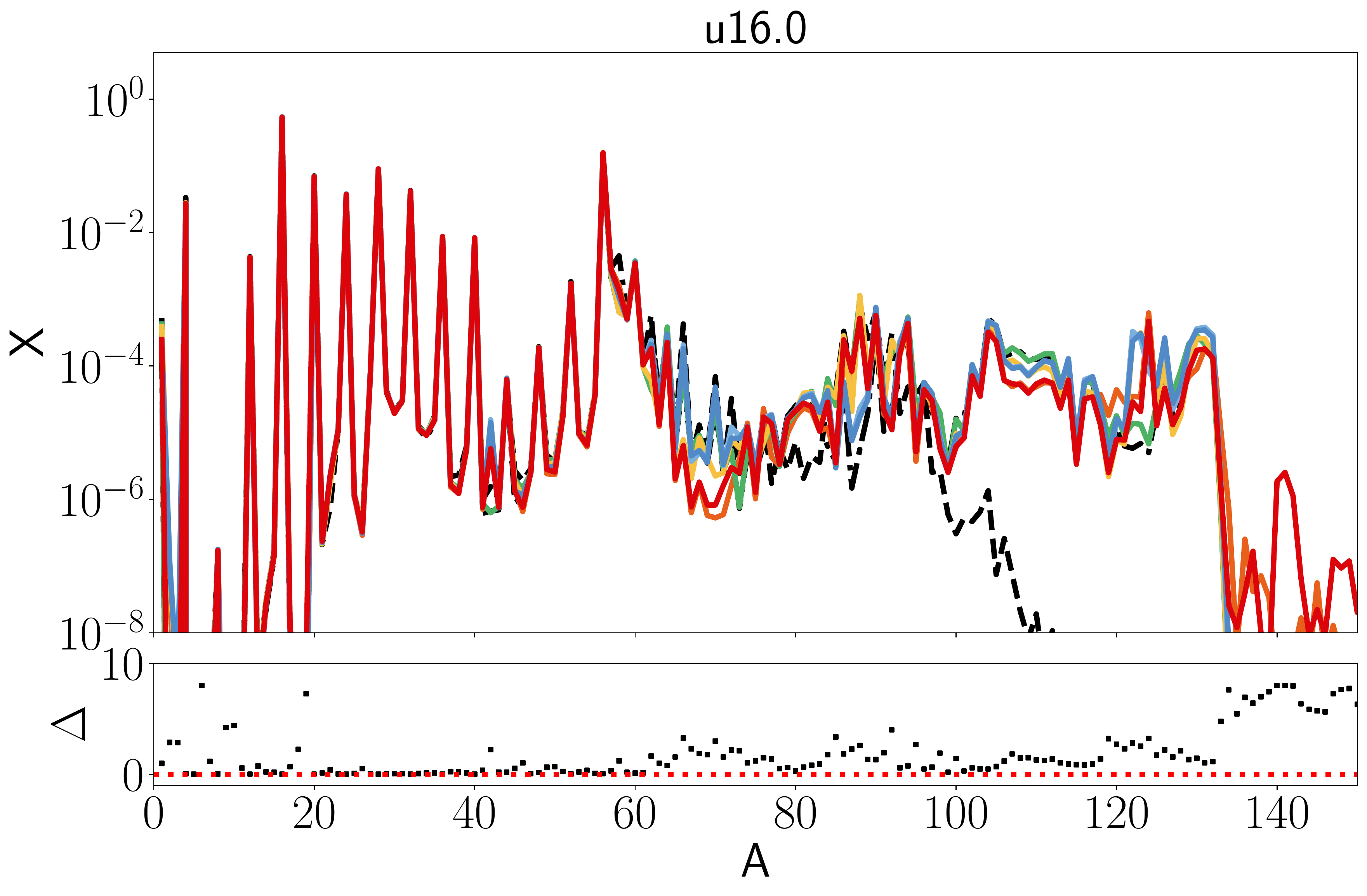}
    \includegraphics[width=0.5\textwidth]{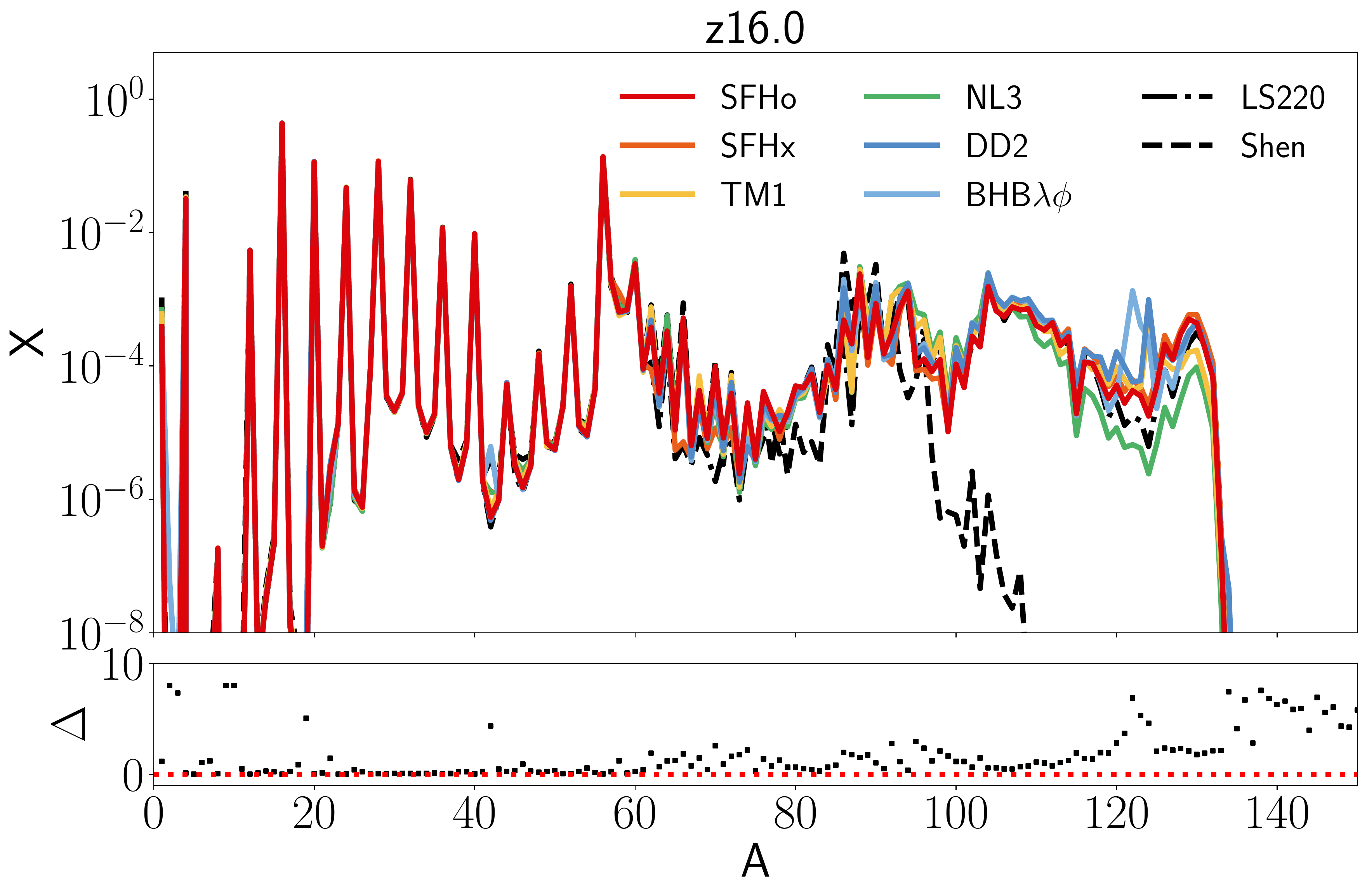}
    \caption{Final yields after decay for the 16 M$_{\odot}$ progenitor of the s (top), u (middle) and z-series (bottom) as a function of the mass number with eight different EOS models. The bottom insert shows $\Delta = |X_{\mathrm{max}} - X_{\mathrm{min}}|/X_{\mathrm{avg}}$,  the difference between the highest and the lowest abundance for each mass number normalized by the average over all eight cases.
    \label{fig:finalyieldsvsA}
     }
\end{figure}

For all three progenitors shown, the innermost $\sim 0.01 M_{\odot}$ of the ejecta are neutron-rich ($Y_e \simeq 0.43 - 0.44$). The variations among six of the EOS models (SFHo, SFHx, TM1, NL3, DD2, and BHB$\lambda\phi$) are smaller than the different between this group and the two reference EOS models (LS220 and Shen). In particular, the Shen EOS is quite different and in one case does not reach $Y_e$ values below 0.46. These layers of relatively low $Y_e$ are where mostly the isotopes with mass numbers $A \gtrsim 130$ are synthesized (see insert in top panel of figure \ref{fig:ye_evol} for the abundances from a representative tracer at $Y_e = 0.434$). This can also be seen in the final overall abundances of u16.0 (see figure \ref{fig:finalyieldsvsA}) where the calculations with the Shen EOS have a distinct lack of material with $A\gtrsim 130$ in the final abundances. 

Outside of the innermost neutron-rich layers there is a region of proton-rich ejecta, seen as peak in the $Y_e$-evolution. This occurs for every EOS and every progenitor model in our study. The peak $Y_e$-value that is reached and the amount of proton-rich material is slightly different for different EOS models. However, overall these proton-rich layers undergo alpha-rich and proton-rich freeze-out. 
The differences in final mass fractions are about 3-8\% (see bottom inserts of figure \ref{fig:finalyieldsvsA} which show the difference between the highest and lowest mass fraction for a given \added{mass number} $A$ normalized to the average mass fraction at this $A$).

\begin{table*}   
\begin{center}
	\caption{Explosion properties of s16.0, u16.0 and z16.0 models with different EOS. 
    	\label{tab:model_prop}
	}
	\begin{tabular}{l*{9}{c}}
	\tableline \tableline 
Model & EOS & $\xi_{2.0\mathrm{,b}}$ & E$_{\mathrm{expl}}$ & M$_{\mathrm{cut}}$ & Layer of M$_{\mathrm{cut}}$ & $^{56}$Ni & $^{57}$Ni & $^{58}$Ni \\

 &   &   & (B) & (\msun)  &   & (\msun) & (\msun) & (\msun)  \\
	\tableline
s16.0 & SFHo & 0.234048 & 1.3650 & 1.5244 & Si-O & 8.00$\times 10^{-2}$ & 3.13$\times 10^{-3}$ & 7.95$\times 10^{-3}$ \\

 & SFHx & 0.234293 & 1.2413 & 1.5300 & Si-O & 7.78$\times 10^{-2}$ & 3.10$\times 10^{-3}$ & 7.90$\times 10^{-3}$ \\

 & TM1 & 0.235293 & 1.4071 & 1.5293 & Si-O & 7.83$\times 10^{-2}$ & 2.96$\times 10^{-3}$ & 7.59$\times 10^{-3}$ \\
       
 & NL3 & 0.235675 & 1.4607 & 1.5263 & Si-O & 7.98$\times 10^{-2}$ & 2.99$\times 10^{-3}$ & 7.29$\times 10^{-3}$ \\
 
 & DD2 & 0.232932 & 1.2365 & 1.5372 & Si-O & 7.54$\times 10^{-2}$ & 2.96$\times 10^{-3}$ & 7.41$\times 10^{-3}$ \\

 & BHB$\lambda \phi$ & 0.232932 & 1.2362 & 1.5372 & Si-O & 7.55$\times 10^{-2}$ & 2.96$\times 10^{-3}$ & 7.41$\times 10^{-3}$ \\

 & LS220 & 0.240236 & 1.4177 & 1.5183 & Si-O & 7.87$\times 10^{-2}$ & 3.17$\times 10^{-3}$ & 8.80$\times 10^{-3}$ \\

 & Shen & 0.233976 & 1.5861 & 1.5142 & Si-O & 7.59$\times 10^{-2}$ & 2.80$\times 10^{-3}$ & 7.11$\times 10^{-3}$ \\

\tableline

u16.0 & SFHo & 0.407177 & 1.8681 & 1.7433 & Si-O & 1.49$\times 10^{-1}$ & 2.66$\times 10^{-3}$ & 1.32$\times 10^{-3}$ \\

 & SFHx & 0.407626 & 1.7126 & 1.7525 & Si-O & 1.45$\times 10^{-1}$ & 2.64$\times 10^{-3}$ & 1.50$\times 10^{-3}$ \\

 & TM1 & 0.409084 & 1.8459 & 1.7560 & Si-O & 1.43$\times 10^{-1}$ & 2.21$\times 10^{-3}$ & 0.59$\times 10^{-3}$ \\

 & NL3 & 0.412725 & 1.8548 & 1.7559 & Si-O & 1.43$\times 10^{-1}$ & 2.07$\times 10^{-3}$ & 0.59$\times 10^{-3}$ \\

 & DD2 & 0.403058 & 1.6606 & 1.7689 & Si-O & 1.36$\times 10^{-1}$ & 2.38$\times 10^{-3}$ & 0.96$\times 10^{-3}$ \\

 & BHB$\lambda \phi$ & 0.403042 & 1.6575 & 1.7690 & Si-O & 1.36$\times 10^{-1}$ & 2.38$\times 10^{-3}$ & 1.00$\times 10^{-3}$ \\

 & LS220 & 0.415827 & 2.0133 & 1.7322 & Si-O & 1.53$\times 10^{-1}$ & 2.86$\times 10^{-3}$ & 4.36$\times 10^{-3}$ \\

 & Shen & 0.405708 & 1.9770 & 1.7388 & Si-O & 1.42$\times 10^{-3}$ & 2.01$\times 10^{-3}$ & 0.72$\times 10^{-3}$ \\

\tableline

z16.0 & SFHo & 0.140235 & 0.9396 & 1.4917 & Si-O & 5.61$\times 10^{-2}$ & 0.79$\times 10^{-3}$ & 0.26$\times 10^{-3}$ \\

 & SFHx & 0.140364 & 0.8507 & 1.4959 & Si-O & 5.44$\times 10^{-2}$ & 0.81$\times 10^{-3}$ & 0.49$\times 10^{-3}$ \\

 & TM1 & 0.140896 & 1.0493 & 1.4884 & Si-O & 5.71$\times 10^{-2}$ & 0.75$\times 10^{-3}$ & 0.23$\times 10^{-3}$ \\

 & NL3 & 0.141000 & 1.0973 & 1.4859 & Si-O & 5.78$\times 10^{-2}$ & 0.72$\times 10^{-3}$ & 0.33$\times 10^{-3}$ \\

 & DD2 & 0.139895 & 0.9201 & 1.4949 & Si-O & 5.41$\times 10^{-2}$ & 0.76$\times 10^{-3}$ & 0.26$\times 10^{-3}$ \\
 
 & BHB$\lambda \phi$ & 0.139895 & 0.9011 & 1.4961 & Si-O & 5.41$\times 10^{-2}$ & 0.76$\times 10^{-3}$ & 0.25$\times 10^{-3}$ \\

 & LS220 & 0.138217 & 0.9372 & 1.4871 & Si-O & 5.73$\times 10^{-2}$ & 0.91$\times 10^{-3}$ & 0.41$\times 10^{-3}$ \\

 & Shen & 0.140313 & 1.1802 & 1.4784 & Si-O & 5.57$\times 10^{-2}$ & 0.68$\times 10^{-3}$ & 0.54$\times 10^{-3}$ \\

	\tableline
	\end{tabular}
\end{center}
\end{table*}

\section{Comparison with observations}
\label{sec:observations} 

In this section, we compare the results of our simulations to observations of local supernovae (section \ref{subsec:sne}) and of metal-poor stars (section \ref{subsec:emps}). Observations of CCSNe provide us with important information about the explosion energy and the amount of $^{56}$Ni produced. Metal-poor stars carry in their atmospheres the signature of one or a few previously exploded core-collapse supernovae.

\subsection{Local supernovae}
\label{subsec:sne}

\begin{figure*}
    \centering
    \includegraphics[width=0.8\textwidth]{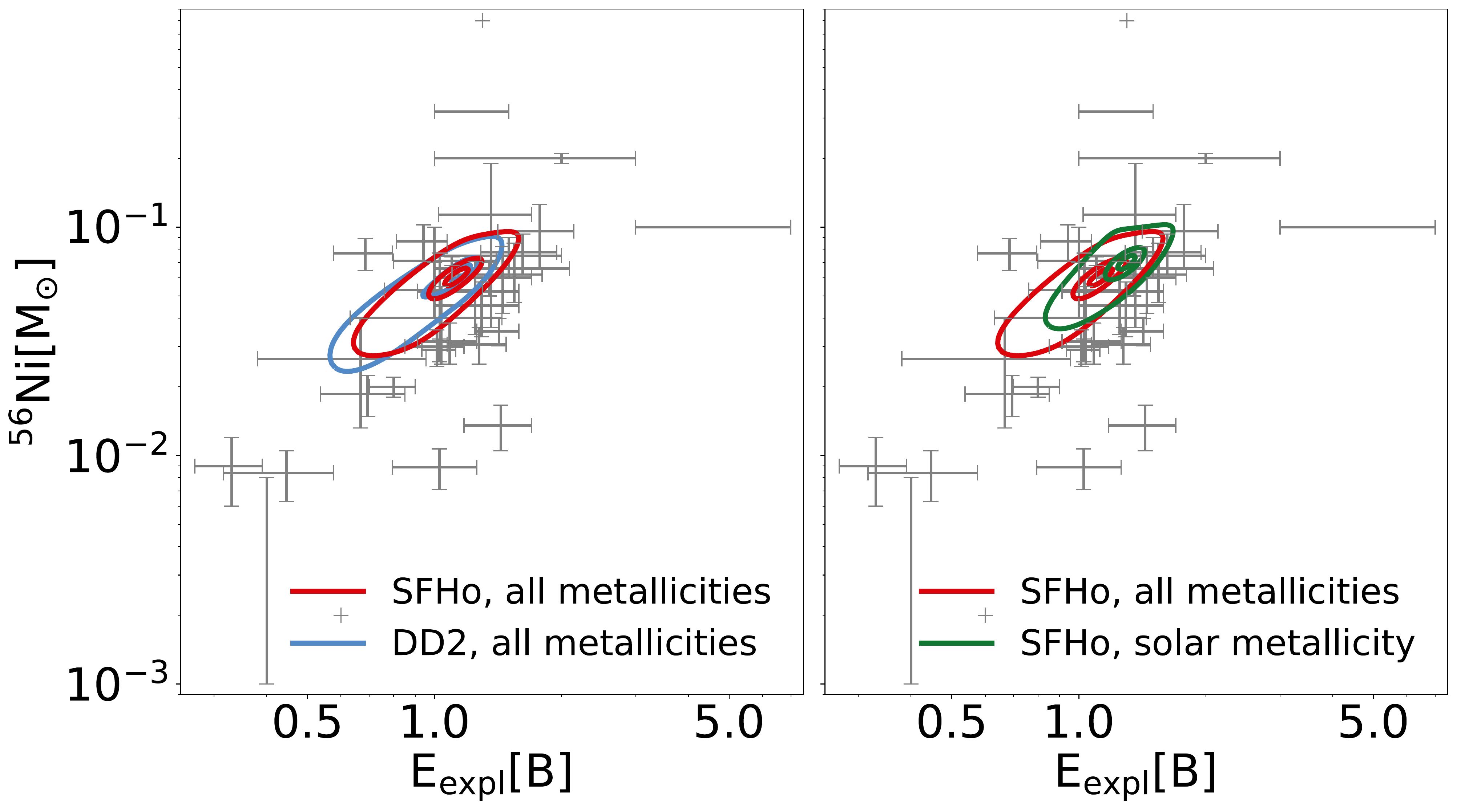}
    \caption{Left panel: Kernel density estimate (KDE) of the explosion energy - $^{56}$Ni mass IMF-weighted distributions for SFHo (red) and DD2 (blue) for all models combined.  The contour lines represent $1\sigma$, $2\sigma$ and $3\sigma$ confidence levels. The observed SNe are shown with gray crosses. 
    \edit1{Right panel: Same for SFHo only, comparing the KDE for all models (red, same as on the left) to the KDE for solar metallicity models only (green). }
    \label{fig:heatmaps}
    }
\end{figure*}

In this section, we compare the results of our simulations to the observations of supernovae in the local universe ($z<0.01$). The observational data are the same as in \paperIV~ (cf.\ Table 4 in \paperII~ and \citet{muller.prieto.ea:2017} for the original references). 
\edit1
{The left panel of Figure~\ref{fig:heatmaps} shows the $1\sigma$, $2\sigma$, and $3\sigma$ confidence intervals of the Kernel Density Estimate (KDE) for the $^{56}$Ni yields and explosion energy (both IMF-weighted) for simulations using the SFHo EOS (blue, this work) and for simulations using the DD2 EOS (red, \paperII~and \paperIV). The confidence intervals are computed for the combined results from all four progenitor series (solar, low and zero metallicity). 
In the right panel, we show -- for the SFHo EOS only -- the impact of excluding low and zero metallicity models (u- and z-series) from the KDE (green) compared to including all metallicities in the KDE (red). Note that the red contour lines are the same in both panels. }

The KDE for the combined the $^{56}$Ni yields and explosion energies \added{(left panel)} is centered around $m(^{56}\mathrm{Ni})=0.06$ and $E_{\mathrm{expl}}=1.13$~B. Almost all of the observational data overlap with the $3\sigma$ confidence interval for our simulation results. There are a few outliers with very low and very high $^{56}$Ni masses, originating from low mass ($< 10$ \msun) and very high mass progenitors ($> 40$ \msun) which we do not include in our study.

The KDE confidence levels for the combined $^{56}$Ni masses and explosion energies are very similar between the SFHo and the DD2 case. At the $3\sigma$ confidence level the slightly higher explosion energies obtain with SFHo (together with the corresponding slightly lower explosion energies for DD2) become visible in the contour line.

\edit1
{As mentioned earlier in this section, the observations are of SNe in the local Universe (redshift $z<0.01$). Hence, in the right panel, we compare the KDE for solar metallicity models only to the observations. In this case, the KDE is centered around 0.07~\msun and 1.3~B, which is a shift towards higher $^{56}$Ni mass and higher explosion energy. This is because the low and zero metallicity models (u- and z-series) explode with lower explosion energy and make slightly smaller amounts of $^{56}$Ni. The shift in the distribution is in the direction of where the observational data are more concentrated.}

Overall, we can conclude that the results from our method are consistent with constraints from observations for both the SFHo and the DD2 EOS. Unfortunately, the differences between the two EOS models is too small in this analysis to identify which one is the more favored EOS.

\subsection{Metal-poor stars}
\label{subsec:emps}

The atmospheres of low-mass long-lived, metal-poor stars can provide us with information about the nucleosynthesis processes in CCSNe, originating from short-lived massive stars. These CCSNe would have deposited their yields in the gas from which these low-mass metal-poor stars have formed. Here, we compare the yields predicted from our simulations with the observationally derived abundances in a well-observed metal-poor star. We focus on the Fe-group elements which are formed in primary explosive nucleosynthesis processes in the explosion.

\begin{figure}
    \centering
    \includegraphics[width=0.48\textwidth]{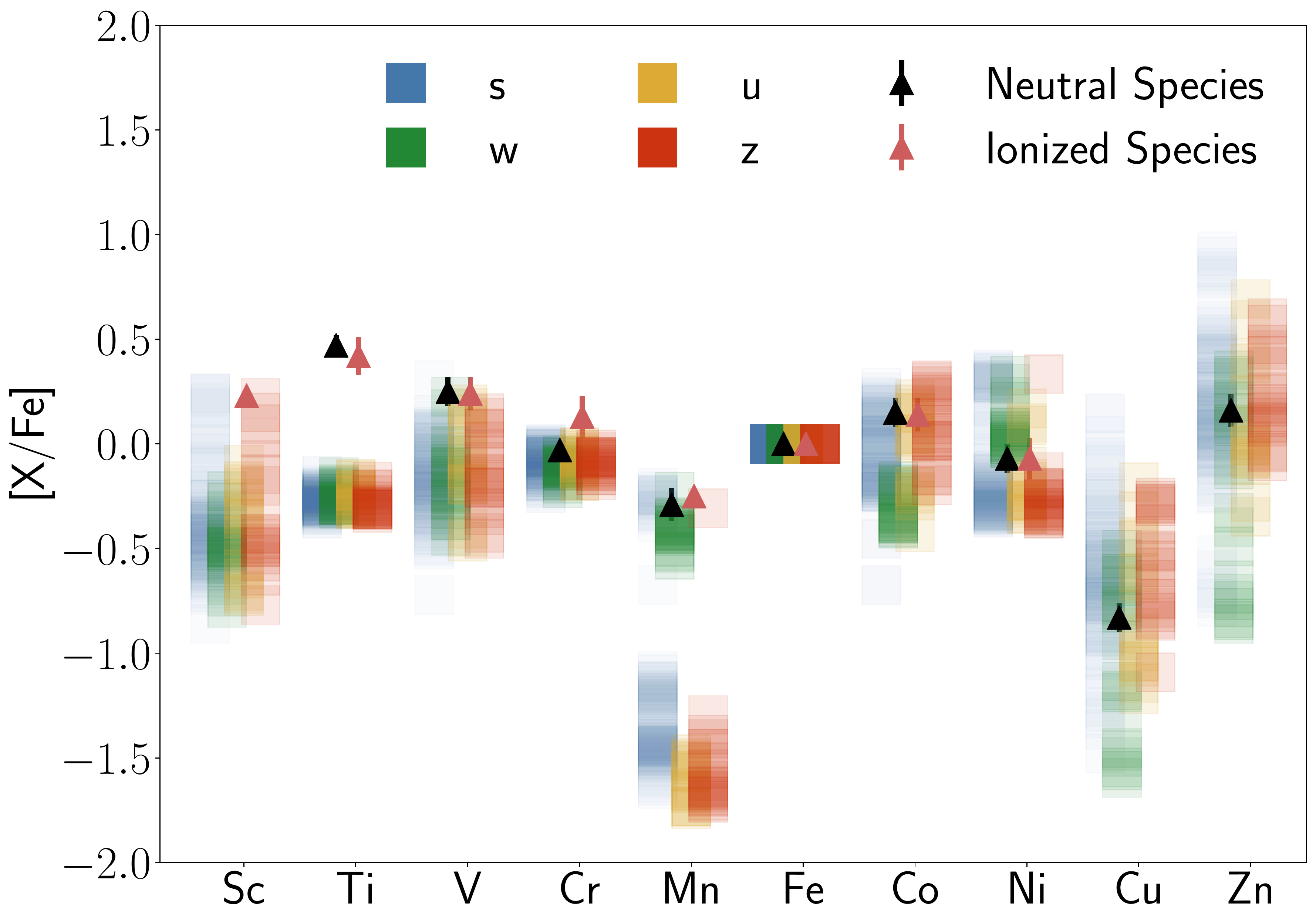}
    \caption{Abundances of Fe-group elements for all the SFHo exploding models of the s- (blue), w- (green), u- (yellow) and z-series (red). The triangles are observationally derived abundances for metal-poor star HD~84937.
    \label{fig:XFe_plots}
    }
\end{figure}

We compare the predicted iron-group elements from our simulations with the observed abundances of metal-poor star HD~84937 in figure \ref{fig:XFe_plots}. The triangles are the abundances derived from neutral and singly-ionized transitions, taken from \cite{MPS}. Each colored square in the figure corresponds to one exploding model (using the SFHo EOS). Note that our results are not IMF-weighted to illustrate the range of yields for each element (and hence how sensitive the results are to different conditions found in different models).

Overall, our data agree with observations. We find that Scandium and Zinc are synthesized at levels comparable to the observed values [Sc/Fe] and [Zn/Fe], respectively. Both of these elements are difficult to produce in sufficient amounts in nucleosynthesis calculations that neglect the neutrino interactions and employ a canonical explosion energy of 1~B, like the traditional piston and thermal bomb set up. In case of hypernovae, the enhanced explosion energies lead to overproduction of Sc and Zn even without the inclusion of neutrino interactions (\citealt{nomoto06}). 
\cite{cf06a} used a careful treatment of neutrino interactions which led to an enhanced production of Sc and Zn at canonical explosion energies of 1~B. Our models presented here co-produce Zn with Fe and we find a smaller spread of [Zn/Fe] for low metallicity and zero metallicity progenitors as compared to solar-metallicity progenitors. We also find that only the w-series (solar metallicity) progenitors consistently produce Mn in agreements with observations. From the other progenitors series, only a handful of s- (solar metallicity) and z-models (zero metallicity) produce similarly high levels of Mn. Most of the s- and z-models and all u-models (low metallicity) strongly underproduce Mn compared to observations. This is due to the size of the network used during stellar evolution, which sets the range of possible $Y_e$-values. In the case of a small network (s-, u-, and z-series) the $Y_e$ cannot take intermediate values required for high Mn production. Since the production of Mn is quite sensitive to the local \Ye value, the local $Y_e$ leaves an imprint on the resulting [Mn/Fe] ratio. Titanium is typically systematically underproduced in spherically symmetric nucleosynthesis models (cf.\ also \paperI~for a detailed discussion of the uncertainties in the Ti yields).

\section{Remnant properties}
\label{sec:remnants}

\begin{figure}
    \centering
    \includegraphics[width=0.45\textwidth]{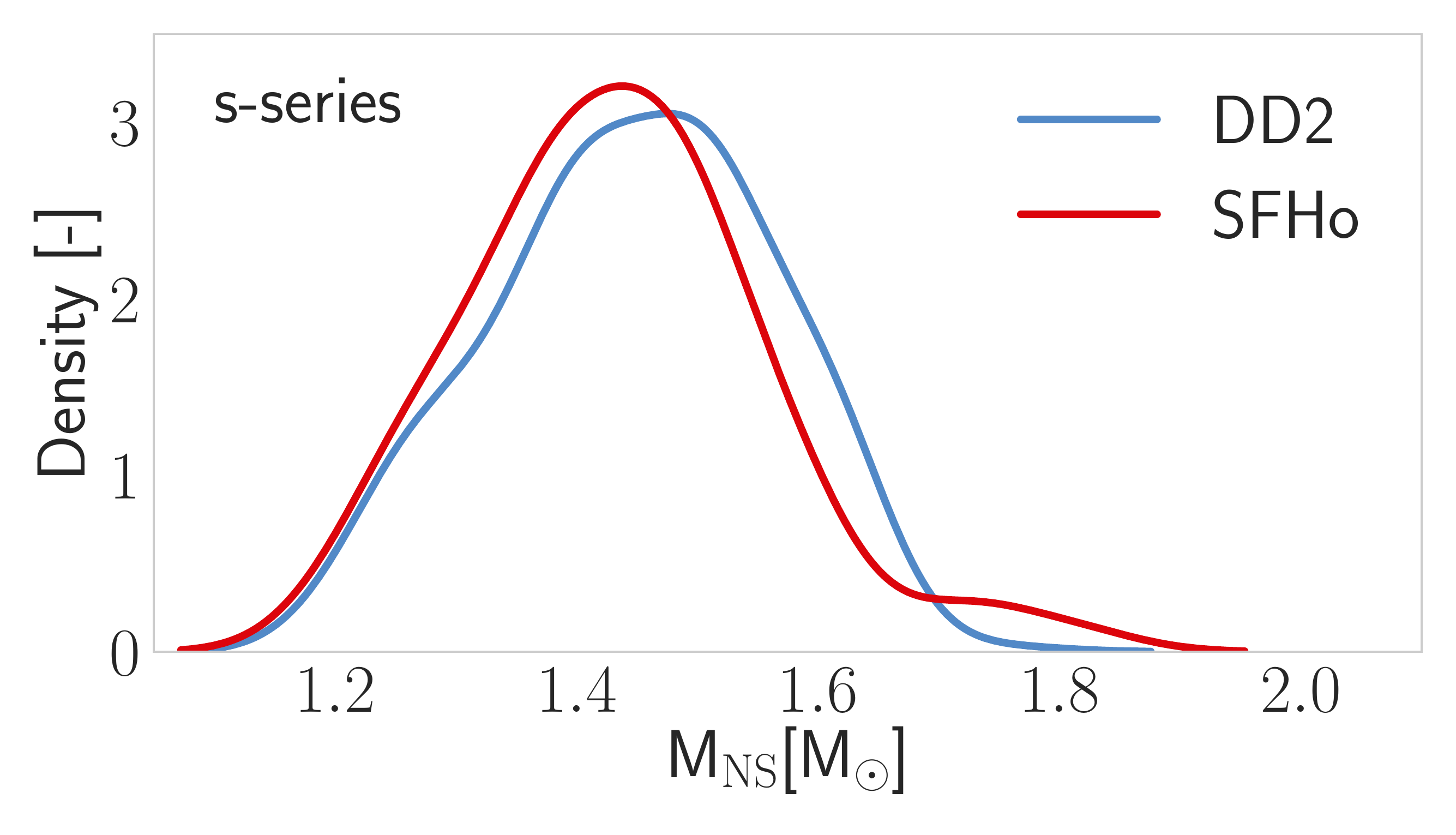}
    \includegraphics[width=0.45\textwidth]{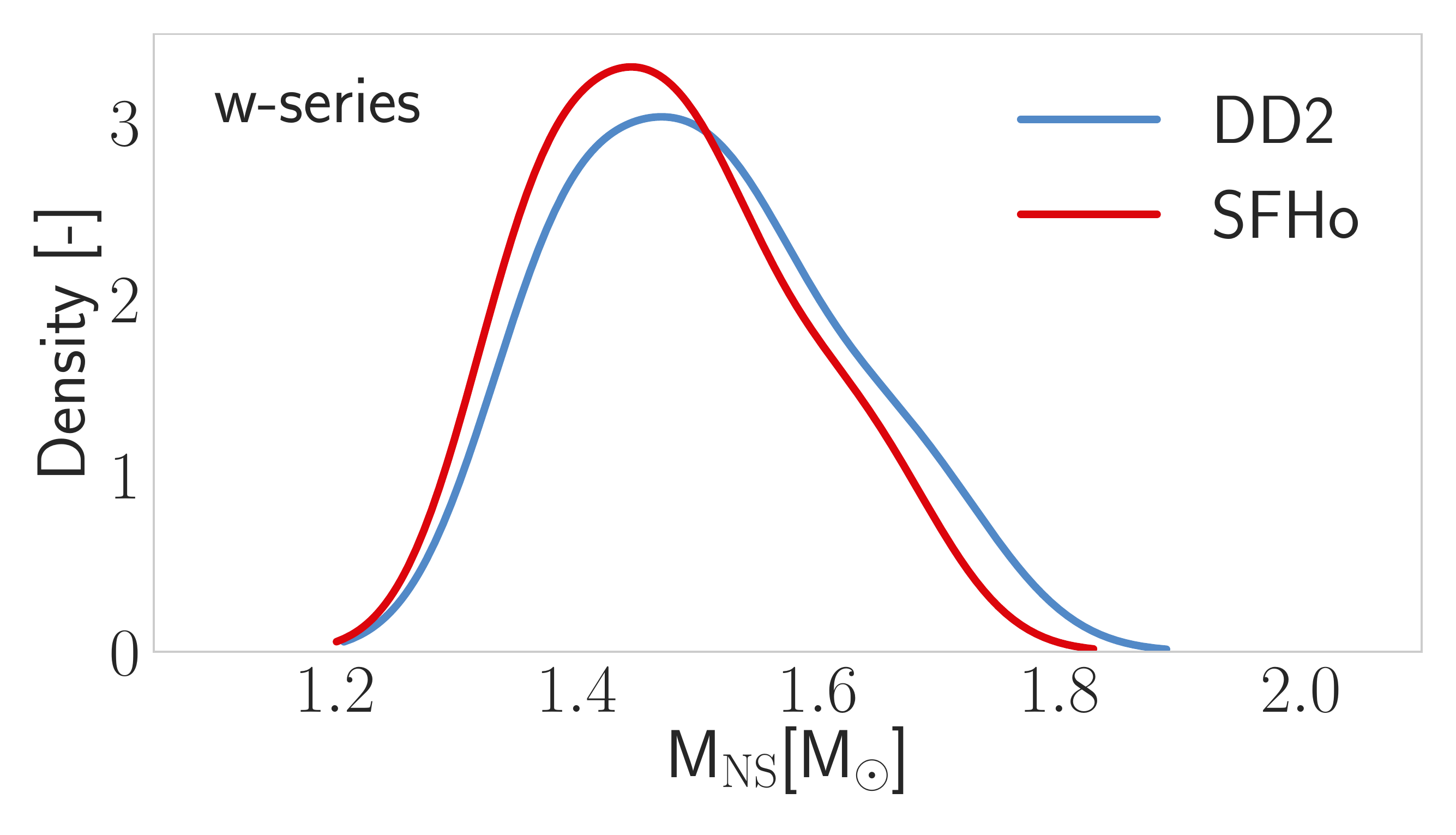}
    \includegraphics[width=0.45\textwidth]{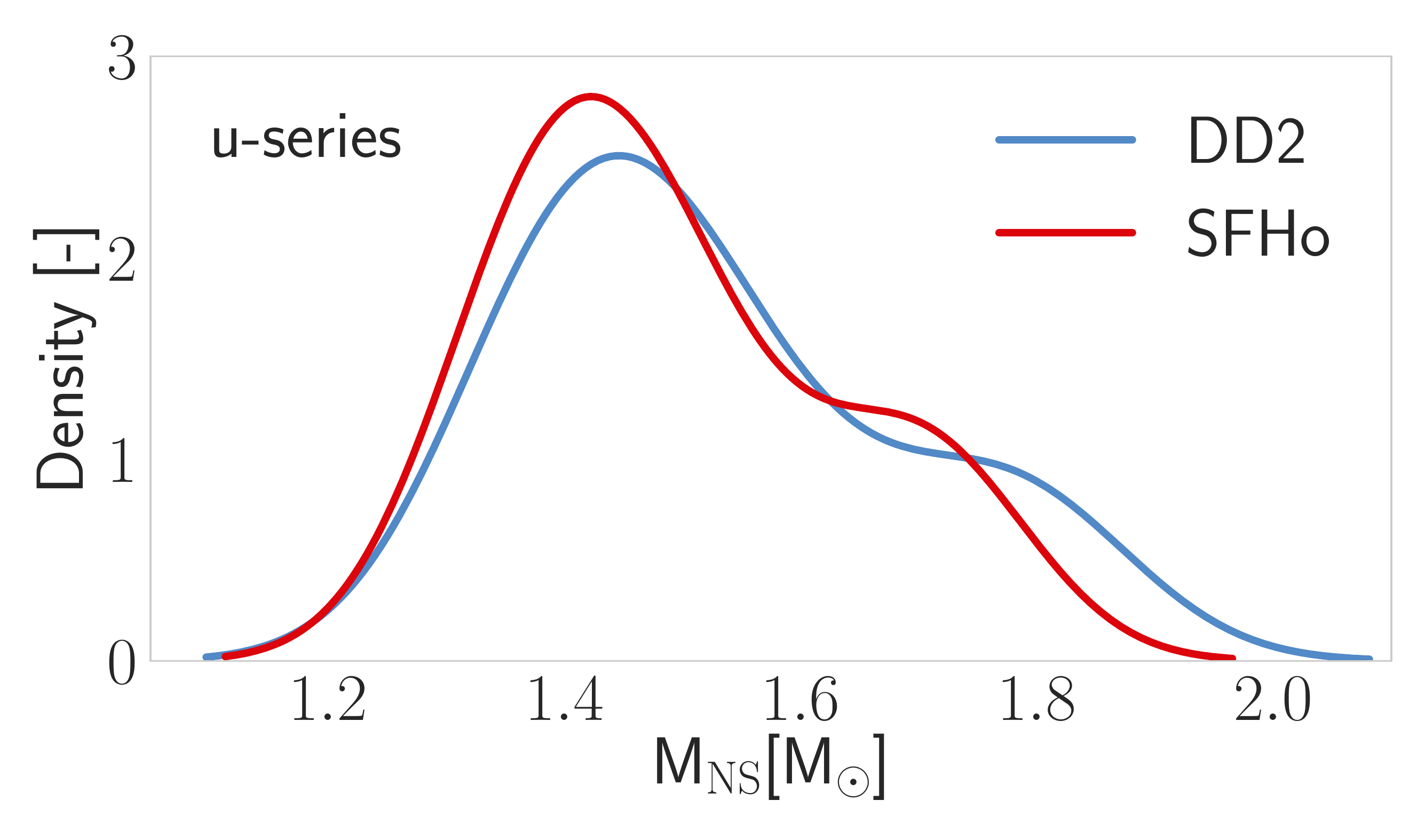}
    \includegraphics[width=0.45\textwidth]{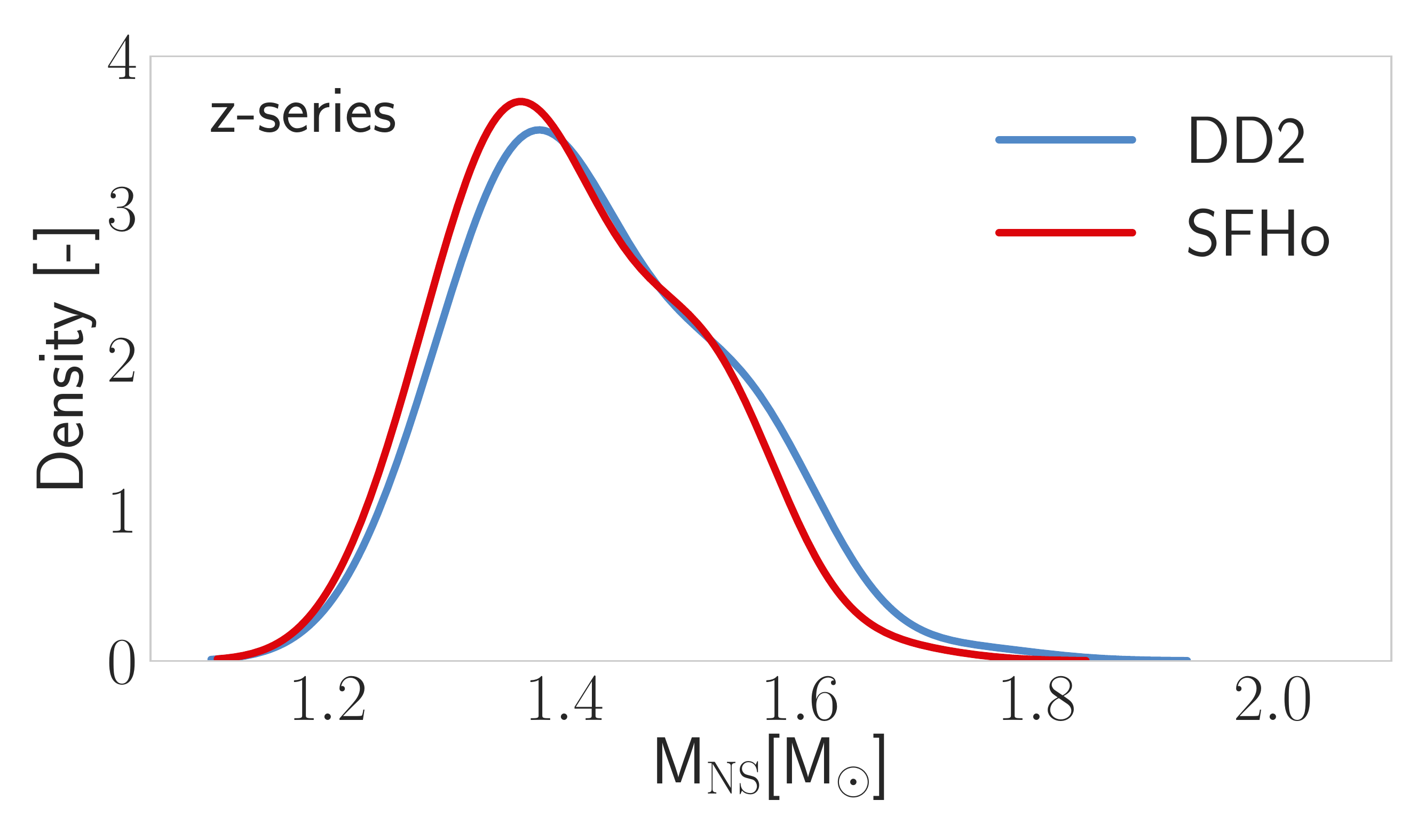}
    \caption{Gravitational birth mass \edit1{KDEs} for cold NSs for the s-series, w-series, u-series, and z-series (from top to bottom). Simulations using the SFHo EOS are shown in red. Simulations using the DD2 EOS are shown in blue (data from \paperII ~and \paperIV).
    \label{fig:NS_mass_distr}
    }
\end{figure}

\begin{figure}
    \centering
    \includegraphics[width=0.4\textwidth]{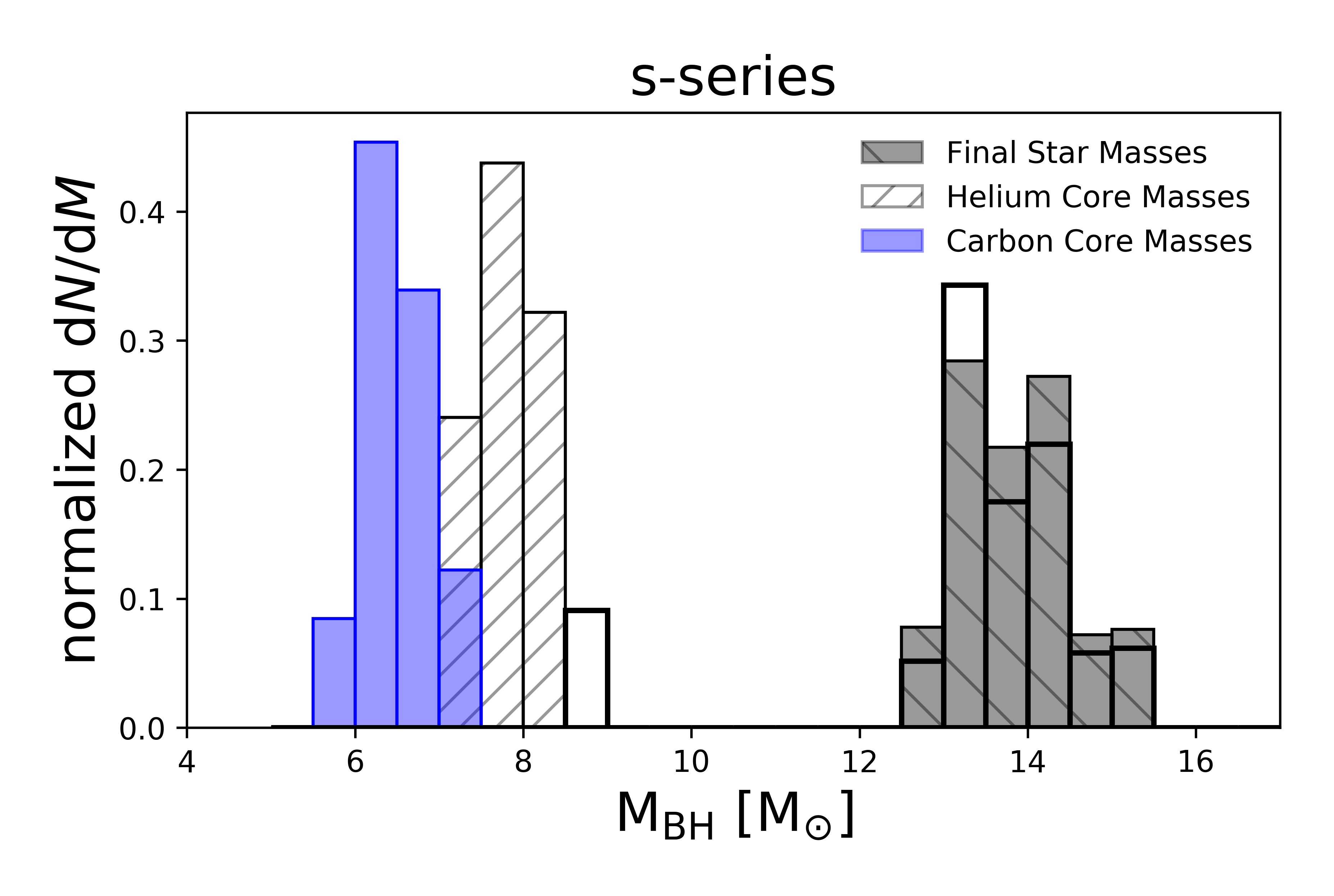}
    \includegraphics[width=0.4\textwidth]{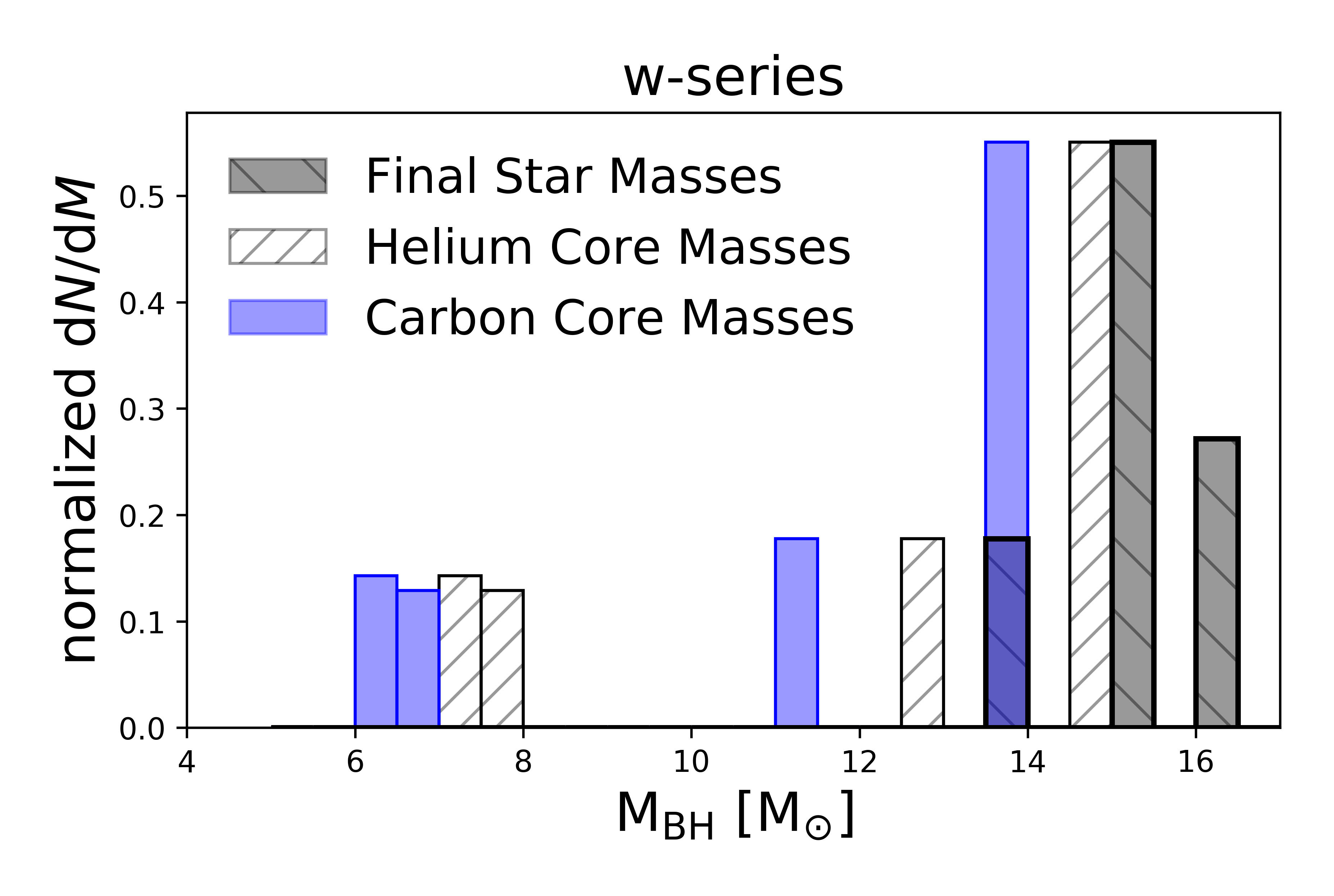}
    \includegraphics[width=0.4\textwidth]{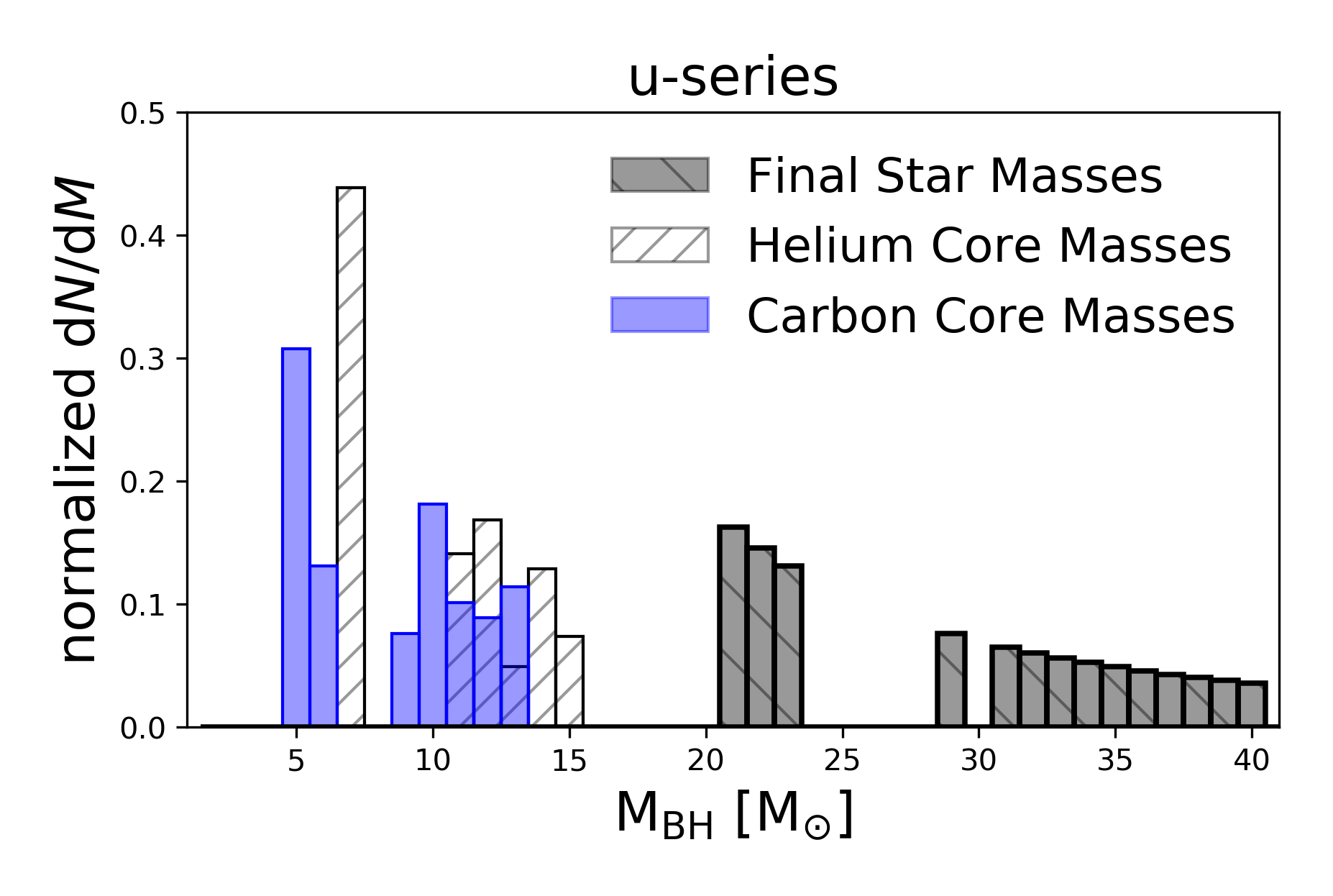}
    \includegraphics[width=0.4\textwidth]{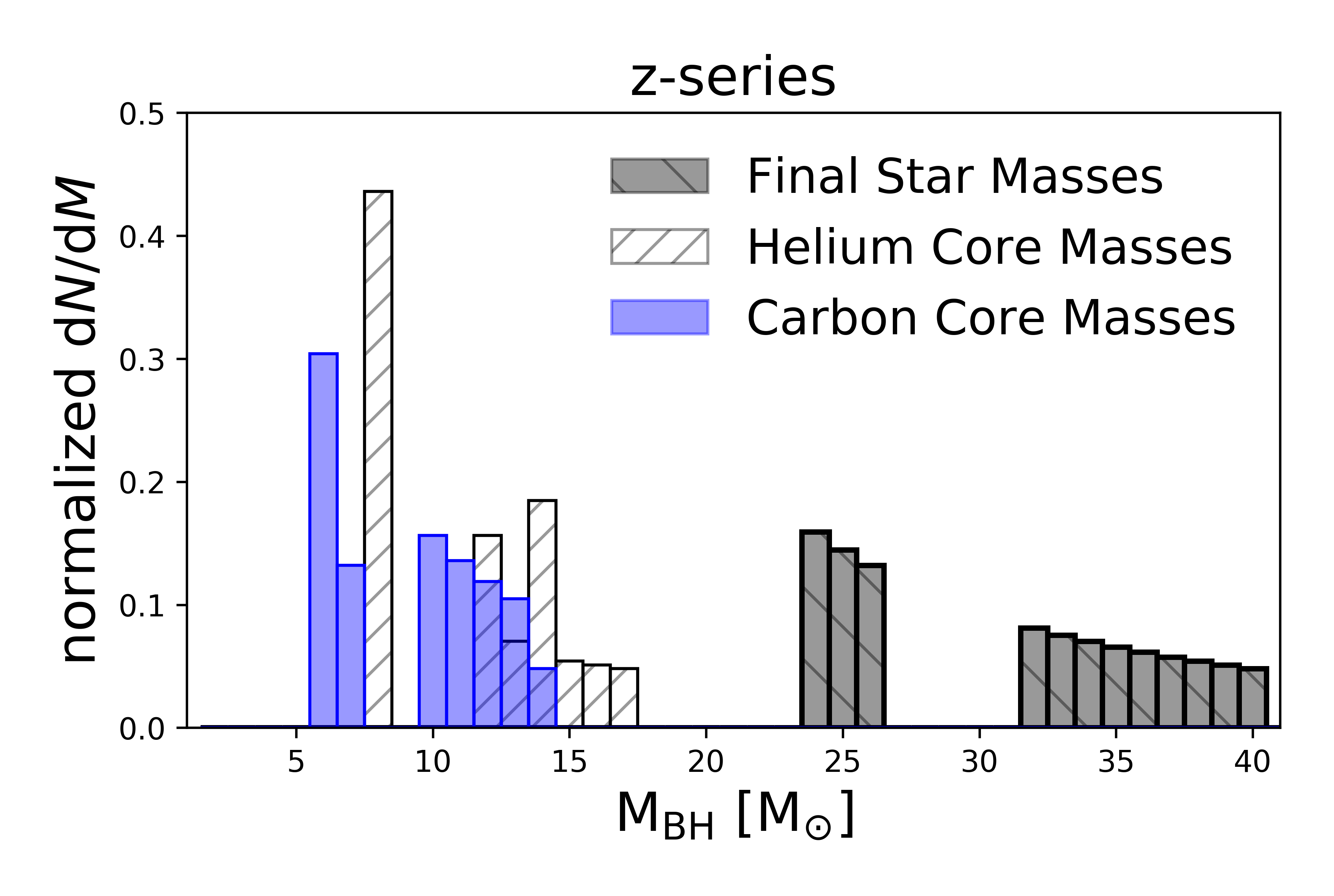}
    \caption{Birth mass distributions of BHs for the s-series, w-series, u-series, and z-series (from top to bottom). The different shaded bars indicate three different cases of possible BH mass distributions depending on how much of the initial stellar mass ultimately contributes to the final BH mass.
    \label{fig:BH_mass_distr}
    }
\end{figure}

The mass distribution of the compact remnants formed in our simulations of core collapse are yet another result -- in addition to the explosion energies, ejecta mass, and elemental yields -- that can be compared with observations.
We use the procedure described in Paper~II and Paper~IV to calculate the remnant mass distribution. We obtain the baryonic mass of the newly formed hot NS by following the evolution of the PNS. We then use the nuclear EOS to calculate the gravitational mass of the corresponding cold NS. 
Finally, we compute the birth mass distribution of the compact remnants by weighting the simulated remnant mass with the Salpeter initial mass function (IMF) for massive stars \citep{salpeter}.

First, we discuss the distribution of NSs. Figure \ref{fig:NS_mass_distr} shows the IMF-weighted \added{KDE of the} mass distribution for cold NSs for all four progenitor series. For each series we have considered pre-explosion masses between 11-40 \msun. The predicted NS masses lie in the range of 1.2--1.8~\msun. 

For the low and zero metallicity series, most of the high ZAMS mass models do not explode and hence are not available to form higher mass NSs. 
 \edit1
 {For all the four series the distribution shows a single peak. For the s-series, the small bump around 1.75~\msun for SFHo comes from the mixed models discussed in section~\ref{subsec:mixedmodels}. For the u-series, most of the exploding models form lower mass NS except for a few high-compactness models which have a lower explosion energy and make more massive NSs (u20.0, u24.0, and u30.0). These models contribute to the plateau around 1.75~\msun. For the z-series, the slight change in shape around 1.55~\msun also comes from exploding models (z17.0 and z18.0) with very high compactness (cf. figure 1 in Paper~IV) which therefore make more massive NSs.}

Comparing the results obtained with the SFHo EOS (this work) to the those obtained with the DD2 EOS (\paperII~and \paperIV), we find that the 
 distribution is shifted to the left, i.e.\ the NSs resulting from the SFHo EOS are less massive than those from the DD2 EOS. This is consistent with the slightly higher explosion energies and earlier explosions found with SFHo as compared to DD2.

Our study is based on single, non-rotating stars. As such, the results do not include the possible effects that are present in binary-star system, like accretion or mass loss. 
Difficulty may arise when we compare the theoretically predicted NS masses with the observed NS masses as the measurements of NS masses are from binary systems. However, \cite{raithel18} argue that we can still make such a comparison as the single star model can be considered as a representative of some close binary scenarios due to the uncertain nature of mass loss. \added{Most available NS mass measurements are between 1.1 - 2.0 \msun \cite{2016NSmass_measurement}, which is similar to what we obtain from our simulations.}

Next, we turn to the simulations that did not explode successfully and instead formed a BH. In our simulation framework, this includes models which run longer than the time when PUSH is active without exploding as well as models which exhibit a very rapid increase in the central density above nuclear saturation density which we interpret as formation of a BH. Note that the Agile-IDSA code cannot follow the formation of the BH due to the metric used. 
Figure \ref{fig:BH_mass_distr} shows the predicted birth mass distribution for all four progenitor series. 
The stellar mass at the time of collapse is determined by the star's mass-loss history. This in turn influences the mass of the BH formed as a result of a failed explosion. Other processes like the loss of the PNS binding energy in a weak shock (\citealt{Lovegrove13}) or the stripping of the envelope by a binary companion before collapse may alter the mass ultimately collapsing to a BH. We study the impact of these effects on the BH mass distribution by considering three different cases covering a range of outcomes.
Our most massive estimate for the BH mass is based on a scenario where the entire stellar mass at the time of core collapse ends up in the BH. 
Our lightest BH mass estimate is for the case where the star is stripped of its outer layers and only the CO-core collapses to a BH. 
In our third (intermediate) case, we follow \citet{kochanek2014} and assume that He-core mass sets the mass of the BH.

Generally, the low and zero metallicity models make more massive BHs than their solar metallicity counterparts. This is a result of the low metallicity progenitors experiencing less mass loss compared to the solar metallicity ones and therefore have retained more mass at the time of collapse. In the BH distribution for the u- and z-series we see two clusters of BH masses separated by a gap. This is a reflection of the `island of non-explodability' near 25~\msun~ (at 21-24 \msun for u-series and 23 - 27 \msun for z-series) found in our work and also in the literature \citep{ertl16,sukhbold16,mueller2016}.
This gap is not visible in the BH distributions for the solar metallicity series. The s-series has a larger number of models, whereas the w-series forms less BHs than the low/zero metallicity series. Interestingly, the w-series has monotonically increasing CO-core masses with increasing ZAMS mass, whereas the s-series has a more complex trend of CO-core mass as a function of the ZAMS mass (cf. figures 2 and 3 in Paper~II). 
We find that the BH mass distributions from the SFHo EOS are very similar to those obtained with the DD2 EOS. Hence, the maximum BH mass is the also the same in both cases.

As sample table of the remnant masses (NSs and BHs) from our simulations is given in Appendix \ref{app:remnantmasses} in table \ref{tab:remnantmasses}. The full table is available as machine-readable data.

\section{Summary} \label{sec:summary}

In this paper we simulate the collapse and explosion of non-rotating massive stars at three different metallicities. We compute the resulting remnants (neutron stars or black holes) and the detailed nucleosynthesis yields in a post-processing approach. We perform two suits of simulations for this. One, a complete suite for all progenitors at all metallicities using the SFHo nuclear EOS. And two, a suite of \edit1{4} progenitors (16 \msun at three metallicities \added{and 25 \msun at solar metallicity}) using eight different EOS: SFHo, SFHx, TM1, NL3, DD2, \BHBlp, LS220 and Shen. 

For the first set of simulations (SFHo EOS), we find that the explosion energies are generally higher with the SFHo EOS when compared to the DD2 EOS. This is due to the higher neutrino and anti-neutrino luminosities for SFHo. Higher explosion energies come hand-in-hand with earlier explosions and less massive remnants. Overall, the outcome of core collapse is similar between the SFHo and DD2 EOS, however, we find that three models (s25.0, s25.2 and s39.0) with very high compactness explode with the SFHo EOS but not with the DD2 EOS.
For our second set of simulations, we find that (for the three progenitors studied) the explosion energy varies by 15\% - 18\% across the different EOS, but the impact on the PNS mass is minimal (about 1\%). 

For the nucleosynthesis yields, we find the elements such as Fe and Ni are correlated with the explosion energy. For example, we find higher yields for example with the SFHo EOS which has higher explosion energies than with DD2, however the difference is small. Other elements, such as Mn, depend strongly on the local electron fraction. And yet other elements (e.g.\ Co) are affected both by the explosion energy and the electron fraction.

For the SFHo EOS and the symmetric isotopes $^{56}$Ni and $^{44}$Ti, we confirm the strong correlation of the final yields with the explosion energy found in our previous work and in other works. Interestingly, we find lower $^{44}$Ti yields for some SFHo models despite their higher explosion energies. The non-symmetric, neutron-rich isotopes ($^{57}$Ni and $^{58}$Ni) on the other hand are strong \Ye dependent. A higher \Ye results in overall less $^{57}$Ni and $^{58}$Ni. 

In the comparison of all the isotopic yields, we find only small differences for $^{56}$Ni and $^{44}$Ti across the eight EOS used. However, for $^{57}$Ni and $^{58}$Ni we find large differences that originate from the different \Ye profiles for each EOS. Both the exact local value of \Ye as well as the amount of material at that \Ye value affect the final yields of these isotopes. 
All our models have some amount of neutron-rich material ($Y_e \sim 0.43)$ in the inner-most ejecta.  In these layers, isotopes up to mass number 130 are produced through a weak r-process. Our models also have proton-rich ejecta (also in the innermost ejecta, adjacent to the neutron-rich ejecta in mass coordinate). The peak \Ye values can be as high as 0.52. In these layers, we find the products of proton-rich and alpha-rich freeze-out. 

We also compare our explosion results with observed supernovae. We find that our models are (within the $3\sigma$level) in agreement with the combined explosion energies - $^{56}$Ni observations of local CCSNe. While we find a difference between simulations using the SFHo EOS and simulations using the DD2 EOS, they are too small to rule out one of these EOS. 
When comparing our predicted iron-group yields with the observationally derived abundances in a metal-poor star, we also find a general agreement, except for Mn and for Ti. This is no surprise, as Ti is traditionally underproduced in spherically symmetric simulations.

Finally, we compute the IMF-weighted distributions of neutron star masses and black hole masses, which can be compared to the observed distributions of neutron stars and black holes. For the SFHo EOS, we find lower neutron star masses than with the DD2 EOS, which slightly shifts the NS distribution to lower masses. For the BH distribution, we find the same results as in our previous work using the DD2 EOS. The three models, which explode with SFHo and fail to explode with DD2, do not have a significant impact on the distributions.

\begin{acknowledgments}
The authors would like to thank Matthias Hempel for providing the EOS tables and associated routines, and especially for fruitful and interesting discussions. 
We acknowledge useful discussions with Sanjana Curtis, Albino Perego, Kevin Ebinger, and Friedel Thielemann.
This work was supported by United States Department of Energy, Office of Science, Office of  Nuclear Physics (award number DE-FG02-02ER41216). 
N.E.W. acknowledges support from the Park Scholarship Program at NC State University.

We are grateful to the countless developers contributing to software
projects on which we relied in this work, including Agile \citep{Liebendoerfer.Agile}, CFNET \citep{cf06a}, 
Python \citep{rossumPythonWhitePaper}, 
numpy and scipy\citep{numpy,scipyLib}, and 
Matplotlib \citep{matplotlib}. 
This research made extensive use of the SAO/NASA Astrophysics Data System (ADS).
\end{acknowledgments}

\software{Agile \citep{Liebendoerfer.Agile}, %
          CFNET \citep{cf06a}, %
          Matplotlib \citep{matplotlib}
          }

\clearpage
\bibliography{references}

\appendix

\section{Machine-readable data tables: Remnant masses} \label{app:remnantmasses}

\begin{deluxetable*}{llllll}    
\tablecaption{
    Remnant masses (NSs and BHs) for simulations with the SFHo EOS and for simulations with the DD2 EOS.
    \label{tab:remnantmasses}
}
\tablewidth{0pt}
\tablehead{
    \colhead{Model} & 
    \colhead{EOS} & 
    \colhead{$m_{\mathrm{NS}}$} & 
    \colhead{$m_{\mathrm{BH}}^{\mathrm{CO}}$} & 
    \colhead{$m_{\mathrm{BH}}^{\mathrm{He}}$} & 
    \colhead{$m_{\mathrm{BH}}^{\mathrm{collapse}}$} \\ 
    %
    %
    \colhead{} &
    \colhead{(\msun)} &
    \colhead{} &
    \colhead{(\msun)} &
    \colhead{(\msun)} &
    \colhead{(\msun)} \\
}
\startdata
    s10.8 &
    SFHo &
    1.49 &
    - &
    - &
    - \\
    s11.0 &
    SFHo &
    1.41 &
    - &
    - &
    - \\
    ... &
    ... &
    ... &
    ... &
    ... &
    ... \\
    s40.0 &
    SFHo &
    1.81 &
    - &
    - &
    - \\
    u11.0 &
    SFHo &
    1.49 &
    - &
    - &
    - \\
    ... &
    ... &
    ... &
    ... &
    ... &
    ... \\
    u40.0 &
    SFHo &
    - &
    13.36 &
    15.29 &
    39.96 \\
    z11.0 &
    SFHo &
    1.49 &
    - &
    - &
    - \\
    ... &
    ... &
    ... &
    ... &
    ... &
    ... \\
    z40.0 &
    SFHo &
    - &
    13.85 &
    16.51 &
    40.00 \\
    s10.8 &
    DD2 &
    1.50 &
    - &
    - &
    - \\
    ... &
    ... &
    ... &
    ... &
    ... &
    ... \\
    s40.0 &
    DD2 &
    1.82 &
    - &
    - &
    -  \\
    u11.0 &
    DD2 &
    1.50 &
    - &
    - &
    - \\
    ... &
    ... &
    ... &
    ... &
    ... &
    ... \\
    u40.0 &
    DD2 &
    - &
    13.36   &
    15.29 &
    39.96 \\
    z11.0 &
    DD2 &
    1.49 &
    - &
    - &
    - \\
    ... &
    ... &
    ... &
    ... &
    ... &
    ... \\
    z40.0 &
    DD2 &
    - &
    13.85 &
    16.51 &
    40.00 \\
\enddata
\tablecomments{
    For models with successful explosions, the mass of the resulting NS is listed. For models which failed to explode, the mass of the BH for all three cases is given. 
    The results with the SFHo EOS are from this work. The results with the DD2 EOS model are taken from \paperII~and \paperIV.
}
%
\end{deluxetable*}


\end{document}